\documentclass[a4paper,11pt]{article}
\pdfoutput=1 
\usepackage{jcappub}
\usepackage{amsmath}
\usepackage{mathrsfs} 
\usepackage{amsfonts}
\usepackage{amssymb}
\usepackage{amsthm}
\usepackage{mathtools}
\usepackage{graphicx}
\usepackage{latexsym}
\usepackage{xspace}
\usepackage{hyperref} 
\usepackage{bm}
\usepackage{relsize}

\usepackage{tabularx}
\usepackage{multirow}
\usepackage{amssymb}
\usepackage{lscape}
\usepackage[most]{tcolorbox}
\usepackage{braket}
\usepackage{comment}
\usepackage[utf8]{inputenc}
\usepackage[english]{babel}
\usepackage{slashed}
\usepackage{empheq}
\usepackage[makeroom]{cancel}
\newcolumntype{P}[1]{>{\centering\arraybackslash}p{#1}}

\newcommand{\bfx}{{\mathbf{x}}}
\newcommand{\bfk}{{\mathbf{k}}}

\newcommand{\e}{\epsilon}

\usepackage{floatrow}
\usepackage{graphicx}
\usepackage[label font=bf,labelformat=simple]{subfig}
\usepackage{upgreek}

\usepackage[margin=1in,includefoot]{geometry}
\usepackage{xfrac} 
\usepackage{tikz-feynman}
\usepackage{tikz}	
\tikzfeynmanset{compat=1.0.0}

\allowdisplaybreaks

\makeatletter
\gdef\@fpheader{}
\g@addto@macro\bfseries{\boldmath}
\makeatother

\makeatletter
\DeclareRobustCommand*\uell{{\mathpalette\@uell\relax}}
\newcommand*\@uell[2]{
\setbox0=\hbox{#1\ell#1\ell}
\setbox1=\hbox{\rotatebox{10}{#1\ell#1\ell}}
\dimen0=\wd0 \advance\dimen0 by -\wd1 \divide\dimen0 by 2
\mathord{\lower 0.1ex \hbox{\kern\dimen0\unhbox1\kern\dimen0}}
}

\def\pia{\pi_{a}}
\def\pir{\pi_{r}}
\def\dpir{\dot{\pi}_{r}}
\def\dpia{\dot{\pi}_{a}}

\def\phia{\phi_{a}}
\def\phir{\phi_{r}}

\def\gr{g^{00}}
\def\ga{a^{00}}

\def\bfk{\textbf{k}}

\renewcommand{\O}{\mathcal{O}}

\usepackage{colortbl}
\definecolor{summersky}{cmyk}{0.71,0.33,0,0.5}
\definecolor{flamingo}{cmyk}{0,0.51,0.71,0.5}
\definecolor{rp}{cmyk}{0.2, 1, 0.6, 0}
\definecolor{pacificblue}{cmyk}{0.95,0.3,0, 0.5}
\definecolor{gray60}{cmyk}{0.4,0.4,0,0.8}

\newcommand{\ex}[1]{\langle #1 \rangle}

\renewcommand{\O}{\mathcal{O}}

\newcommand{\Mpl}{M_{\mathrm Pl}}
\newcommand{\cs}{c_{s}^{2}}

\newcommand{\stuck}{St\"uckelberg }

\newcommand{\gtt}[1]{\gamma^{tt}_{#1, \ell}}
\newcommand{\gts}[1]{\gamma^{ts}_{#1, \ell}}
\newcommand{\gss}[1]{\gamma^{ss}_{#1, \ell}}

\newcommand{\gat}[1]{\gamma^{at}_{#1, \ell}}

\newcommand{\gtto}[1]{\gamma^{tt}_{#1, 0}}

\newcommand{\gtso}[1]{\gamma^{ts}_{#1, 0}}
\newcommand{\gsso}[1]{\gamma^{ss}_{#1, 0}}

\newcommand{\gato}[1]{\gamma^{at}_{#1, 0}}

\newcommand{\gtti}[1]{\gamma^{tt}_{#1, 1}}

\newcommand{\gssi}[1]{\gamma^{ss}_{#1, 1}}

\newcommand{\gssii}[1]{\gamma^{ss}_{#1, 2}}

\newcommand{\bt}[1]{\beta_{#1,\ell}}

\newcommand{\btt}[1]{\beta^{tt}_{#1, \ell}}
\newcommand{\bts}[1]{\beta^{ts}_{#1, \ell}}
\newcommand{\bss}[1]{\beta^{ss}_{#1, \ell}}

\newcommand{\ie}{\textsl{i.e.}~}

\newcommand{\eg}{\textsl{e.g.}\xspace}

\newcommand{\dd}{\mathrm{d}}
\newcommand{\ee}{e}
\newcommand{\po}{\mathrm{P.O.}}

\newcommand{\bmx}{\boldmathsymbol{x}}

\newcommand{\Ima}{\Im \mathrm{m}\,}

\newcommand{\beq}{\begin{equation}}
	\newcommand{\eeq}{\end{equation}}
\newcommand{\bea}{\begin{equation}\begin{aligned}}
		\newcommand{\eea}{\end{aligned}\end{equation}}

\newlength{\wsingfig}
\newlength{\wdblefig}
\newlength{\wquadfig}
\newlength{\wtriplefig}
\newcommand{\Eq}[1]{Eq.~(\ref{#1})}
\newcommand{\Eqs}[1]{Eqs.~(\ref{#1})}
\newcommand{\Fig}[1]{Fig.~{\ref{#1}}}

\newcommand{\Sec}[1]{Sec.~\ref{#1}}

\newcommand{\App}[1]{Appendix~\ref{#1}}

\usepackage{dsfont}

\def\bmx{{\boldsymbol{x}}}

\def\bmk{{\boldsymbol{k}}}

\subheader{}

\title{An Open System Approach to Gravity}

\author[a]{Santiago Ag\"u\'i Salcedo,}
\author[a]{Thomas Colas,}
\author[a]{Lennard Dufner,}
\author[a]{and Enrico Pajer}

\affiliation[a]{Department of Applied Mathematics and Theoretical Physics, University of Cambridge, Wilberforce Road, Cambridge, CB3 0WA, UK}
\emailAdd{sa2013@cam.ac.uk}
\emailAdd{tc683@cam.ac.uk}
\emailAdd{ld689@cam.ac.uk}
\emailAdd{enrico.pajer@gmail.com}

\begin{document}
\sloppy

\abstract{
Several major open problems in cosmology — including the nature of inflation, dark matter, and dark energy — share a common structure: they involve spacetime-filling media with unknown microphysics, and can be probed so far only through their gravitational effects. This observation motivates a systematic open-system approach to cosmology, in which gravity evolves in the presence of a generic, unobservable environment. In this work, we develop a general framework for open gravitational dynamics based on general relativity and the Schwinger-Keldysh formalism, carefully addressing the nontrivial constraints imposed by diffeomorphism invariance. At the quantum level, our path integral formulation computes the gravitational density matrix in perturbation theory around a semi-classical spacetime.

As illustrative applications, we study inflation and the propagation of gravitational waves in classical regimes where environmental interactions are non-negligible. In the inflationary context, our framework reproduces the known Open Effective Field Theory of Inflation in the decoupling limit and extends it to include gravitational interactions.  
For gravitational waves, we derive the most general conservative and dissipative corrections to propagation. 
Remarkably, we find that the leading-order gravitational birefringence is dissipative in nature, whereas conservative birefringence appears only at higher derivative order — opposite to the electromagnetic case. Our results pave the way to modeling dissipative effects in the late universe.

} 

\maketitle


\section{Introduction}

The Lambda Cold Dark Matter ($\Lambda$CDM) model, supplemented by the inflationary generation of primordial perturbations, has established itself as the standard model of cosmology. In the nearly 30 years since the discovery of the universe's accelerated expansion, this model has been repeatedly tested by increasingly precise and expansive cosmological datasets. While many so-called ``tensions" have emerged over the years, it is fair to say that $\Lambda$CDM still provides an excellent fit to observations~\cite{ACT:2025fju,ACT:2025tim,SPT-3G:2025bzu,Wright:2025xka, DESI:2025zgx}, and the evidence supporting it remains at least as strong as that for the best competing models involving new physics.

Nevertheless, the phenomenological success of $\Lambda$CDM stands in stark contrast to its theoretical foundations. The essential ingredients of the model — dark matter, dark energy, and inflation — remain poorly understood. For inflation, we lack knowledge of how many degrees of freedom are involved, how they interact, and what symmetries they obey. One may also question to what extent inflation is necessarily described by local quantum field theory, as opposed to more speculative frameworks such as string theory. For dark matter, we often assume the existence of a weakly interacting particle, possibly composite, to explain the data, but even its mass is uncertain by at least 80 orders of magnitude, underscoring the limitations imposed by the universality of hydrodynamics. Dark energy is arguably even more puzzling due to the old and new cosmological constant problems, as well as the coincidence problem.

We revisited these well-known open problems\footnote{In passing, it is worth mentioning that we also don't have a detailed understanding of the period of reheating, which connects the inflationary epoch to the hot big bang, and baryogenesis, from which the current baryonic content of the universe should emerge.} to underscore three features inflation, dark matter and dark energy all share:
\begin{itemize}
    \item They involve spacetime-filling media;
    \item The microscopic physics of these media is completely unknown;
    \item They can only be probed gravitationally, at least until now.
\end{itemize}
We will now argue that the three features above have profound implications for how we should tackle open problems in cosmology.


\paragraph{The medium: particle physics, cosmology and condensed matter.}

Let us begin with the first commonality. The central open questions in cosmology pertain to substances — or perhaps entire ``dark sectors" — that fill spacetime and dominate the energy content of the universe during different epochs. ``Inflation" and ``dark energy" are merely labels for whatever is responsible for generating primordial perturbations and driving late-time acceleration.

\begin{figure}[tbp]
    \centering
    \includegraphics[width=\linewidth]{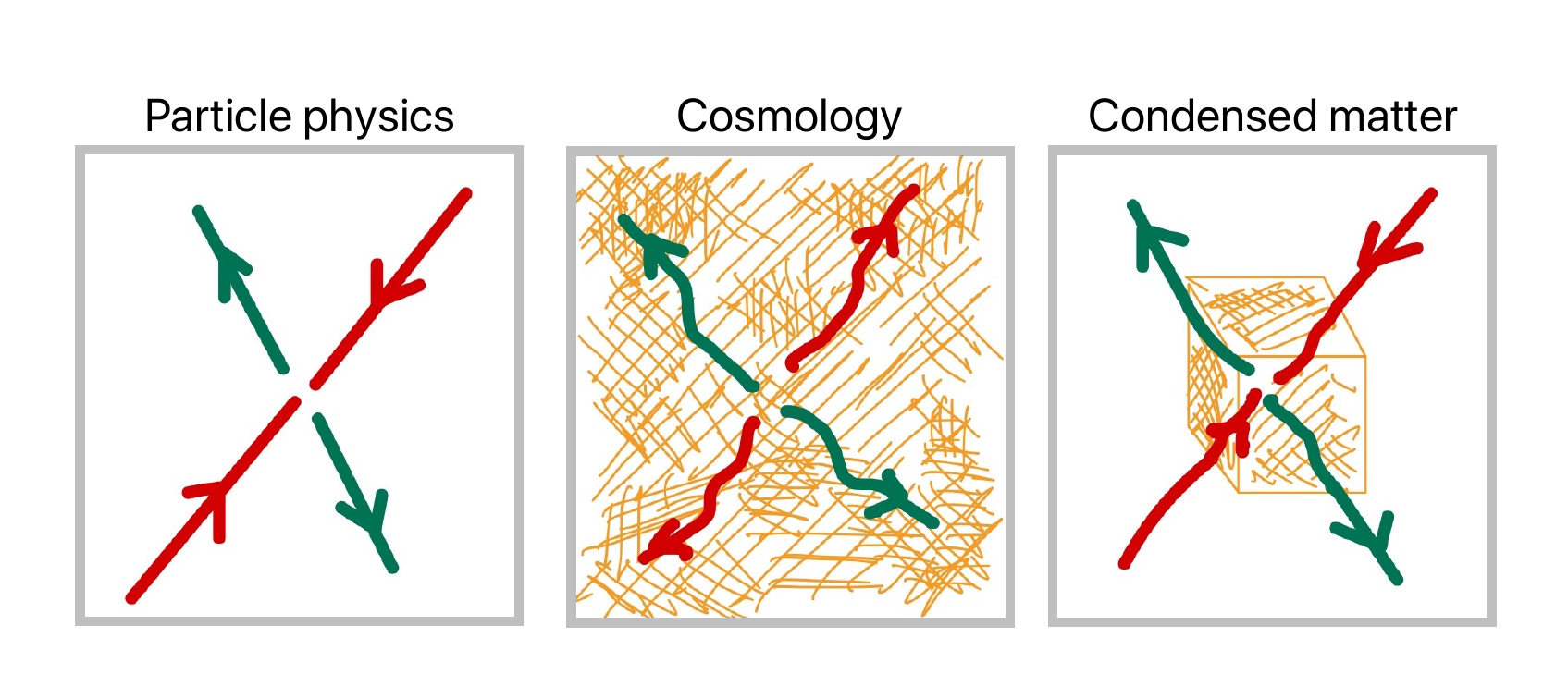}  
    \caption{Some distinctions between particle physics, cosmology and condensed matter. \textit{Left:} Particle physics studies particles and their interactions atop a of clean and isolated vacuum. \textit{Middle:} Cosmology reconstructs the properties of the universe by analyzing perturbations that propagate through an unknown medium to our detectors. \textit{Right}: Condensed matter creates and probes engineered media, which can be observed externally and under controlled conditions. 
    }
    \label{fig:boxes}
\end{figure}

This is in stark contrast to the paradigm of particle physics. There, experimentalists spend decades engineering detectors such that the initial quantum state — the vacuum prior to a scattering event — is as clean and isolated as possible. The focus is not on the vacuum itself, but on the particles and their interactions atop it. This situation is illustrated in the left panel of Fig.~\ref{fig:boxes}.

In cosmology, however, we do not prepare the state of the system. This is not only because we cannot manipulate the entire universe, but also because cosmology is fundamentally the study of the universe's state over time. If we were able to prepare the ``system universe" in a desired state, we would forfeit the possibility of studying its cosmology. Here, the goal is rather to understand what the universe is made of — a setup shown in the central panel of Fig.~\ref{fig:boxes}.

For example, during inflation, the perturbations that seed large scale structures are small local changes in the state of the spacetime-filling inflaton. More precisely, they are small advances and delays in the clock whose ticks measure the duration of the inflationary phase. As such, perturbations do not exist in isolation, but only in relationship to the underlying clock itself. In this regard, cosmology bears a closer resemblance to condensed matter physics than to particle physics, which also studies properties and dynamics of media. One sociological observation motivating this work is that cosmologists are often trained as high-energy theorists rather than as condensed matter physicists — a fact that may limit our conceptual toolkit. Yet cosmology differs from condensed matter physics too in at least two ways: we do not engineer the medium we want to study, and we observe from within it, not from without. This is depicted in the right panel of Fig.~\ref{fig:boxes}.

These considerations motivate our development of an \textit{open-system} approach to cosmology — one that treats the gravitational sector as evolving in the presence of a generic, unobservable medium. In a closed system approach, we attempt to describe all constituents of a given \textit{full system}. This is theoretically the best we can do. However, in practice this limits our investigation to a small class of simple, calculable models. Conversely, in an open system approach, we separate the full system into \textit{system} and \textit{environment} and we only attempt to describe the former. By acknowledging the presence of an environment, whose details are not of interest, we achieve two goals: we focus on the relevant variables of the system and we manage to model the system's phenomena without the burden to develop the detailed dynamics of the environment that accompany them. Giving up control of the full system is a big departure from the traditional particle physics perspective and comes at a cost. Energy and other charges may appear not to be conserved because we can only track them within the system but not the environment. Likewise, information is inevitably lost and the flow of time becomes irreversible. Nevertheless, we argue that this change of paradigm is worthy of consideration because it allows us to describe a whole range of \textit{qualitatively new phenomena} such as dissipation, noise, and, at the quantum level, decoherence.

%


\paragraph{Unknown microphysics.}

The second shared feature of open problems in cosmology is the lack of a known microscopic description. This is in contrast to condensed matter systems, where the microscopic theory (e.g., electrons and nuclei) is known but often intractable. In cosmology, we lack even that starting point. Although a vast array of models has been proposed — spanning scalar fields, modified gravity, and exotic particles — most rely on a handful of weakly coupled degrees of freedom. This is not because such simplicity is well-motivated, but because tractable models dominate the discourse. By deriving our predictions exclusively from these simple models, we are likely only scratching the surface of the possible phenomenology — even within the domain of local field theories.

This motivates a shift of focus: in addition to speculating about possible high-energy completions, we should also adopt the perspective of a material scientist and characterize the macroscopic properties of the unknown cosmological media. This includes questions like: What is their most general effect on the dynamics of gravity? Do they dissipate? What perturbations can they in principle support? 

Effective field theory (EFT) provides a natural framework for this approach: it allows one to systematically parametrize the behavior of a system in a given regime, without detailed knowledge of its microscopic constituents. Here, we construct an EFT for cosmological media interacting with gravity as an open system. This gives both a direct handle on computing observables, which are gravitational, and a way to constrain or match candidate microphysical models. The use of EFT techniques in cosmology has consistently grown over the past two decades, and the well-known arguments for their usefulness also applies to our case. We are certainly not the first to develop EFTs with an open system approach. In fact, the EFT of Large-Scale Structure \cite{Baumann:2010tm,Carrasco:2012cv} is a beautiful example of an open system EFT. Our contribution here is to include dynamical gravity in the relativistic regime into the description, which is essential for studying inflation, dark energy and the large scale dynamics of dark matter. Moreover, we prefer to work at the level of an action principle, or more precisely an open functional, as opposed to the equations of motion, as done in the mainstream EFT of large scale structures (see however \cite{Blas:2015qsi,Blas:2016sfa}). This functional formalism provides several distinctive advantages: (i) conserved charges can be derived from symmetries, as for example in Noether's theorem; (ii) a classical theory can be directly extended to a quantum theory via the path integral; (iii) the rich physics associated with total derivatives and topology can be captured; (iv) a straightforward coupling to gravity can be achieved via minimal coupling; (v) it's a natural space for renormalization and the formulation of dualities. 


\paragraph{Gravitational observables.}

As mentioned above, up to this date, all our knowledge of inflation, dark matter, and dark energy is ultimately associated with gravitational observables. For inflation, the key observables are curvature perturbations, which in comoving gauge are exactly a fluctuation of the spatial metric. For dark energy, we observe the expansion history — i.e., the zero-momentum mode of the metric — through standard candles and rulers. For dark matter, we track structure formation, which is encoded in the metric’s Newtonian potentials.

Hence, a theory of dynamical gravity in a medium is necessary and sufficient to compute the relevant cosmological observables. An increasingly important probe for cosmology is gravitational waves. For example, the detection of a binary neutron star merger with an electromagnetic counterpart has tightly constrained the speed of gravity~\cite{LIGOScientific:2017vwq}, with profound implications for modified gravity and dynamical dark energy~\cite{Creminelli:2017sry,Creminelli:2018xsv,Creminelli:2019kjy,deRham:2018red}. Our perspective is that gravitational waves play a role analogous to light in a medium: their propagation through the dark sector can reveal the medium’s properties. Just as light in a material may experience dispersion, dissipation and birefringence, gravitational waves may exhibit similar effects in the presence of unknown cosmological media.

Previous work~\cite{Flauger:2017ged,Flauger:2019cam} studied this idea for cold dark matter modeled as a non-interacting gas. Our approach complement this line of work by remaining agnostic about the nature of the medium, asking instead: what corrections to gravitational wave propagation are allowed by symmetry, locality, and causality in a generic medium? We anticipate that we will find an interesting answer to this question in the phenomenon of dissipative birefringence. 


\paragraph{An open effective field theory for gravity.}

When combined, the three common features we just discussed point towards a unifying framework: an \textit{open effective field theory of gravity in a medium}. To this end, we build upon tools from the theory of open quantum systems~\cite{Hu:1991di,breuerTheoryOpenQuantum2002,Boyanovsky:2015xoa,2016RPPh...79i6001S} and nonequilibrium physics~\cite{Calzetta:1986cq, Berges:2004yj, Calzetta:2008iqa, kamenev_2011, Berges:2012ty, Aron:2017spi, Liu:2018kfw, Radovskaya:2020lns, Rosenhaus:2025bgy}, rooted in the seminal work of Schwinger, Keldysh, Feynman and Vernon~\cite{FEYNMAN1963118, Schwinger:1960qe, Keldysh:1964ud, Galley:2012hx}. These tools have already been instrumental in developing a systematic study of dissipation and noise in cosmological environments \cite{Calzetta:1995ys, Koks:1996ga, Lombardo:2004fr, Lombardo:2005iz, Burgess:2006jn, Burgess:2014eoa, Burgess:2015ajz, Boyanovsky:2015jen, Boyanovsky:2015tba, Nelson:2016kjm, Liu:2016aaf, Hollowood:2017bil, Shandera:2017qkg, Martin:2018zbe, Kaplanek:2019dqu, Brahma:2020zpk, Burrage:2018pyg, Colas:2021llj, Zarei:2021dpb, Sou:2022nsd, Brahma:2022yxu, Colas:2022hlq, Colas:2022kfu, DaddiHammou:2022itk, Burgess:2022nwu, Burgess:2022rdo, Colas:2023wxa, Sharifian:2023jem, Ning:2023ybc, Burgess:2024eng, Colas:2024xjy, Bowen:2024emo, Tinwala:2024wod, Brahma:2024yor, deKruijf:2024ufs, Salcedo:2024nex, Colas:2024lse, Colas:2024ysu, Burgess:2024heo, Lau:2024mqm, Launay:2024trh, Brax:2024obu, Pueyo:2024twm, Burrage:2025xac, Bhattacharya:2025fxw, Takeda:2025cye, Wang:2025hlz, Kading:2025cwg, DuasoPueyo:2025lmq, Lopez:2025arw, Sano:2025ird, Kranas:2025jgm, Mahajan:2025iuz, Li:2025azq}, and our work builds upon these earlier results. Parallel developments in related fields to account for environmental effects such as dissipation have taken place in dissipative hydrodynamics \cite{Endlich:2012vt, Crossley:2015evo, Glorioso:2016gsa, Glorioso:2017fpd, Liu:2018crr, Hongo:2018ant, Chen-Lin:2018kfl, Landry:2019iel, Hongo:2019qhi, Armas:2020mpr, Landry:2021kko, Baggioli:2023tlc, Vardhan:2024qdi, Grosvenor:2024vcn, Huang:2024rml, Ota:2024yws, Ota:2024mps, Akyuz:2025bco}, black hole physics \cite{Danielson:2022tdw, Danielson:2022sga, Biggs:2024dgp, Danielson:2024yru, Danielson:2025iud} and holography \cite{Haehl:2016pec, Ghosh:2020lel, Geng:2023ynk, Haehl:2024pqu, Anninos:2024wpy, Liu:2024tqe}.

This work is the third in a series of papers \cite{Salcedo:2024smn,Salcedo:2024nex} devoted to develop open EFTs for cosmology within the Schwinger-Keldysh formalism. In our first work \cite{Salcedo:2024smn}, we studied a scalar field theory in de Sitter spacetime to capture the decoupling limit during inflation. The novelty compared to the traditional calculation of primordial perturbations is that here they are sourced by the environment noise as opposed to the quantum fluctuations of the ground state, a possibility proposed long ago \cite{Berera:1995ie}. Because of this, this class of open models predicts primordial non-Gaussianity in the bispectrum of a very specific shape. Depending on the amount of dissipation, the signal in the bispectrum peaks either on equilateral triangles, as for self-interactions of the inflation, or on so-called ``folded" triangles with vanishing area. This is expected for models where the ground state contains physical particles \cite{Chen:2006nt,Holman:2007na,Green:2020whw}. One of the goals of this work is to develop the full theory of scalar fluctuations coupled to a dynamical metric and test under what conditions the decoupling assumption hold (see \cite{LopezNacir:2011kk} for an earlier study of this setup within effective field theories). To include gravity in an open system using the Schwinger-Keldysh formalism, we encountered several obstacles (see Sec. \ref{sec:OpenEFT}). To make progress we developed a theory of light in a medium \cite{Salcedo:2024nex}, namely open electromagnetism, and carefully studied the role of Abelian gauge transformations in the double branch path integral. This gave us as a blueprint for describing gravity and the associated non-Abelian diffeomorphism group. 


\paragraph{Outline.} 

In \Sec{sec:OpenEFT}, {after reviewing the basics of the Schwinger-Keldysh formalism, we present the extension of the formalism to gauge theories, following \cite{Salcedo:2024nex}}. In \Sec{sec:unitgauge}, we define a notion of unitary gauge in the Schwinger-Keldysh contour. Following a construction analogous to the effective field theory of inflation, we exploit the unitary gauge formulation to write down the most generic theory invariant under 3d retarded spatial diffs. The key difference is that now the theory incorporates dissipation and noise, arising from interactions with an unknown environment. The background equations are studied in \Sec{Sec:Background}, where we find modifications to the Friedmann equations and continuity equation typical of dissipative and viscous cosmology. The dynamics of gravitational waves during inflation is studied in \Sec{sec:tensor}, where we derive the tensor-to-scalar ratio for this class of theories. As expected, the amplitude of primordial tensor modes does not fix the scale of inflation. At last, performing St\"uckelberg tricks in \Sec{sec:Stuck}, we recover the decoupling limit results from \cite{Salcedo:2024smn} and assess the mixing with gravity.  Concluding remarks are gathered in \Sec{sec:conclu}, followed by a series of technical appendices.

\paragraph{Reader's guide.}

\Sec{sec:OpenEFT} {may be skipped by readers familiar with gauge symmetries in the Schwinger-Keldysh formalism}. Readers interested in formal developments of Open EFTs in cosmology may focus on \Sec{sec:unitgauge}, which develops the tools to describe dynamical gravity in the presence of dissipation and noise, and on \Sec{sec:Stuck}, which investigates the mixing with gravity and the decoupling of the Pseudo-Goldstone boson. Readers interested in the phenomenological aspects for cosmology may opt for \Sec{Sec:Background}, where we discuss modifications to the Friedmann and continuity equations. Finally, readers interested in gravitational waves should focus on \Sec{sec:tensor}, where we discuss the general dissipative and conservative effects on propagation and derive explicit solutions in de Sitter spacetime as a concrete example. 

\paragraph{Notation and conventions.} We use mostly plus signature. We use the short-hand notation 
\begin{align}
    A^{(\mu\nu)}= \frac{1}{2} \left(A^{\mu\nu} + A^{\nu\mu}  \right), \qquad A^{[\mu\nu]}= A^{\mu\nu} - A^{\nu\mu} .
\end{align}
The trace of rank-2 tensors taken with the retarded metric $g_{\mu\nu}$ is indicated with no indices: $K = K^\mu_{~\mu} = g_{\mu\nu}K^{\mu\nu}$.


\subsection{Summary of the main results}

For the convenience of the reader, we summarize in the following our main results.

\paragraph{A heuristic approach to open gravity.}

Before working out the details of our theory, we would like to set expectations with a less precise but more intuitive qualitative discussion. In the full ``closed" theory, where one knows the details of the environment and the evolution is unitary and conservative (i.e. for ``closed'' dynamics), the Einstein equations should take the usual form
\begin{align}
    \Mpl^2 {G}_{\mu\nu} = {T}^{(\text{all})}_{\mu\nu}  \qquad \Rightarrow \qquad {\nabla}^\mu  {G}_{\mu\nu}  = 0 = {\nabla}^\mu  {T}^{(\text{all})}_{\mu\nu}, 
\end{align}
where ${T}^{(\text{all})}_{\mu\nu} $ is the full energy-momentum tensor of the theory. In our open system approach, we separate ${T}^{(\text{all})}_{\mu\nu} $ into an environment ${T}^{(\text{end})}_{\mu\nu} $ and a system ${T}^{(\text{sys})}_{\mu\nu} $. In the single-clock case we focus on in this paper, the system would be the sector housing the single-clock, whose dynamics can be described by our open EFT. Conversely, all the other degrees of freedom in the environment are assumed to be non-dynamical in the regime we are working, and have hence been integrated out. This would be true if there was a hierarchy of scales between the characteristic time and length scales we want to study and those of the environment. 
As a consequence of this, ${T}^{(\text{all})}_{\mu\nu} $ should be substituted with its expectation value, which in turn can be written in terms of ``light" degrees of freedom, namely the metric, and the single clock. As a result, Einstein's equations take the schematic form
\begin{align}\label{eq:mainidea}
    \Mpl^2 {G}_{\mu\nu} - \ex{{T}^{(\mathrm{env})}_{\mu\nu}}= {T}^{(\text{sys})}_{\mu\nu}  \qquad \Rightarrow \qquad  \Mpl^2 {G}_{\mu\nu} +\text{modifications}= {T}^{(\text{sys})}_{\mu\nu} \,,
\end{align}
where the ``modifications'' are terms built out of the metric and its derivatives that are fixed by diff-invariance and in terms of the couplings appearing in our $M_{\mu\nu}$ (see \eqref{eq:S1v1} for an explicit expression). Since the Einstein tensor obeys the contracted Bianchi identities, $\nabla^\mu G_{\mu\nu} = 0$, but the modifications in general do not, one finds the surprising result that $ \nabla^\mu {T}^{(\text{sys})}_{\mu\nu}\neq 0$.
This apparent non-conservation of ${T}^{(\text{sys})}_{\mu\nu}$, implied by our modified Einstein equations, is actually simply the statement that the \textit{full} energy-momentum tensor is conserved
\begin{align}
    \nabla^\mu{T}^{(\text{sys})}_{\mu\nu}=-\nabla^\mu{T}^{(\text{env})}_{\mu\nu}\,.
\end{align}
This is what one generically expects when the dynamics of gauge bosons, such as the graviton or the photon, is modified by the presence of an environment. Indeed, this is exactly the same structure we encounter for electromagnetism in a medium, see \eqref{eq:noiseconstEM}. {A hierarchy of scales of this kind can naturally arise in dark-energy scenarios. In models such as quintessence, the dynamics of the dark-energy sector occurs at the Hubble scale, while additional fields or sectors with parametrically faster dynamics, e.g. massive fields with $m \gg H_0$ or rapidly relaxing degrees of freedom — may act as an environment. When such a separation of scales is present, these fast modes can be consistently integrated out, yielding a local open EFT for the dark-energy sector with dissipative corrections encoded in the modified Einstein equations above.}

To familiarize ourselves with this rich theory, it is interesting to compute the modified Einstein equations assuming homogeneity and isotropy. Direct calculation (see \eqref{bkgdsummary}) shows that the modified Friedmann equations read 
\begin{align}
 3\Mpl^{2}H^2 &= \alpha_{1}+\alpha_{2}H \,, &
   2\Mpl^{2}\dot{H} &= \alpha_{3}+\alpha_{4}H\,,
\end{align}
where $H$ is the Hubble parameter and the $\alpha$'s are linear combinations of EFT coefficients, which are in general time dependent. For $\alpha_1=\rho$, $\alpha_3=-(\rho+p)$ and $\alpha_2 = \alpha_4 = 0$, these reduce to the standard Friedmann equations. More generally, these describe the evolution induced in the presence of a dissipative matter sector. For example, $\alpha_4$ can be seen to be related to the fluid's bulk viscosity $\zeta$ by $\alpha_4=3\zeta$. Another example is $\alpha_2\neq 0$, which can capture the leakage of gravity into the 5-dimensional bulk in the Dvali-Gabadaze-Porrati model \cite{Dvali:2000hr}. To obtain a solution, the above equations have to be supplemented by three equations of state relating the four $\alpha$'s to each other. This additional information specifies the macroscopic properties of the environment. In the following, we summarize the more precise construction of our theory. 


\paragraph{The symmetry story.} Here we present an overview of the role that various symmetries play in the construction and analysis of a theory of open gravity. We remove all details for clarity and focus on the main steps. 

Let's begin with a reminder of the symmetry structure in the original Effective Field Theory of Inflation (EFToI) \cite{Cheung:2007st}. The starting point is the most generic theory of just the metric fields $g^{\mu\nu}$ that is invariant under (time-dependent) 3d spatial diffeomorphisms (3d-diffs). This \textit{defines} the theory in what is known as the ``unitary gauge", i.e. the choice of time coordinates in which the single scalar degree of freedom in the theory is eaten by the metric. Having defined the theory, we can choose any convenient way to analyse it. A clever way to proceed is to make the theory momentarily more complicated but eventually much simpler. One performs the St\"uckelberg trick, which simultaneously introduces a ``Goldstone boson" of time translations $\pi$ and enlarges the symmetry group to 4d diffs. The simplification arises when one notices that at high energies $\pi$ decouples from the dynamical metric and we can exclusively focus on its self interactions in a fixed, unperturbed spacetime. These steps can be schematically summarised by
\begin{align}
    S_\text{3-diff}[g_{\mu\nu}] \overset{\text{St\"uck.}}{\longrightarrow} S_\text{\text{4-diff}}[g_{\mu\nu},\pi] \overset{\text{decoup.}}{\longrightarrow} S[\pi]\,,
\end{align}
where the decoupled theory lives on a fixed background and hence enjoys no diff invariance.

To construct an \textit{open theory of gravity and a single clock}, we start with doubling the fields. This gives us two copy of diff invariance in each branch of the Schwinger-Keldysh path integral, which we denote as diff$_+$ and diff$_-$. It is convenient to work with diffs that act in the same (retarded) or opposite (advanced) direction in each branch, which we denote as diff$_r$ and diff$_a$ respectively. Since an open theory by definition includes couplings between fields in the $+$ and $-$ branch, the 4d advanced diffs are explicitly broken by dissipative effects. The 4d retarded diffs are broken to 3d spatial retarded diffs by the clock foliation. {
\begin{align}\label{eq:symintro}
   ( \text{4d-diff}_+ \times \text{4d-diff}_- )\overset{\text{open}}{\longrightarrow} (\text{4d-diff}_r ) \overset{\text{clock}}{\longrightarrow} (\text{3d-diff}_r)\,.
\end{align}
}
The open theory of gravity plus a single clock is defined in the unitary gauge as the most generic theory of an advanced and retarded metric plus the advanced clock field $t_a(t,\bfx)$ that is invariant under 3d-diff$_r$. The advanced clock $t_a(t,\bfx)$ encodes stochastic fluctuations of the foliation due to the presence of an unknown environment. It can be removed by a field-redefinition in single-clock cosmologies, but should be kept in more general models with many matter sectors, such as the late universe. 

Having defined our theory, we can choose any convenient way to analyse it. Because we envisage that some appropriately defined Goldstone boson decouples from the metric, we choose to perform \textit{two} St\"uckelberg tricks, which simultaneously introduce the new fields $\pi_r$ and $\pi_a$ and make the theory retarded and advanced time-diff invariant. We don't see any advantage in enforcing advanced 3d spatial diff invariance with additional St\"uckelberg transformations, so {3d-diff$_+ \times$ 3d-diff$_-$ remain explicitly broken to 3d-diff$_r$}. In summary, for single-clock cosmologies we have
\begin{align}
      S_{\text{3d-diff}_r}[g_{\mu\nu},a_{\mu\nu} ] \overset{\text{St\"uck.}^2}{\longrightarrow} S_{\text{3d-diff}_r \times {\text{t-diff}_+ \times \text{t-diff}_-}} [g_{\mu\nu},a_{\mu\nu}, \pi_r,\pi_a] \overset{\text{decoup.}}{\longrightarrow} S[\pi_r,\pi_a]\,,
\end{align}
where the last step corresponds to the Open Effective Field Theory of Inflation (Open EFToI) in the decoupling limit constructed and studied in \cite{Salcedo:2024nex}.

\paragraph{Open gravity.} Following the above steps we arrive at the open functional that defines our theory in the retarded unitary gauge in which the single clock is chosen to define a preferred foliation of spacetime. The classical and deterministic dynamics up to second order in derivatives is derived from (see \eqref{eq:S1v1})
\begin{align} 
    S_1 &= \int \dd^4 x \sqrt{-g} \, \sum_{\ell=0} \, \left(g^{00} + 1 \right)^\ell \bigg\{ \, a^{00} \Big[ \gtt{1} + \gtt{2} K + \gtt{3}  K^2  + \gtt{4} K_{\alpha\beta} K^{\alpha\beta} \nonumber \\
     &\quad+ \gtt{5} \nabla^0 K + \gtt{6} R + \gtt{7} R^{00}  \Big] \bigg.  + a^{0\mu} \Big[ \gts{1} R^{0}{}_\mu + \gts{2} \nabla_\mu K + \gts{3} \nabla_\beta K^{\beta}{}_\mu \Big] \bigg.\nonumber \\
     & \quad + a^{\mu\nu} \Big[ g_{\mu\nu} \Big( \gss{1} + \gss{2} K + \gss{3}  K^2 + \gss{4} K_{\alpha\beta} K^{\alpha\beta} + \gss{5} \nabla^0 K+ \gss{6} R + \gss{7} R^{00} \Big)  \bigg.\nonumber \\ 
    &\quad  + \gss{8} K_{\mu\nu} + \gss{9} \nabla^0 K_{\mu\nu} + \gss{10} K_{\mu\alpha} K^{\alpha}{}_{\nu} \nonumber \bigg. + \gss{11} K K_{\mu\nu} + \gss{12} R_{\mu\nu} + \gss{13} R_\mu{}^0{}_\nu{}^0 \\
    &\quad  + \gamma^{\po}_{1,\ell} \epsilon_\mu{}^{\alpha\beta0} \nabla_\alpha K_{\beta\nu} + \gamma^{\po}_{2,\ell} \epsilon_\mu{}^{\alpha\beta0} R_{\alpha\beta}{}^0{}_\nu \Big] + t_a \Big[\gat{1} + \gat{2} K + \gat{3}  K^2 \nonumber \\
    &\quad+ \gat{4} K_{\alpha\beta} K^{\alpha\beta} + \gat{5} \nabla^0 K + \gat{6} R + \gat{7} R^{00}
    \Big] + \dots\bigg\} ,
\end{align}
by varying with respect to the ``auxiliary" advanced metric $a^{\mu\nu}$. Here the many $\gamma$'s are arbitrary time-dependent effective couplings, $K_{\mu\nu}$ is the extrinsic curvature of the spatial hypersurfaces at constant time, and $R$, $R_{\mu\nu}$ and $R^{\mu}_{\,\nu\rho\sigma}$ are the familiar Ricci scalar, Ricci tensor and Riemann tensor, respectively. All these tensors are defined with respect to the retarded metric $g_{\mu\nu}$. The dots denote higher derivative terms that are expected to give subleading effects. A few field redefinitions can be performed to simplify the $\ell=0$ part of $S_1$ as detailed around \eqref{eq:S1fin}. 

{To recover a unitary (closed) theory from the open-system effective action, one may furthermore require that $S_1$ is invariant under
\begin{align}\label{eq:symnew}
   ( \text{4d-diff}_+ \times \text{4d-diff}_- )  \overset{\text{clock}}{\longrightarrow} (\text{3d-diff}_+ \times \text{3d-diff}_- )\,.
\end{align}
We demonstrate explicitly in \App{app:ope} that all operators in the universal part of the EFT of inflation \cite{Cheung:2007st} respect \Eq{eq:symnew} when expressed in the Schwinger-Keldysh contour. By contrast, operators that are inherently linked to the openness of the system, such as dissipative or noise-induced terms, necessarily break $\text{4d-diff}_+ \times \text{4d-diff}_-$ to its diagonal subgroup $\text{4d-diff}_r$ (and eventually to $\text{3d-diff}_r$, as in \Eq{eq:symintro}) and can thus be systematically identified through this non-invariance. Hence, the closed, unitary limit corresponds to sending to zero all EFT coefficients of operators that break the symmetry of \Eq{eq:symnew}.} 

The stochastic fluctuations of the unknown environment are determined by the functional $S_2$, which is now quadratic in the advanced fields $a^{\mu\nu}$ and $t^a$, 
\begin{align}\label{eq:S2intro}
    S_2 &= i \int \dd^4 x \sqrt{-g} \sum_{\ell=0} \left(g^{00} + 1 \right)^\ell \Big[ \bt{1} \left( t_a \right)^2 + \bt{2} a^{00} t_a + \bt{3} a^{\mu\nu} g_{\mu\nu} t_a + \beta_{4,\ell} \left( a^{00} \right)^2 \nonumber \bigg. \\
    &\quad + \beta_{5,\ell} (a^{\mu\nu} g_{\mu\nu})^2 + \beta_{6,\ell} a^{\mu\nu} g_{\mu\rho} g_{\nu\sigma} a^{\rho\sigma} 
    + \beta_{7,\ell} a^{\mu\nu} g_{\mu\nu} a^{00} + \beta_{8,\ell} a^{0\mu} a^{0\nu} g_{\mu\nu} \Big] . \bigg. 
\end{align}
Here, the many $\beta$'s are arbitrary time-dependent effective couplings and we limited ourselves to zeroth order in derivatives, but higher order terms are straightforward to write down. The noise contributions to the equations of motion (EOMs) are obtained by performing a Hubbard-Stratonovich trick \cite{Hubbard:1959ub,Stratonovich1957} to make $S_2$ linear in $a^{\mu\nu}$ and then taking the variation with respect to $a^{\mu\nu}$. 

\paragraph{The general propagation of gravitational waves.} Our theory has the same degrees of freedom as general relativity plus a scalar field. If we linearized our theory of open gravity and focus on the dynamics of the transverse traceless component of the metric, we find a modified equation for the propagation of gravitational waves, (see \eqref{eq:S_2_coefficients})
\begin{align}
    \ddot{h}_{ij} - c_T^2 \frac{\nabla^2}{a^2} h_{ij} +\left(\Gamma_T + 3 H\right) \dot{h}_{ij}  + \frac{\chi}{a} \Tilde{\epsilon}_{imn} \partial_m \dot{h}_{nj}  = \xi_{ij}.
\end{align}
This contains as a special case the familiar ``conservative" evolution equation of gravitational waves, which in a medium can take place at a generic speed $c_T$. To this lowest order in derivatives, we find three dissipative effects: an arbitrary dissipation term $\Gamma_T \dot h_{ij}$, which is independent of the Hubble parameter, a dissipative birefringence term\footnote{Note that our dissipative birefringent term already appeared in the closed-system model studied in \cite{Alexander:2004wk}, although there, this two-derivative term was accompanied by a three-derivative term, because of the structure of the non-dissipative theory. Moreover, our dissipative birefringence survives in the flat space limit.}  proportional to the (flat-space) Levi-Civita symbol $\tilde \e_{imn}$ of size determined by $\chi$, and finally the presence of a stochastic source $\xi_{ij}$, whose amplitude is parameterized by the EFT coefficient $\beta_{6,0}$ in \eqref{eq:S2intro}. In the simplest setup, this noise is white and Gaussian, but our formalism can accommodate arbitrary colored and non-Gaussian noises. 

In de Sitter spacetime we have computed the retarded/advanced and Keldysh propagators for the graviton. Setting birefringence to zero, we have found the primordial tensor power spectrum (see \eqref{eq:gws_ps_Gamma})
\begin{equation}
    \Delta_h^2(k) =  \frac{4 \beta_{6,0}}{\Mpl^4} 2^{2\nu_\Gamma} \frac{\Gamma(\nu_\Gamma-1 ) \Gamma(\nu_\Gamma)^2}{\Gamma(\nu_\Gamma-\tfrac{1}{2} )\Gamma(2\nu_\Gamma-\tfrac{1}{2})}\,,
\end{equation}
where $ \nu_\Gamma \equiv \frac{3}{2} + \frac{\Gamma_T}{2H} $. This can be large or small, both in the large, $\Gamma_T \gg H$, or small, $\Gamma \ll H$, dissipation regimes, as shown in \Fig{fig:gw_ps_nobf}. Our finding hence suggest that the tensor power spectrum is not necessarily small in models with environmental effect, although it may certainly be. This should be contrasted with the results in several concrete models of warm inflation, e.g. \cite{Berera:1995ie,Berera:2008ar}, where the tensor-to-scalar ratio is typically found to be small. In contrast to warm inflation, where the environment is typically assumed to be in thermal equilibrium, in our setup we don't make this more restrictive assumption. This implies that our dissipation and noise are unrelated and hence we don't find any obstruction to having a noise with a large amplitude. In the presence of birefringence, we find that the response to a stochastic noise features an early-time divergence for one of the two polarization of the graviton. 


\paragraph{The decoupling limit.} Using our theory of open gravity for inflation, we have studied the gravitational mixing of curvature perturbation with other metric components (see Sec. \ref{Sec:minimal}). We have computed and solved the constraint equations that follow from the modified Einstein equations to linear order in perturbations. By plugging back these solutions into our open functional, we were able to determine the regime of parameter for which curvature perturbations decouple from the metric and one can simply focus on the dynamics of a Goldstone scalar field in an unperturbed inflationary background, which was studied in \cite{Salcedo:2024smn} (a much earlier study was performed in \cite{LopezNacir:2011kk}). Instead of studying the most general open theory, we have focussed on one choice of parameters that does reproduce the decoupling limit studied in \cite{Salcedo:2024smn}. For that theory, we have calculated the ``mixing energy scale" $E$, i.e. the energy below which the mixing of curvature perturbations with gravity becomes important. Neglecting noise but accounting for dissipation, we found that the mixing scale is $\sqrt{\gamma \epsilon H}$ for small dissipation $\gamma\ll H$ and $\e H$ for large dissipation $\gamma \gg H$. This means that if one uses our open effective theory to model the dynamics of perturbations, which in single-clock inflation freeze out at energy scale of order $H$, gravitational effects can be neglected to leading order, in agreement with the findings in \cite{LopezNacir:2011kk}. The above result is robust to the inclusion of noise, see Table \ref{tab:noisedecoupv0}, where we find that around Hubble (energy) crossing the gravitational effects are negligible.


\section{Learnings from open electromagnetism}\label{sec:OpenEFT}    

Open EFTs aim at describing a \textit{system} that interacts with an unspecified \textit{environment}. The environment can either affect the system's dynamics in a unitary manner by generating an effective Hamiltonian, or in a non-unitary manner, inducing dissipation and noise. In this work, we construct a bottom-up open EFT for a scalar and massless tensor. These will eventually be interpreted as an inflaton field and the graviton and we will use our theory to make prediction for cosmological observables. Our approach extends our previous construction of the Open EFToI in the decoupling limit \cite{Salcedo:2024smn} by incorporating dynamical gravity.  Our construction can equivalently be thought of as the most general local EFT for gravity propagating in a material which features a single light scalar degree of freedom. A main conceptual hurdle is how to use the Schwinger-Keldysh formalism to describe the dissipative regime of gauge theories and in particular gravity. We addressed this problem in a companion paper \cite{Salcedo:2024nex} where we discuss and open EFT for a massless spin-one field, namely electromagnetism. {Here, we present our main findings, providing a road-map to the study of dynamical gravity.}

\begin{figure}[tbp]
    \centering
    \includegraphics[width=1\textwidth]{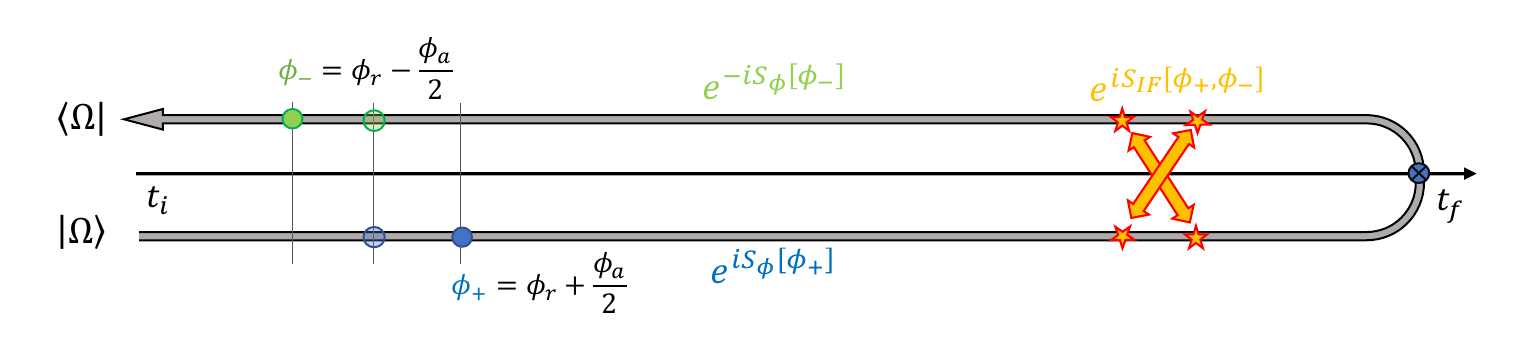}
\caption{Illustration of the closed-time-path, where time is running from left to right in both contours and the arrow represent path ordering (time ordering in $\ket{\Omega}$ and anti-time-ordering in $\bra{\Omega}$). {Memory of the initial conditions is quickly lost due to dissipation.}
}
    \label{fig:contour}
\end{figure}

\paragraph{Schwinger-Keldysh formalism.} The Schwinger-Keldysh formalism \cite{Schwinger:1960qe, Keldysh:1964ud} is a formulation of QFT that allows the description of non-equilibrium systems and/or open systems in which energy and information can be exchanged between the system and the environment \cite{kamenev_2011, breuerTheoryOpenQuantum2002, 2016RPPh...79i6001S}. Also known as in-in or closed-time-path formalism, this construction relies on a path integral contour represented in \Fig{fig:contour} that departs from the familiar in-out contour used in particle-physics (see \cite{Donath:2024utn} for a comparison). In the Schwinger-Keldysh case, the path integral is doubled, with one direction going forward in time denoted as the ``$+$" contour, and one direction going backward in time denoted as the ``$-$" contour. This reflects that common situation in which one is interested in the expectation value of observables in the state of the system, as opposed to for example transition amplitudes.
In the perturbative picture, interactions involving field operators $\phi$ can then be inserted either on the $+$ contour or on the $-$ contour, hence the notation $\phi_\pm$. 

{The doubling of the path integral contour is crucial for the formalism to describe dissipative phenomena. 
Indeed, mixing between the two branches of the path integral accounts for dissipation and noise in a path integral language \cite{2016RPPh...79i6001S}.} The dynamical evolution is specified by an open effective functional
\begin{align}\label{eq:Seffdef}
   S_{\mathrm{eff}}[\phi_+,\phi_-] =  S_\phi[\phi_+] - S_\phi[\phi_-] + S_{\mathrm{IF}}[\phi_+,\phi_-],
\end{align}
whose first two terms correspond to the unitary (\ie Hamiltonian) part, a.k.a. the action, while the third term is the so-called Feynman-Vernon influence functional \cite{FEYNMAN1963118}, which cannot be written as a sum of terms that depend on $\phi_+$ and $\phi_-$ separately. 
{The Keldysh basis provides a convenient basis \cite{Kamenev:2009jj}
\begin{align}
    \phi_r = \frac{\phi_+ + \phi_-}{2},\qquad \phi_a = \phi_+ - \phi_- \qquad \Leftrightarrow \qquad \phi_+ = \phi_r + \frac{\phi_a}{2}, \qquad \phi_- = \phi_r - \frac{\phi_a}{2} ,
\end{align}
where the retarded and advanced components $\phi_r$ and $\phi_a$, which respectively correspond to the mean and the difference of the field inserted along each branch of the path integral, turn out to have a simpler physical interpretation than $\phi_{\pm}$.} Observables can be extracted from functional derivatives with respect to external currents of the generating functional
\begin{align}\label{eq:PIref}
\mathcal{Z}[J_{r},J_{a}]=\int_{\Omega}^\pi \mathcal{D}\pir  \int_{\Omega}^{0}\mathcal{D}\pia \ee^{iS_{\mathrm{eff}}[\phi_r, \phi_a] + \int d^{4}x \sqrt{-g}\left(J_{r}\phia+J_{a}\phir\right)},
\end{align}
where $\Omega$ represents the choice of initial state and can be in principle a mixed density matrix \cite{2016RPPh...79i6001S}. The boundary conditions in the path integral are crucial role in ensuring the right number of propagating degrees of freedom, as discussed below. 

To originate from a unitary UV theory, \Eq{eq:Seffdef} must satisfy a set of UV-unitarity constraints (also referred to as non-equilibrium constraints, as they represent non-perturbative conditions valid even far from equilibrium) \cite{Crossley:2015evo, Glorioso:2016gsa, Liu:2018kfw}
\begin{align}
S_{\mathrm{eff}} \left[\phi_r,\phi_a = 0\right] &= 0 \,,\label{eq:norm} \\
S_{\mathrm{eff}} \left[\phi_r,\phi_a\right] &= - 	S^*_{\mathrm{eff}} \left[\phi_r,-\phi_a\right] \,,\label{eq:herm} \\
\Ima S_{\mathrm{eff}} \left[\phi_r,\phi_a\right] &\geq 0. \label{eq:pos}
\end{align} 
which restrict the freedom in constructing bottom-up open EFTs \cite{Hongo:2019qhi, Salcedo:2024smn}. These constraints also ensure the convergence of the path integral specified in \Eq{eq:PIref}.

    
\paragraph{Open electromagnetism.} In \cite{Salcedo:2024nex}, we studied Abelian gauge theories within the Schwinger-Keldysh formalism. Specifically, we constructed the most general open effective field theory for electromagnetism in a dielectric medium. We begin by doubling the gauge fields, $A^\mu \to A^\mu_\pm$, and expressing the result in the Keldysh basis
\begin{align}
\text{retarded: } A^{\mu}&=\frac{1}{2}\left( A_{+}^{\mu}+A_{-}^{\mu}\right) \,, & 
 \text{advanced: } a^{\mu}=A_{+}^{\mu}-A_{-}^{\mu}\,.
\end{align}
In the dissipative theory, there is a gauge invariance under the transformation acting on both $A_+$ and $A_-$ corresponding to direction $ \e_{+}=\e_{-}=\e$. Under this \textit{retarded gauge transformation}, $ A^{\mu}$ transforms as the usual gauge field, while $ a^{\mu}$ does not transform:
\begin{align}\label{eq:gaugetrans}
\text{retarded gauge transformation: } A^{\mu}&\to A^{\mu}+\partial^{\mu}\e\,, & a^{\mu}&\to a^{\mu}\,.
\end{align}
Ensuring retarded gauge invariance guarantees the theory only propagates two helicities. A simple implementation consists in building the effective functional out of the retarded field strength $F^{\mu\nu}=\partial^{\mu}A^{\nu}-\partial^{\nu}A^{\mu}\,$. 
To all order in derivatives and in the free theory, the open effective functional $S= S_1 + S_2$ of open electromagnetism is \cite{Salcedo:2024nex}
\begin{align}\label{S1old}
S_1=\int_{\omega,\bmk}\left[ a^{0}  i k_{i}F^{0i}+a_{i}\left(  \gamma_{2}F^{0i}+\gamma_{3}ik_{j}F^{ij}+\gamma_{4}\e^{i}_{\,jl}F^{jl}\right)\right]\,, 
\end{align}
and 
\begin{equation}
S_2= i \int_{\omega,\bmk} a^{\mu}N_{\mu\nu}a^{\nu},
\end{equation}
where $N_{\mu\nu}$ is any $4\times4$ positive definite matrix and $\gamma_{2,3,4}$ are model-dependent analytic functions of $\omega$ and $k^2$. In position space, the action can also be written suggestively as 
\begin{align}\label{MAxpos}
    S_1=-\int_{t,\bfx} a_\mu \left[\partial_\nu F^{\mu\nu}+\delta^\mu_i \left(n_\nu F^{\nu i}-\gamma_4 \e^i_{\,jl}F^{jl}\right) \right]\,,
\end{align}
where we introduced the notation 
\begin{align}\label{Gamma}
\gamma_2& = \partial_0 +\Gamma\,, \qquad \text{and} \qquad n_\nu\equiv(-\Gamma,(\gamma_3+1)\partial_j),
\end{align}
with $\Gamma$ a local function of $\partial_t$ and $(\partial_i)^2$ that represents the corrections beyond Maxwell's theory. Recall that in our Fourier conventions $\partial_0=-i\omega$ and $\partial_i=i k_i$. In the form \eqref{MAxpos} the four dynamical Maxwell's equations can be read off directly as the factor multplying $a_\mu$.

Let us now consider the \textit{advanced gauge transformation}, which transforms the $+$ and the $-$ branch of the path integral in opposite directions, that is $\epsilon_+  = \epsilon_a/2$ while $\epsilon_-  = - \epsilon_a/2$. The retarded and advanced components transform as 
\begin{align}
    \text{advanced gauge transformation: } A^\mu \rightarrow A^\mu, \qquad a^\mu \rightarrow a^\mu + \partial^\mu \epsilon_a.
\end{align}
One of the main findings of \cite{Salcedo:2024nex} (see also \cite{Lau:2024mqm} for a complementary discussion) is to relate the lack of unitarity due to the openness of the dynamics to the transformation of the effective functional under the above advanced gauge transformation. Let us first focus on $S_1$. Under the advanced gauge transformation, it transforms as\footnote{The operators proportional to $\gamma_3$ and $\gamma_4$ do not transform under advanced gauge transformations, the former being manifestly advanced gauge invariant and the latter being conserved by Bianchi identity, that is
\begin{align}
    i \gamma_3 a_i k_j F^{ij} \qquad &\rightarrow \qquad- \gamma_3 \epsilon_a k_i k_j F^{ij} = 0,\\
    \gamma_4 a_i \widetilde{F}^{0i} \qquad &\rightarrow \qquad i \gamma_4 k_i\widetilde{F}^{0i} = 0,
\end{align}
where we defined $\widetilde{F}^{0i} \equiv \e^{i}_{\,jl}F^{jl}$.
}
\begin{align}
    \Delta S_1 = \int_{\omega, \bmk} \epsilon_a \left( i \omega + \gamma_2\right) i k_i F^{0i}.
\end{align}
The unitary limit is reached when 
\begin{align}
    \gamma_2 &= - i \omega, &\gamma_3&=c^2=1 \,, &\gamma_4&=0\,,
\end{align}
in which case we recover Maxwell's theory in vacuum \cite{Salcedo:2024nex}. Conversely, whenever the theory is open, one finds that $\Delta S_1 \neq 0$ under advanced gauge transformation.

\paragraph{Advanced St\"uckelberg trick. } Following \cite{Lau:2024mqm}, one may choose to recover manifest advanced gauge invariance by introducing a St\"uckelberg field $X_a$ that non-linearly realises advanced gauge transformation 
\begin{align}
    X_a \rightarrow X_a - \epsilon_a,
\end{align}
such that the combination $\mathcal{A}_a^\mu \equiv a^\mu + \partial^\mu X_a$    is manifestly advanced gauge invariant. The action constructed from the promotion of $a^\mu \rightarrow \mathcal{A}_a^\mu$ is then invariant under both retarded and advanced gauge transformations. Let us consider
\begin{align}
    S_1^{\mathrm{new}} &= \int_{\omega,\bmk}\left[ \mathcal{A}_a^{0}  i k_{i}F^{0i}+\mathcal{A}_{ai}\left(  \gamma_{2}F^{0i}+\gamma_{3}ik_{j}F^{ij}+\gamma_{4}\e^{i}_{\,jl}F^{jl}\right)\right], \\
    &= S_1^{\mathrm{old}} + \int_{\omega, \bmk} X_a \left(  i \omega + \gamma_2 \right) i k_i F^{0i}.
\end{align}
where $S_1^{\text{old}}$ was given in \eqref{S1old}. Now deriving the equation of motion for $X_a$ one finds that the on-shell relations impose the system must be closed through
\begin{align}
    \frac{\delta S_1}{\delta X_a} = 0 \quad \Rightarrow \quad \gamma_2 = - i \omega,
\end{align}
This result for electromagnetism is the equivalent of the nice result found by \cite{Lau:2024mqm} in the context of gravity.

The constraint $\gamma=-i\omega$ naively appears to prevent any deviation from the unitary theory. The issue is that we are asking for a medium that dissipates photons but we are not coupling the system to an external current. To obtain a non-trivial result, we need to allow for an external current $j^\mu$, which will also contain noise contributions $\xi_\mu$. Applying the St\"uckelberg transformation to the coupling to an external current we find
\begin{align}\label{jpxi}
    S_1 \supset - \int_{\omega, \bmk} (j_\mu + \xi_\mu) a^\mu \rightarrow -\int_{\omega, \bmk} (j_\mu + \xi_\mu) \mathcal{A}_a^{\mu}.
\end{align}
The new equation of motion for $X_a$ now becomes
\begin{align}\label{eomXa}
     \frac{\delta S_1}{\delta X_a} = 0 \quad \Rightarrow \quad \Gamma i k_i F^{0i} = i \omega (j_0+\xi_0) + i k_i (j^i+\xi^i) .
\end{align}
We can now make use of the on-shell equations of motion 
\begin{align}
    \frac{\delta S_1}{\delta a^0} &= 0 \quad \Rightarrow \quad i k_i F^{0i} =  j^0 + \xi^0, \\
    \frac{\delta S_1}{\delta a^i} &= 0 \quad \Rightarrow \quad \gamma_{2}F^{0i}+\gamma_{3}ik_{j}F^{ij}+\gamma_{4}\e^{i}_{\,jl}F^{jl} =  j^i + \xi^i,
\end{align}
which yields a modified Maxwell equation 
\begin{align}
  \partial_\mu F^{\mu\nu} + \delta^\nu_{~i} \left[\Gamma F^{0i}+(\gamma_{3}+1)ik_{j}F^{ij}+\gamma_{4}\e^{i}_{\,jl}F^{jl}\right] =  j^\nu + \xi^\nu ,
\end{align}
Plugging this into \eqref{eomXa} we obtain the \textit{non-standard current conservation}, 
\begin{align}\label{nonstand}
    - \gamma_2 (j_0 + \xi_0)+ i k^i (j_i + \xi_i) = 0\,,
\end{align}
Making explicit the dissipative term, $\gamma_2 = \Gamma - i \omega$, we recover the \textit{noise constraint} of \cite{Salcedo:2024nex}, that is 
\begin{align}\label{eq:noiseconstEM}
\boxed{\partial^\mu (j_\mu + \xi_\mu) = \Gamma (j_0 + \xi_0) \Big.}\, . 
\end{align}
The present derivation employed the St\"uckelberg trick for explicitly broken advanced gauge transformations. However, this result can be straightforwardly derived simply taking the gradient of the dissipative and stochastic Maxwell equations, as was done in \cite{Salcedo:2024nex}. Here we have emphasized the St\"uckelberg derivation to show the equivalence between our approach in \cite{Salcedo:2024nex} and that presented in \cite{Lau:2024mqm}. Later on, when developing an open theory of inflation, the St\"uckelberg approach will be particularly convenient.        

\paragraph{Why is the current not conserved?} The fact that the current $(j+\xi)^\mu$ does not satisfy the standard conservation equation might appear concerning. Intuitively, the electric charge should still be conserved even if we separate system and environment. Moreover, in standard electrodynamics, charge conservation is necessary for gauge invariance, and gauge invariance is necessary for unitarity. Here we want to show in a more transparent and intuitive way that the non-standard current conservation in \eqref{nonstand} is precisely what is expected from the fact that the full electric charge is conserved. 

Consider the standard Maxwell equation
\begin{align}
    \partial_\mu F^{\mu\nu} = J^\nu\,,
\end{align}
with $J^\mu $ the total conserved current $\partial_\mu J^\mu = 0$. By our assumption of separation of scales, the environment is gapped and so $J^\mu$ can be written as a fixed external current $j^\mu$ plus a function of the light fields in the Open EFT, namely the photons. Since $J^\mu$ is gauge invariant, its expectation value must be written in terms of $F_{\mu\nu}$. To lowest order in derivatives the simplest possibility is\footnote{To keep the discussion as transparent as possible here we focus just on the dissipation $\Gamma$ and neglect $\gamma_{3,4}$.} 
\begin{align}
    J^\mu=j^\mu - \Gamma \delta^\mu_{~i} F^{0i}+\dots\,,
\end{align}
where $\Gamma$ is a model dependent local function of time and space derivatives, $\Gamma=\Gamma(\omega,k^2)$ and the minus sign has been chosen for later convenience. Since part of the environment's current is now proportional to $F^{\mu\nu}$, it feels natural to re-write the Maxwell's equestions as
\begin{align}\label{eq:dissipMax}
    \partial_\mu F^{\mu\nu}+ \Gamma \delta^\nu_{~i} F^{0i} = j^\nu\,.
\end{align}
We recognize $\Gamma$ as the deviations from Maxwell's theory that we encountered in \eqref{Gamma}, which plays the role of a friction in the equations of motion for $A_i$. The time-component imposes
\begin{align}
    \partial_\mu F^{\mu0}  = j^0\,.
\end{align}
Taking the gradient of \eqref{eq:dissipMax} and making use of the anti-symmetry of the field strength, we find
\begin{align}
     \Gamma j^0  = \partial_\mu j^\mu \,,
\end{align}
which is precisely the non-standard current conservation we found in \eqref{nonstand}. This illustrates how the conservation of the total current induces a non-trivial relation in the system when the latter is open.

\paragraph{Retarded St\"uckelberg trick.} One may wonder what happens when we not only break the advanced gauge transformation but also the retarded one. To study this in a concrete example, we add a mass term to the above effective action, i.e. we study a \textit{dissipative Proca theory}. Explicitly, we consider
\begin{align}
    S_1 \supset - \int_{\omega, \bmk} m^2 a^\mu A_\mu.
\end{align}
As a consequence of this new term, $S_1$ is not invariant under both advanced and retarded gauge transformations. One can straighforwardly study this theory as is. However, in certain situation one is interested in the simplification that may happen at high energies where the longitudinal degree of freedom is expected to decouple from the transverse one, a result know as the decoupling theorem in the context of particle physics. In that case, it is useful \cite{Stueckelberg:1938hvi,Coleman:1969sm,Arkani-Hamed:2002bjr} to restore invariance under both gauge transformations, we perform a retarded St\"uckelberg trick to addition to the advanced St\"uckelberg trick we just discussed. More in detail, we introduce a St\"uckelberg field $X_r$ that non-linearly realises retarded gauge transformations 
\begin{align}
    X_r \rightarrow X_r - \epsilon_r.
\end{align}
It follows that the combination $\mathcal{A}_r^\mu \equiv A^\mu + \partial^\mu X_r$ is manifestly retarded gauge invariant, basically by construction. The functional constructed from the promotion of $a^\mu \rightarrow \mathcal{A}_a^\mu$ and  $A^\mu \rightarrow \mathcal{A}_{r}^\mu$ is then invariant under both retarded and advanced gauge transformations. Under this promotion, 
\begin{align}
    S^{\mathrm{new}}_1 \supset S^{\mathrm{old}}_1 - m^2\int \dd^4x  \left[\partial^\mu X_a \partial_\mu X_r - a^\mu \partial_\mu X_r - \partial^\mu X_a A_\mu \right],
\end{align}
where we wrote the expression in real space to improve the readibility. In the high energy limit $E \gg m$, we recover the familiar \textit{decoupling} limit where the mixing between the the scalar $(X_{r}, X_a)$ and the gauge vector $(A^\mu, a^\mu)$ becomes negligible. The take-home message is that breaking retarded gauge invariance triggers new degrees of freedom, in this case $X_r$, while the breaking of advanced gauge invariance caused by cross-branch interactions representing open effect does \textit{not}. Moreover, in electrodynamics, the presence or absence of the additional scalar degree of freedom $X_r$ is a model dependent choice. The theory with just the two photon polarizations exists and can be dissipative. This will change in gravity where the presence of dissipation selects a reference frame of the medium hence breaking retarded diff invariance. Depending on the medium in question, e.g. superfluid versus solid/jelly, this in turn triggers the appearance of one or more additional degrees of freedom.   


\paragraph{A difficulty with gauge theories in the Schwinger-Keldysh formalism.}

Before concluding, we take a brief detour to clarify what the difficulty is in describing gauge theories in the Schwinger-Keldysh formalism and how one can proceed in electromagnetism. This gives us a (non comprehensive) menu of options to choose from when we tackle gravity.

Let's begin with a simple observation. When the number of advanced fields is larger than the number of retarded fields, an issue arises unless additional structure is taken into account. Indeed consider the toy model with one retarded field $\phi_r$ and two advanced fields $\phi_{a1}$ and $\phi_{a2}$,
\begin{align}\label{nabiggernr}
    S_{SK}=\int \, \dd^4x \left[\phi_{a1}F_1(\phi_r)+\phi_{a2}F_2(\phi_r) \right]\,,
\end{align}
where $F_{1,2}$ are some generic functionals of $\phi_r$. The classical deterministic equations of motions are $F_{1}(\phir)=0$ and $F_{2}(\phir)=0$. Unless $F_{1}$ and $F_2$ are related to each other in such a way to admit the same solutions, the classical equations of motion have no solutions whatsoever. Now let's see why this observation is relevant for a gauge theory with a symmetry group $G$. Because of the doubling of the path integral contour in the Schwinger-Keldysh formalism, the group is naively doubled to $G_+\times G_-$. Now one expects that generic dissipative effects couple the two branches and break the anti-diagonal combination leaving only the retarded diagonal symmetry $G_r$ unbroken\footnote{Actually, each individual dissipative interaction may break only a few generators of the full advanced symmetry group $G_a$. It is only when all possible dissipative interactions are included that one expects $G_a$ to be fully broken. For example, in the example discussed in Sec. \ref{sec:Stuck} we will encounter dissipative interactions that break spatial advanced diffs, but not time advanced diff.}. If one fixes the retarded gauge, e.g. by setting to zero some retarded fields, the number of retarded fields generically decreases by one but the number of advanced fields remains unchanged. One hence worries about the problem pointed out above, namely that there are more classical equations than fields and there may be no solutions. Of course the theory does not change under gauge fixing, so it must be that non-trivial relations are present among the operators linear in the different advanced fields. In other words, additional structure must be accounted for. A similar situation is encountered in a different case where one performs an \textit{advanced} \stuck transformation. This increases the number of advanced fields by one, but does not change the number of retarded fields. Once again, the number of classical equations is larger than the number of retarded fields. Of course, yet again, non-trivial structure among the different terms linear in advanced fields ensure that the equations of motions admit non-trivial solutions. 

Accounting for this additional structure may at times render the construction of a theory considerably more involved. In \cite{Salcedo:2024nex}, we discuss three different ways to deal with this issue, which give each different strategies to account for the additional structure:
\begin{itemize}
    \item One doubles the fields only \textit{after} having already fixed the gauge in the unitary theory so that the number of retarded and advanced fields matches by construction. In some sense in this case we are simply \textit{not} dealing with a gauge theory because we have fixed the gauge to begin with. A shortcoming of this approach is that the construction of the dissipative theory is gauge dependent from the start and to express the theory in a different gauge one needs to start back from the beginning.
    \item One fixes the retarded gauge but then notices that a new, \textit{deformed advanced gauge} is automatically present in the linear theory. Fixing this deformed advanced gauge brings the number of retarded and advanced fields to match again. We have not yet investigated what happens to this deformed advanced gauge transformation beyond the Abelian theory studied in \cite{Salcedo:2024nex}. Hence in this work we will not discuss this possibility further, even though we have noticed that such a deformed advanced symmetry does arise in gravity at linear order in retarded and advanced fields. A brief discussions of these is included in Sec. \ref{App:noise_constraints}.
    \item One never fixes the retarded gauge and proceeds with gauge invariant quantization, as often done in electromagnetism or in BRST quantization. 
\end{itemize}


\section{The construction of open gravity}\label{sec:unitgauge}

Now that we have introduced the Schwinger-Keldysh formalism, we want to construct an open EFT that captures local dissipation and noises for a light scalar coupled to dynamical gravity. This goes beyond our previous construction in \cite{Salcedo:2024smn}, where we only considered the decoupling limit, hence neglecting the possibly dissipative dynamics of gravitational perturbations. The additional scalar field is a Goldstone boson of (approximate) time translations as in the EFTs of \cite{Cheung:2007st, Gubitosi:2012hu}. Those constructions make extensive use of the so-called \textit{unitary gauge}\footnote{Beware of the misleading terminology. We call it the ``unitary" gauges in agreement with the previous literature, but this does not mean that the dynamical evolution is unitary. Rather this gauge would reduce to that chosen in \cite{Cheung:2007st, Gubitosi:2012hu} if we removed all dissipative effects.} to formulate the theory. Here, we perform an analogous construction in the Schwinger-Keldysh formalism.


\subsection{{Closed effective field theory setup}}

{We begin by presenting the original construction introduced in \cite{Cheung:2007st, Gubitosi:2012hu, Piazza:2013coa}, which forms the basis for the present development}. We consider a single scalar field $\phi(t,\bfx)$ evolving unitarily in a perturbed Friedmann–Lema\^itre–Robertson–Walker (FLRW) geometry. Both the scalar field and the metric are expanded around their background value 
\begin{align}
    \phi(t,\bfx) = \bar \phi(t) + \delta \phi(t, \bfx), \qquad\qquad g_{\mu\nu}(t,\bfx) = \bar{g}_{\mu\nu}(t) + \delta g_{\mu \nu}(t,\bfx),
\end{align}
and we aim at understanding the dynamics of the fluctuations. The theory is conveniently constructed in the unitary gauge, which is defined by the condition\footnote{In this subsection, we use $\doteq$ to denote an equality that holds in unitary gauge.} $\delta \phi \doteq 0$. In this case, all the perturbations are absorbed into the metric and the homogeneous scalar field $\phi(t, \bfx) \doteq \phi(t)$ can be used as a clock. Indeed, one can parametrize the time slicing in terms of the homogeneous value of the scalar field $t = t(\phi(t,\bfx))$ and construct geometrical objects based on this time foliation of spacetime. For instance, the future-pointing unit vector $n^\mu$ that is normalised, $n^\mu n_\mu=-1$, and perpendicular to the foliation is defined by the unit co-vector 
\begin{align}\label{eq:n_mu_def}
    n_\mu \equiv - \frac{\partial_\mu \phi}{\sqrt{-g^{\mu\nu} \partial_\mu \phi \partial_\nu \phi}} \doteq - \frac{\delta^0_{~\mu}}{\sqrt{-g^{00}}}\,, 
\end{align}
where the second equality holds in the unitary gauge. Related useful expressions are\footnote{The attentive reader will have noticed that the ``unitary gauge" coordinates are nothing but the Arnowitt-Deser-Misner \cite{Arnowitt:1962hi} decomposition of spacetime into space and time, $4=3+1$. So all these formulae can be borrowed from standard textbooks. For example, in these coordinates the ADM lapse is $N=-n_0=1/n^0$, the ADM shift is $N_i=g_{0i}=(n_0)^2g^{0j}g_{ji}$ or $N^i=(n_0)^2g^{0i}$. Concerning the ADM spatial metric, one can define the covariant 2-tensor $h_{\mu\nu}\equiv g_{\mu\nu}+n_\mu n_\nu$. This tensor is perpendicular to $n^\mu$ by construction. However note that $h_{0\mu}\neq 0$ even in unitary gauge. Instead, it is $h^{\mu\nu}$ that takes a particular simple value in unitarity gauge, namely $h^{0\mu}\doteq 0$ and $h^{ij}=g^{ij}$.}
\begin{align}
    n^\mu & \doteq -\frac{g^{\mu 0}}{\sqrt{-g^{00}}}\,, \qquad g^{0\mu} \doteq \frac{n^\mu}{n_0}\,, 
\end{align}
and
\begin{align}
     \sqrt{-g^{00}}&\doteq n^0 \doteq -\frac{1}{n_0}\,, \qquad  g_{00}=(n_0)^2(g_{0i}g^{0i}-1).
\end{align}
Two comments are in order. First, note that $n_i=0$, but in general $n^i\neq 0$, which is why it was more natural to specify the co-vector $n_\mu$ in \eqref{eq:n_mu_def}. Because of this, by contracting with $n_\mu$ we can in effect write an \textit{upper} zero index on any 4d-diff covariant tensor, such as $g^{\mu\nu}$, and the result is a 3d-diff covariant tensor, such as $g^{00}$ or $g^{0i}$. Conversely, a \textit{lower} zero index on a 4d-diff covariant tensor does \textit{not} in general give a 3d-diff covariant tensor. For example $g_{0i}$ does not transform as a covariant tensor under 3d-diffs. Second, note our signs are chosen in such a way that $n^\mu$ is a future-pointing time-like vector.

This formulation allows us to write down the most general (unitary) EFT compatible with the symmetries of the problem \cite{Cheung:2007st}. The presence of the inflaton background $\bar \phi(t)$ spontaneously breaks time-translation symmetry, such that the resulting action is made of terms that are invariant under \textit{time-dependent spatial diffs} only. To understand why this is the residual symmetry, we can simply ask what 4d-diffs preserve the gauge fixing condition $n_i=0$  that defines the preferred foliation ($n_0$ is simply fixed by requiring that $n_\mu$ has unit norm). To find out, we simply recall that the transformation of tensor under the change of coordinates $x^\mu\to x^\mu+\epsilon^\mu$ is minus its Lie derivative in the $\e^\mu$ direction, 
\begin{align}
    \Delta n_\mu=-(\e^\rho \partial_\rho n_\mu+n_\nu \partial_\mu \e^\nu)\,.
\end{align}
The unitary gauge condition $n_i=0$ is therefore preserved by any diff of the form $\e^\mu=(0,\e^i(\bfx,t))$, i.e. time-dependent 3d-diffs.

Now we wish to write down all terms invariant under time-dependent 3d-diffs. Of course allowed terms include $4d$ covariant operators such as $R$, which are \textit{a fortiori} $3d$ covariant. But one should also allow time-dependent functions ($\Lambda(t),~\cdots$), contractions with $n_\mu$ such as $g^{00}$, or geometrical objects constructed out of the foliation such as the extrinsic curvature 
\begin{align}\label{eq:extrinsiccurv}
    K_{\mu\nu} \equiv \left(\delta_{\mu}^{\sigma} + n_\mu n^\sigma \right) \nabla_\sigma n_\nu \doteq \delta_\mu^i \delta_\nu^j \Gamma_{ij}^0 (-g^{00})^{-1/2}\,.
\end{align}
The most generic action takes the form \cite{Cheung:2007st}
\begin{align}
    S = \int \dd^4 x \sqrt{-g} F(R_{\mu\nu\rho\sigma}, g^{00}, K_{\mu\nu}, \nabla_\mu; t),
\end{align}
where $F$ is an arbitrary function. Expanding the metric in powers of the perturbations we obtain \cite{Gubitosi:2012hu}
\begin{align}\label{eq:universal}
    S = \int \dd^4 x \sqrt{-g} \left[ \frac{M^2_{\mathrm{Pl}}}{2} R - \Lambda(t) - c(t) g^{00}\right] + S^{(2)},
\end{align}
where $R$ is the Ricci scalar and $f, ~\Lambda$ and $c$ are functions of time, and $S^{(2)}$ starts at second order in perturbations, and consequently does not affect the background equations. The first three terms in \Eq{eq:universal} constitute the \textit{universal part} of the EFToI \cite{Cheung:2007st}. 
Details about $S^{(2)}$ can be found in \eg \cite{Creminelli:2017sry}. 
The space-diff invariant unitary-gauge action is the starting point to derive a fully diff-invariant theory for the Goldstone boson $\pi$ of time translations via the St\"uckelberg trick \cite{Cheung:2007st}. We don't review this here and instead turn to the problem of defining unitary gauges for the dissipative theory.

    
\subsection{Open gravity in retarded unitary gauge}\label{subsec:functional}

We now move on to the construction of the most general open theory of general relativity and a single scalar clock. 

\paragraph{Field content and retarded unitary gauge.} Just like any field is doubled in the Schwinger-Keldysh formalism, so is the metric. One then must consider two rank-2 tensors $g_{\mu\nu}^\pm$. Expressed in the Keldysh basis, the retarded and advanced metric read
\begin{align}\label{eq:retadvmetricsdef}
    g_{\mu\nu} &= \frac{\left(g_+\right)_{\mu\nu} + \left(g_-\right)_{\mu\nu}}{2}, \qquad \mathrm{and} \qquad 
    a^{\mu\nu} = \left(g_+\right)^{\mu\nu} - \left(g_-\right)^{\mu\nu} \,.
\end{align}
Since physical fields are associated to the retarded sector, we will later find it convenient to construct geometrical objects based on the retarded metric $g_{\mu\nu}$. 

In addition to the metric we have to consider the clock of the theory, which is a scalar field we will call $\phi$. Since we work with the Schwinger-Keldysh contour we have to double this field to distinguish insertion of $\phi$ in the plus and minus branch, so we must have $\phi_\pm(t,\bfx)$. The corresponding retarded and advanced fields are
\begin{align}
    \phi_r(t,\bfx)&\equiv\frac12 [\phi_+(t,\bfx)+\phi_-(t,\bfx)]\,, &
     \phi_a(t,\bfx)&\equiv \phi_+(t,\bfx)-\phi_-(t,\bfx)\,.
\end{align}
If generic dissipative effects are present, at this point the theory is invariant under 4d-diff$_r$ while we expect all advanced diffs to be broken
\begin{align}
    S=S_{\text{4d-diff}_r}[g_{\mu\nu},a^{\mu\nu},\phi_r,\phi_a]\,.
\end{align}
To construct the theory following the EFToI recipe \cite{Cheung:2007st}, it is convenient to fix retarded time diffs with some gauge prescription. In the EFToI we would simply choose coordinates such that $\phi=\bar{\phi}(t)$. Here however we have only one retarded gauge transformation but two perturbed fields $\phi_\pm(t,\bfx)$, so we have to make a choice. Since it is the retarded/symmetric combination of $\phi_\pm(t,\bfx)$ that contains the background of the field, the first natural option is to choose the gauge such that
\begin{align}
    \phi_r(t,\bfx)\doteq \bar\phi(t)\,.
\end{align}
In this gauge, which we dub \textit{clock retarded unitary gauge}, the advanced field $\phi_a(t,\bfx)$ is not homogeneous in general but perturbed. There is actually a second natural choice. To see this, let's go back to the fully 4d-diff$_r$ invariant theory and consider the following \textit{field redefinition}
\begin{align}
    \phi_+(t,\bfx)&\equiv \bar\phi(t_+(t,\bfx))\,, &\phi_-(t,\bfx)&\equiv \bar\phi(t_-(t,\bfx))\,,
\end{align}
where we have simply traded the two fields $\phi_\pm(t,\bfx)$ for the two fields $t_\pm (t,\bfx)$, which can always be done as long as $\bar \phi$ is a monotonic function. The name ``$t$'' for these new fields suggest the interpretation of maps of the spacetime to the doubled fluid space as suggested in \cite{Liu:2018kfw}. We stress however that $t_\pm(t,\bfx)$ are dynamical fields in the action, as opposed to coordinates that are integrated over. The second gauge that fixes retarded time diffs is then
\begin{align}\label{tr=t}
   \boxed{ t_r(t,\bfx)\equiv\frac12 \left[t_+(t,\bfx)+t_-(t,\bfx)\right]\doteq t}\,.
\end{align}
In this second gauge, which we dub \textit{(time) retarded unitary gauge}, the advanced combination $t_a$ is still perturbed
\begin{align}
     t_a(t,\bfx)\equiv t_+(t,\bfx) - t_-(t,\bfx)\,.
\end{align}
 In a general gauge, the coupling constants in the EFT of perturbations are functions of the fields $\phi_{\pm}(t, \bfx)$, that is of both the retarded and advanced clocks $t_r(t, \bmx)$ and $t_a(t, \bmx)$. Hence a generic coupling constant $\gamma$ will have the form $ \gamma(t_r; t_a)$. During inflation, the expansion in $t_r$ is known to be slow-roll suppressed \cite{Cheung:2007st}, however the dependence of the open functional on $t_{a}$ is \textit{a priori} arbitrary and un-restricted (up to the limitation enforced by the unitarity constraints given in \Eqs{eq:norm}, \eqref{eq:herm} and \eqref{eq:pos}). In other words, in contrast to the original EFToI \cite{Cheung:2007st}, there are no constraints relating the derivatives of the coupling constants to the coupling constants of terms containing $t_{a}^{n}$ (see \cite{Finelli:2018upr}). As a consequence, we should write down terms with all possible powers of $t_{a}$.\\

Before proceeding we point out that the existence of this advanced clock $t_a$ introduces a few subtleties. First, one now has more advanced fields ($a^{\mu\nu}$ and $t_a$) than retarded fields ($g_{\mu\nu}$) so to avoid the problems discussed around \eqref{nabiggernr}, additional structure must be present. Indeed, we anticipate that, in the specific case of a single-clock cosmology and at the classical stochastic order, we will be able to remove the $t_a$ dependent terms with a field redefinition. For the time being though, let us just forge on.


\paragraph{Building the action.} Following the approach of \cite{Cheung:2007st}, we want our effective functional to be invariant under spatial diffeomorphisms, but which ones? All advanced diffs are broken by generic open systems effects, which couple the two branches of the Schwinger-Keldysh contour, so it is natural to build our theory requiring invariance under \textit{retarded} 3d-diffs, under which fields on both branches of the Schwinger-Keldysh contour transform in the same way:
\begin{align}
    S=S_{\text{3d-diff}_r}[g_{\mu\nu},a^{\mu\nu},t_a]\,.
\end{align}
It is important to bear in mind that since $g_{\mu\nu}$ and $a^{\mu\nu}$ are made of $g_{\mu\nu}^\pm$, they \textit{both transform under retarded diffs}. The construction above leads to a theory that describes the two degrees of freedom of the graviton plus a scalar field. An alternative symmetry-breaking possibility, which we leave for future studies, would be to consider the case in which retarded spatial diffs may also broken, as in solid and supersolid inflation \cite{Endlich:2012pz,Celoria:2021cxq}. 

The unitarity constraints \Eqs{eq:norm}, \eqref{eq:herm} and \eqref{eq:pos} provide a convenient expansion scheme in powers of the advanced components \cite{Liu:2018kfw, Hongo:2018ant, Salcedo:2024smn}
\begin{align}
    S  = \sum_{n=1}^{\infty} S_n \qquad \mathrm{with} \qquad S_n = \mathcal{O}(\mathrm{adv}^n), 
\end{align}
where we notice that $S$ starts linear in the advanced fields. Beware that, as in \cite{Salcedo:2024smn}, operators with more powers of the advanced fields may not be suppressed compared to their retarded counterpart, see \cite{Salcedo:2024nex} for a detailed power counting in the decoupling limit. Instead, we are just using the organization in powers of the advanced fields $a^{\mu\nu}$ to group the many different terms. Restricting ourselves to order $\O((a^{\mu\nu})^2)$ for practical applications, we focus on $S_1$ and $S_2$, for which we illustrate the general procedure. $S_1$ encodes how many and what degrees of freedom the EFT describes while $S_2$ models the noise. The next order $S_3$ is cubic or higher in the advanced fields. Those terms in $S_3$ that survive in the decoupling limit were studied in detail in \cite{Salcedo:2024smn}.  {We should stress that in this work we are agnostic about the cosmological epoch for which our theory is used. Because of this, we don't specify a priori what FLRW background we choose. As a result, we don't state the precise power counting between $a_{\mu\nu}$ and $g_{\mu\nu}$, which in general may depend on the details of the model under consideration. An extensive discussion of this power counting was given for example in Sec. 2.3 of \cite{Salcedo:2024smn} for the case of inflation and will be given elsewhere for applications to dark energy.}

A simple procedure to guarantee that the effective functional is invariant under retarded \textit{spatial} diffeomorphisms is to use geometric objects built out of $g_{\mu\nu}$. Since we are already expanding in $a^{\mu\nu}$ and $t_a$, we can define these objects for $a^{\mu\nu}=t_a=0$. Then, we can define a normal vector $n_\mu$ to the time foliation as in \eqref{eq:n_mu_def}, 
\begin{align}\label{eq:n_mu_defv2}
    n_\mu \equiv - \frac{\partial_\mu \phi_\pm}{\sqrt{-g^{\mu\nu} \partial_\mu \phi_\pm \partial_\nu \phi_\pm}}=- \frac{\partial_\mu \bar\phi(t_r)}{\sqrt{-g^{\mu\nu} \partial_\mu \bar\phi(t_r) \partial_\nu \bar\phi(t_r)}} \doteq - \frac{\delta^0_{~\mu}}{\sqrt{-g^{00}}} \quad \text{for } t_a=0\,,
\end{align}
where the last equality is true in retarded unitary gauge $t_r=t$. From this we can define the induced metric $h_{\mu\nu} \equiv g_{\mu\nu} + n_\mu n_\nu$, the extrinsic curvature $K_{\mu\nu}$ in \eqref{eq:extrinsiccurv}, the covariant derivative%
\footnote{This choice has several practical consequences, such as the fact that $a^{\mu\nu}$ is a simple rank-$2$ tensor for $\nabla_\mu$, leading to the non-vanishing result
\begin{align}
    \nabla_\lambda a^{\mu\nu} = \partial_\lambda a^{\mu\nu} + \Gamma^\mu_{\lambda\sigma} a^{\sigma\nu} + \Gamma^\nu_{\lambda\sigma} a^{\mu\sigma},
\end{align}
where the Christoffel symbols are constructed from the retarded metric $g_{\mu\nu}$.}
$\nabla_\mu$ and the Riemann tensor $R_{\mu\nu\rho\sigma}$ are all built from the retarded metric $g_{\mu\nu}$. 

Hence, in retarded unitary gauge, the first contribution to the action takes the form 
\begin{align}\label{eq:S1try}
    S_1 = \int \dd^4 x \sqrt{-g}& \Big[M_{\mu\nu} (R_{\mu\nu\rho\sigma}, g^{00}, K_{\mu\nu}, \nabla_\mu; t) a^{\mu\nu} +  M(R_{\mu\nu\rho\sigma}, g^{00}, K_{\mu\nu}, \nabla_\mu; t) 
    t_a \Big] , 
\end{align}
where $M_{\mu\nu}$ can break retarded time diffeomorphisms but has to transform as a rank-$2$ cotensor under retarded spatial diffeomorphisms. Notice that it is the determinant of the retarded metric $g$ that appears in the volume measure $\sqrt{-g}$. In contrast to $\sqrt{\det(a_{\mu\nu})}$, the choice of using the retarded metric has the feature of having a well defined background value and signature.

Similarly, given that $t_a(t, \bfx)$ transforms as a scalar under retarded spatial diffeomorphisms, it is enough to enforce that $M$ also transforms as a scalar to guarantee the invariance of $S_{\mathrm{eff}}$. Under this construction, the deterministic equations of motion in the unitary gauges can be simply obtained from
\begin{align}
    \frac{\delta S_1}{\delta a^{\mu\nu}} \Big|_{a^{\mu\nu}=t_a=0}= 0 \quad \Rightarrow \quad M_{\mu\nu}  = 0 \qquad \& \qquad \frac{\delta S_1}{\delta t_a}\Big|_{a^{\mu\nu}=t_a=0} = 0 \quad \Rightarrow \quad M = 0. \quad
\end{align}
Two comments are in order. First, note that specifying the deterministic equations of motion is equivalent to specifying the linear effective functional $S_1$. Second, we should think of $M_{\mu\nu}=0$ as the modified Einstein equations  in the presence of a medium. Combining these comments we conclude that for gravity in a medium we get to write directly the modified Einsten's equations without having to derive them from the variation of a conservative action, which allows one to evade Lovelock theorem \cite{Lovelock:1971yv}.

We can also construct the noise functional 
\begin{align}\label{eq:S2try}
    S_2 &= i \int \dd^4 x \sqrt{-g} \Big[N_{\mu\nu\rho\sigma} (R_{\mu\nu\rho\sigma}, g^{00}, K_{\mu\nu}, \nabla_\mu; t) a^{\mu\nu}  a^{\rho\sigma} \\
    &\qquad +  N_{\mu\nu} (R_{\mu\nu\rho\sigma}, g^{00}, K_{\mu\nu}, \nabla_\mu; t) a^{\mu\nu} t_a  + N(R_{\mu\nu\rho\sigma}, g^{00}, K_{\mu\nu}, \nabla_\mu; t) t_a^2 \Big], \nonumber 
\end{align}
where $N_{\mu\nu\rho\sigma}$ has to transform as a rank-$4$ cotensor under retarded spatial 3d-diffs. Similarly, $N_{\mu\nu}$ has to be a rank-$2$ cotensor under retarded spatial 3d-diffs and $N$ a scalar. As always, the Hubbard-Stratonovich trick \cite{Hubbard:1959ub,Stratonovich1957} replaces the quadratic terms in $a^{\mu\nu}$ with a path integral over an auxiliary field $\xi_{\mu\nu}$ 
\begin{align}
    &\text{exp}\left\{-\int \dd^{4}x\sqrt{-g}N_{\mu\nu\rho\sigma} a^{\mu\nu}  a^{\rho\sigma}\right\}  \nonumber \\
    &\qquad =\int[\mathcal{D}\xi_{\mu\nu}]\;\text{exp}\left\{-\int \dd^{4}x \sqrt{-g}\left[\frac{1}{4}(N^{-1})^{\mu\nu\rho\sigma}\xi_{\mu\nu}\xi_{\rho\sigma}+i\xi_{\mu\nu}a^{\mu\nu}\right]\right\},
\end{align}
leading to the stochastic Einstein-Langevin equations in the unitary gauges
\begin{align}
    \frac{\delta S_1}{\delta a^{\mu\nu}}  + \frac{\delta S_2}{\delta a^{\mu\nu}} = 0 \qquad \Rightarrow \qquad M_{\mu\nu}  = \xi_{\mu\nu}.
\end{align}
The equations of motion are now stochastic and their solutions should be averaged over the distribution of the noise. The construction can be extended to include terms such as $N_{\mu\nu} a^{\mu\nu}  t_{a}$ and $Nt_{a}^{2}$, as explicitly shown in \App{app:mixedHS}. However, in the single-clock case, this extension is not required, as we now demonstrate.


\paragraph{Single-clock simplification.} A remarkable simplification takes place when we focus our attention on a single-clock cosmology, where the only degrees of freedom are the metric and a scalar field. In this case, we know that the modified Einstein equations  $M_{\mu\nu}=0$ fix the metric $g_{\mu\nu}$ up to spatial diffs. To see this, simply perform a retarded St\"uckelberg transformation with field $\pi_r$ (see Sec. \ref{subsec:stuck} for more details) to obtain a fully covariant set of Einstein's equations for $g_{\mu\nu}$ and $\pi_r$. These equations fix both $g_{\mu\nu}$ and $\pi_r$ without the need to derive the equation of motion for $\pi_r$ because these should be proportional to the constraint $\nabla^{\mu}M_{\mu\nu}$, which in the unitary case is simply the conservation of the energy-momentum tensor of the matter sector. This is very specific to a single-clock cosmology because in the general case the conservation of the \textit{total} energy-momentum tensor, or equivalently $\nabla^{\mu}M_{\mu\nu}$ in the dissipative case, is \textit{not} sufficient to fix the dynamics of all degrees of freedom. Because of this special feature of the single-clock case, it must be that the other equation of motion we found, namely $M=0$, is compatible with the Einstein's equations, otherwise no solution would be left. But since the modified Einstein equations  are already sufficient to fix a solution, up to (retarded) spacetime diffs, then is must be that $M$ is simply proportional to the Einstein's equations, otherwise the theory would not admit any solutions. Hence, in the single-clock case, $M$ must take the form
\begin{align}\label{singleclockstructure}
    \text{Single-clock: \quad}M=X^{\mu\nu}M_{\mu\nu}\,,
\end{align}
for some matrix $X^{\mu\nu}$ built out of retarded fields and derivatives. This is an example of the additional structure discussed below \eqref{nabiggernr}. Thanks to this structure, we can actually remove $t_a$ from $S_1$ with the field redefinition
\begin{align}
    a^{\mu\nu}\to a^{\mu\nu}-t_a X^{\mu\nu}\,,
\end{align}
which respects the boundary condition of the advanced fields. A similar simplification takes place when we also include the noise terms from $S_2$. There we can perform the Hubbard-Stratonovich trick \cite{Hubbard:1959ub,Stratonovich1957}, after which the equations of motion become stochastic differential equations, which are the gravitational analog to the Langevin equation,
\begin{align}
    M_{\mu\nu}&=\xi_{\mu\nu}\,, \qquad \text{and} \qquad M=\xi_t\,.
\end{align}
Because of the additional structure in \eqref{singleclockstructure}, the second of these equations should not change the set of solutions for $g_{\mu\nu}$ and $\pi_r$. Instead, it is interpreted as a constraint on what the noise $\xi_t$ must be. Because of this, without loss of generality, we can neglect $t_a$ completely in $S_{1}$ and $S_{2}$. In this paper we will restrict ourselves to the linear and quadratic order in advanced fields and postpone a discussion of higher order terms to future investigation. 

In summary, if one is interested in the classical stochastic equations of motion for a single-clock cosmology one may safely neglect $t_a$ and construct the theory exclusively using $a^{\mu\nu}$ and $g_{\mu\nu}$. The construction in the rest of this section is not specific to single-clock cosmology and will discuss the structure of $M$ in detail. While we do not make use of this in this paper explicitly, these results will be useful when we apply our formalism to the late universe and dark energy, where many different matter sectors are present. 


\paragraph{Construction of $S_1$.}

We now proceed to expanding the action $S_1$ up to second order in derivatives. 
Before we start, notice that any EFT operator can be multiplied by arbitrary powers of $g^{00}$, which transforms as a scalar under spatial retarded diffeomorphisms and does not change the order in derivatives. Just as in the original EFToI \cite{Cheung:2007st}, it is convenient to work with $(1+g^{00})$, as this expression vanishes on an FLRW background and starts explicitly at linear order. Hence, we expand both terms in \Eq{eq:S1try} according to\footnote{In unitary gauge, $g^{00}$ and $n^\mu$ are fixed in terms of each other, but we nevertheless find it convenient to use sometimes one and sometimes the other, where this helps intuition.} 
\begin{align}
    M &= \sum_{\ell=0} (g^{00} +1)^\ell M_{\ell}(R_{\mu\nu\rho\sigma}, K_{\mu\nu}, n_\mu, \nabla_\mu; t),  \label{eq:M_g00_expansion} \\
    M_{\mu\nu} &= \sum_{\ell=0} (g^{00} +1)^\ell M_{\mu\nu, \ell}(R_{\mu\nu\rho\sigma}, K_{\mu\nu}, n_\mu, \nabla_\mu; t)   \, ,
\end{align}
where the index $\ell$ here is not a spacetime-index, but rather denotes the fact that $M_{\mu\nu,\ell}$ will in general have different EFT coefficients at different orders in $(1+g^{00})$.
Each individual $M_{\ell}$ is given by the most generic scalar under spatial retarded 3d-diffs. Restricting to up to two derivatives, we obtain: 
\begin{align}\label{eq:S1_M_constr}
    M_\ell &= \tfrac{1}{4}\gat{1}  g_{\alpha \beta} g^{\alpha \beta} + \left( \gat{2} g_{\alpha\beta} + \gat{3} g_{\rho\sigma} K^{\rho\sigma} g_{\alpha\beta} + \gat{4} K_{\alpha\beta} + \gat{5} n_\rho  g_{\alpha\beta}g^{\rho\sigma} \nabla_\sigma \right) K^{\alpha\beta} \bigg. \\
    +& \left( \gat{6} g_{\alpha\gamma} g_{\beta \delta} + \gat{7} n_\alpha n_\gamma g_{\beta \delta} \right) R^{\alpha \beta \gamma \delta} \nonumber \bigg. \, .
\end{align}
To derive this expression, we have contracted the tensors $R^{\mu\nu\rho\sigma}$, $K^{\mu\nu}$ and $g^{\mu\nu}$ with all possible combinations of $n_\mu$, $\nabla_\mu$, $g_{\mu\nu}$ and $K_{\mu\nu}$, up to second order in derivatives. Recall that indices are raised and lowered with the \textit{retarded} metric $g_{\alpha \beta}$ and its inverse, $g^{\alpha\beta}$, so that $g_{\alpha \beta} g^{\alpha\beta}=4$. Some terms in this expression vanish due to the metric compatibility of the covariant derivative with the retarded metric, $\nabla_\alpha g_{\mu\nu} = 0$, and due to the orthogonality condition $n_\mu K^{\mu\nu} = 0$. While a term $n_\alpha n_\beta g^{\alpha \beta}$ is in principle allowed as well, it is degenerate with $\gat{1}$ and the expansion in \Eq{eq:M_g00_expansion}. Furthermore, any operator that features a covariant derivative acting on $t_a$ in \Eq{eq:S1try} is redundant, as it can be integrated by parts to yield only terms already displayed in \Eq{eq:S1_M_constr}. The introduced EFT coefficients $\gat{i}$ may in general depend on time $t_r = t$ (the eventual $t_a$ dependence being already accounted for in \Eq{eq:S1try}). In the retarded unitary gauge, where $n_\mu \sim \delta^0_\mu$, such that $M_\ell$ simplifies to
\begin{align}
    M_\ell &= \gat{1} + \gat{2} K + \gat{3}  K^2 + \gat{4} K_{\alpha\beta} K^{\alpha\beta} + \gat{5} \nabla^0 K + \gat{6} R + \gat{7} R^{00}  \, , \label{eq:S1_M}
\end{align} 
where we have re-absorbed minus signs and normalizations of $n_\mu$ in the EFT coefficients for simplicity.  

To construct the most generic $M_{\mu\nu, \ell} (R_{\mu\nu\rho\sigma}, K_{\mu\nu}, n_\mu, \nabla_\mu; t)$ up to second derivatives, we split it along the foliation according to
\begin{equation}\label{eq:decomp}
    M_{\mu\nu, \ell} =  n_\mu n_\nu M^{tt}_\ell + n_{(\mu} M^{ts}_{\nu),\ell} + M^{ss}_{\mu\nu, \ell} \, .
\end{equation}
Here $M^{tt}_\ell$, $M^{ts}_{\rho,\ell}$ and $M^{ss}_{\rho \sigma, \ell}$ denote the most generic scalar, vector and rank-2 tensor under spatial retarded diffs, which will again be constructed using $R_{\mu\nu\rho\sigma}$, $K_{\mu\nu}$ $n_\mu$, $\nabla_\mu$ and $g_{\mu\nu}$.
The rank-2 tensor $M^{ss}_{\mu\nu, \ell}$ can be further decomposed into: 
\begin{equation}\label{eq:decomp_Mmunu}
    M^{ss}_{\mu\nu, \ell} = M^{ss}_{\ell} g_{\mu\nu} + \Tilde{M}^{ss}_{\mu\nu, \ell} \, .
\end{equation}
Now all free indices in $M^{ts}_{\nu,\ell}$ and $\Tilde{M}^{ss}_{\mu\nu, \ell}$ should not be $\sim n_\mu$ or $\sim g_{\mu\nu}$, or else they are redundant with the other terms in \Eqs{eq:decomp} and \eqref{eq:decomp_Mmunu}. 
Since $M^{tt}_\ell$ and $M^{ss}_\ell$ must also be scalars under spatial 3d-diffs, they are constructed in the same way as $M_\ell$, involving precisely the same set of operators. However, to maintain generality, each operator in $M^{tt}_\ell$ and $M^{ss}_\ell$ may come with a different EFT coefficient compared to the ones of $M_\ell$ in \Eq{eq:S1_M}, such that we get 
\begin{align}
    M^{tt}_\ell &= \gtt{1} + \gtt{2} K + \gtt{3}  K^2 + \gtt{4} K_{\alpha\beta} K^{\alpha\beta} + \gtt{5} \nabla^0 K + \gtt{6} R + \gtt{7} R^{00} \, , \bigg. \label{eq:Mtt} \\
    M^{ss}_{\ell} &= \gss{1} + \gss{2} K + \gss{3}  K^2 + \gss{4} K_{\alpha\beta} K^{\alpha\beta} + \gss{5} \nabla^0 K + \gss{6} R + \gss{7} R^{00} \, , \bigg. \label{eq:Mss_scalar} 
\end{align}
in the retarded unitary gauge. To obtain $M^{ts}_{\rho,\ell}$ and $\Tilde{M}^{ss}_{\rho \sigma, \ell}$, we have to construct the most generic vector and rank-2 tensor under retarded spatial 3d-diffs, which follows the procedure of \Eq{eq:S1_M}: we contract $R^{\mu\nu\rho\sigma}$ and $K^{\mu\nu}$ with all possible combinations of $K_{\mu\nu}$, $n_\mu$, $\nabla_\mu$ and $g_{\mu\nu}$, up to second order in derivatives, such that one and two indices remain uncontracted. Note that, as mentioned previously, the free indices may not be $n_\mu$ or $g_{\mu\nu}$. This leads to
\begin{align}
    M^{ts}_{\mu, \ell} &= \gts{1} n_\alpha g_{\mu\gamma} g_{\beta \delta} R^{\alpha \beta \gamma \delta} + \gts{2} \nabla_\mu g_{\alpha\beta} K^{\alpha\beta} + \gts{3} g_{\mu\alpha} \nabla_\beta K^{\alpha\beta} \bigg. \label{eq:Mts} \\
    &= \gts{1} R^{0}{}_\mu + \gts{2} \nabla_\mu K + \gts{3} \nabla_\beta K^{\beta}{}_\mu \, \bigg. , \\
    \Tilde{M}^{ss}_{\mu\nu, \ell} &= \big(\gss{8} g_{\mu\alpha} g_{\nu\beta} + \gss{9} g_{\mu\alpha} g_{\nu\beta} n^\rho \nabla_\rho + \gss{10} g_{\mu\alpha} K_{\nu\beta} + \gss{11} g_{\alpha\beta} K_{\mu\nu} \big) K^{\alpha\beta} \bigg. \label{eq:Mss} \\
    +& \left[g_{\mu\alpha} g_{\nu \gamma} \left( \gss{12} g_{\beta\delta} + \gss{13} n_\beta n_\delta \right)  \right] R^{\alpha\beta\gamma\delta} \bigg. \nonumber \\
    &= \gss{8} K_{\mu\nu} + \gss{9} \nabla^0 K_{\mu\nu} + \gss{10} K_{\mu\alpha} K^{\alpha}{}_{\nu} + \gss{11} K K_{\mu\nu} + \gss{12} R_{\mu\nu} + \gss{13} R_\mu{}^0{}_\nu{}^0 \, \bigg.  ,
\end{align}  
where we have expressed the final results in the retarded unitary gauge. 

We may also construct parity-violating operators by considering additional terms constructed with the totally antisymmetric tensor $\epsilon_{\mu\nu\rho\sigma}$. Up to second order in derivatives, there are only two additional terms contributing to $S_1$:
\begin{align}\label{eq:bir1}
   M^{\mathrm{P.O.}}_{\mu\nu, \ell} &= \epsilon^{\alpha\beta\gamma\delta}  g_{\mu \alpha} n_\beta \left(\gamma^{\po}_{1,\ell}  \nabla_\gamma K_{\nu\delta} + \gamma^{\po}_{2,\ell} g^{\rho\sigma} n_\sigma R_{\gamma\delta\rho\nu}  \right) \, ,
\end{align}
which may be included in $M^{ss}_{\mu\nu, \ell}$. Any other potential term including the totally antisymmetric tensor either vanishes due to its symmetries, is of higher order in derivatives or of higher order in advanced variables.

Eventually, in the retarded unitary gauge the action $S_1$ reads
\begin{align}\label{eq:S1v1}
    S_1 &= \int \dd^4 x \sqrt{-g} \, \sum_{\ell=0} \, \left(g^{00} + 1 \right)^\ell \bigg\{ \, a^{00} \Big[ \gtt{1} + \gtt{2} K + \gtt{3}  K^2  + \gtt{4} K_{\alpha\beta} K^{\alpha\beta} \nonumber \\
     &\quad+ \gtt{5} \nabla^0 K + \gtt{6} R + \gtt{7} R^{00}  \Big] \bigg.  + a^{0\mu} \Big[ \gts{1} R^{0}{}_\mu + \gts{2} \nabla_\mu K + \gts{3} \nabla_\beta K^{\beta}{}_\mu \Big] \bigg.\nonumber \\
     & \quad + a^{\mu\nu} \Big[ g_{\mu\nu} \Big( \gss{1} + \gss{2} K + \gss{3}  K^2 + \gss{4} K_{\alpha\beta} K^{\alpha\beta} + \gss{5} \nabla^0 K+ \gss{6} R + \gss{7} R^{00} \Big)  \bigg.\nonumber \\ 
    &\quad  + \gss{8} K_{\mu\nu} + \gss{9} \nabla^0 K_{\mu\nu} + \gss{10} K_{\mu\alpha} K^{\alpha}{}_{\nu} \nonumber \bigg. + \gss{11} K K_{\mu\nu} + \gss{12} R_{\mu\nu} + \gss{13} R_\mu{}^0{}_\nu{}^0 \\
    &\quad  + \gamma^{\po}_{1,\ell} \epsilon_\mu{}^{\alpha\beta0} \nabla_\alpha K_{\beta\nu} + \gamma^{\po}_{2,\ell} \epsilon_\mu{}^{\alpha\beta0} R_{\alpha\beta}{}^0{}_\nu \Big] + t_a \Big[\gat{1} + \gat{2} K + \gat{3}  K^2 \nonumber \\
    &\quad+ \gat{4} K_{\alpha\beta} K^{\alpha\beta} + \gat{5} \nabla^0 K + \gat{6} R + \gat{7} R^{00}
    \Big] \bigg\} .
\end{align}
We can absorb a few terms by rescaling and redefining the field $a^{\mu\nu}$. First, we identify the operators $\gsso{12} a^{\mu\nu} R_{\mu\nu}$ and $\gsso{6} a^{\mu\nu} g_{\mu\nu} R$ as the two terms appearing in the Einstein tensor. We assume throughout that these two EFT coefficients are non-zero. Consequently we can set $\gsso{12} = \Mpl^2/2$ by rescaling $a^{\mu\nu}$, redefine $\gsso{6} \rightarrow - \Mpl^2/4 + \gsso{6}  $ and rescale all other EFT coefficients by $\gamma_{i, \ell} \rightarrow (2\gsso{12}/\Mpl^2) \gamma_{i, \ell} $. Moreover we can redefine 
\begin{align}\label{eq:amunu_redef}
    a^{\mu\nu} &\rightarrow \Tilde{a}^{\mu\nu} = a^{\mu\nu} + \alpha_1(t) a^{0(\mu} g^{\nu)0} + \alpha_2(t) a^{00} g^{\mu\nu} + \alpha_3(t) a^{00} g^{\mu 0} g^{\nu 0} \, .
\end{align}
to absorb a few EFT operators. This field redefinition preserves boundary conditions of the advanced metric, does not mix the order in advanced fields and transforms covariantly under retarded spatial diffs, hence it is allowed. Under this redefinition $a^{00}$ and $a^{0\mu}$ change according to
\begin{align}
    a^{00} &\rightarrow a^{00} (1-\alpha_1-\tfrac{1}{4}\alpha_2+\tfrac{3}{4}\alpha_3) + \dots \, ,\\
    a^{0\mu} &\rightarrow a^{0\mu}(1-\tfrac{1}{2}\alpha_1) + a^{00} g^{0\mu}(\tfrac{1}{2}\alpha_1+\tfrac{1}{4}\alpha_2-\tfrac{3}{4}\alpha_3) + \dots \, ,
\end{align}
where we have only kept terms at order $\ell=0$ and dots denote terms at higher order in $(g^{00}+1)$. 
Indeed, plugging this redefinition into the action simply shifts the EFT coefficients in $M^{tt}_{\ell}$ and $M^{ts}_{\mu, \ell}$, while the coefficients in $M^{ss}_{\mu\nu, \ell}$ remain unaffected.
For instance, $\gtso{1}$ is shifted by
\begin{equation}
    \gtso{1} \rightarrow \gtso{1}(1-\tfrac{1}{2}\alpha_1) + \tfrac{\Mpl^2}{2} \alpha_1 \, ,
\end{equation}
where the last term originates from the operator $\gtso{12}=\Mpl^2/2$. Consequently $\gtso{1}$ can be set to zero by choosing $\alpha_1 = 2\gtso{1}/(\gtso{1}-\Mpl^2)$, as long as $\gtso{1} \neq \Mpl^2$. 

An appropriate choice of $\alpha_2$ and $\alpha_3$ may further remove two operators in the first three lines of \Eq{eq:S1v1} at a fixed order in $\ell$. 
However, for this procedure to consistently eliminate an operator, the terms in $M^{ts}_{\mu, \ell}$ and $M^{ss}_{\mu\nu, \ell}$ responsible for generating the shift in its coefficient must not all simultaneously be zero. If they were, the redefinition would have no effect and the corresponding operator could not be removed. To avoid this issue, a safe approach is to remove only those operators whose shift is generated by the two terms that also appear in the Einstein tensor.
Based on this criterion, we can also safely remove $\gtto{6}$ and $\gtto{7}$, such that the action at $\ell = 0$ takes the form: 
\begin{align}\label{eq:S1fin}
    S_{1,0} &= \int \dd^4 x \sqrt{-g} \, \bigg\{ \, a^{00} \Big[ \gtto{1} + \gtto{2} K + \gtto{3}  K^2 + \gtto{4} K_{\alpha\beta} K^{\alpha\beta} + \gtto{5} \nabla^0 K \Big] \nonumber \\
 & \quad +a^{0\mu} \Big[ \gtso{2} \nabla_\mu K + \gtso{3} \nabla_\beta K^{\beta}{}_\mu \Big]  + a^{\mu\nu} \Big[ g_{\mu\nu} \Big( \gsso{1} + \gsso{2} K + \gsso{3}  K^2 + \gsso{4} K_{\alpha\beta} K^{\alpha\beta}  \nonumber \Bigg.\\ 
    &\quad + \gsso{5} \nabla^0 K + \gsso{6} R + \gsso{7} R^{00} \Big)  + \gsso{8} K_{\mu\nu} + \gsso{9} \nabla^0 K_{\mu\nu} + \gsso{10} K_{\mu\alpha} K^{\alpha}{}_{\nu} \nonumber \\
    &\quad+ \gsso{11} K K_{\mu\nu} + \frac{\Mpl^2}{2} G_{\mu\nu} + \gsso{13} R_\mu{}^0{}_\nu{}^0+ \gamma^{\po}_{1,0} \epsilon_\mu{}^{\alpha\beta0} \nabla_\alpha K_{\beta\nu} + \gamma^{\po}_{2,0} \epsilon_\mu{}^{\alpha\beta0} R_{\alpha\beta}{}^0{}_\nu \Big] \nonumber \\
    &+ t_a \Big[\gato{1} + \gato{2} K + \gato{3}  K^2 + \gato{4} K_{\alpha\beta} K^{\alpha\beta} + \gato{5} \nabla^0 K + \gato{6} R + \gato{7} R^{00}
    \Big]
    \bigg\} \, ,
\end{align}
whereas all terms at $\ell \geq 1$ remain as displayed in \Eq{eq:S1v1}. 


\paragraph{Construction of $S_2$.}

The construction of $S_2$ proceeds in a manner analogous to that of $S_1$. However, now we also have to include operators which contain covariant derivatives acting on one of the advanced fields. Previously, these terms were redundant as $S_1$ only features one advanced field per term, such that any derivative acting on $t_a$ or $a_{\mu\nu}$ can be integrated by parts. This integration by parts yields terms already considered in $S_{1}$. Conversely, operators in $S_2$ feature two advanced fields, such that this redundancy does not hold anymore and these derivative terms have to be considered explicitly. Moreover, $S_2$ admits a broader class of tensor structures, including higher-rank objects such as the rank-4 tensor $N_{\mu\nu\rho\sigma}$. As a result, the number of terms found in $S_2$ is significantly larger compared to the case of $S_1$. To keep things simple, we restrict the construction of $S_2$ to \textit{zeroth order} in derivatives in this section. A systematic approach for including higher derivative operators is outlined in \App{app:S2_second_order}. 

Just as before, we begin by expanding each term in $S_2$ in powers of $(g^{00}+1)$: 
\begin{align}
    N &= \sum_{\ell=0}\left(g^{00} + 1 \right)^\ell N_{\ell}(R_{\mu\nu\rho\sigma}, K_{\mu\nu}, n_\mu, \nabla_\mu; t)  \, ,\\
    N_{\mu\nu} &= \sum_{\ell=0}\left(g^{00} + 1 \right)^\ell N_{\mu\nu, \ell}(R_{\mu\nu\rho\sigma}, K_{\mu\nu}, n_\mu, \nabla_\mu; t)   \, , \\
    N_{\mu\nu\rho\sigma} &= \sum_{\ell=0}\left(g^{00} + 1 \right)^\ell N_{\mu\nu\rho\sigma, \ell}(R_{\mu\nu\rho\sigma}, K_{\mu\nu}, n_\mu, \nabla_\mu; t)  \, .
\end{align}
Restricting to zeroth order in derivatives, the scalar building block in \Eq{eq:S1_M} simplifies to just the first operator, a function of time. Higher tensorial objects, such as vectors and tensors, may only be build out of $n_\mu$ and $g_{\mu\nu}$ multiplying arbitrary EFT coefficients, as all other objects are of higher order in derivatives (see e.g. $M^{ss}_{\mu, \ell}$ and $\Tilde{M}^{ss}_{\mu\nu, \ell}$ in \Eqs{eq:Mts} and \eqref{eq:Mss}). As a result, we obtain: 
\begin{align}
    N_\ell &= \bt{1} \bigg. \, , \\
    N_{\mu\nu, \ell} &= \bt{2} n_\mu n_\nu + \bt{3} g_{\mu\nu} \bigg. \, , \\
    N_{\mu\nu\rho\sigma, \ell} &= \bt{4} n_\mu n_\nu n_\rho n_\sigma + \bt{5} g_{\mu\nu} g_{\rho\sigma} + \bt{6} g_{\mu(\rho} g_{\sigma)\nu} \bigg. \\
    +& \frac{1}{2} \bt{7} \left(g_{\mu\nu}n_\rho n_\sigma + n_\mu n_\nu g_{\rho\sigma} \right) + \frac{1}{2} \bt{8} \left(g_{\mu(\rho} n_{\sigma)} n_\nu + g_{\nu(\rho} n_{\sigma)} n_\mu \right) \bigg. \, .
\end{align}
In the retarded unitary gauge, the noise functional thus takes the form: 
\begin{align}\label{eq:S2v1}
    S_2 &= i \int \dd^4 x \sqrt{-g} \sum_{\ell=0} \left(g^{00} + 1 \right)^\ell \Big[ \bt{1} \left( t_a \right)^2 + \bt{2} a^{00} t_a + \bt{3} a^{\mu\nu} g_{\mu\nu} t_a + \beta_{4,\ell} \left( a^{00} \right)^2 \nonumber \bigg. \\
    &\quad + \beta_{5,\ell} (a^{\mu\nu} g_{\mu\nu})^2 + \beta_{6,\ell} a^{\mu\nu} g_{\mu\rho} g_{\nu\sigma} a^{\rho\sigma} 
    + \beta_{7,\ell} a^{\mu\nu} g_{\mu\nu} a^{00} + \beta_{8,\ell} a^{0\mu} a^{0\nu} g_{\mu\nu} \Big] . \bigg. 
\end{align}
Finally, one can trade these quadratic terms in the advanced variables by linear terms in said variables through the Hubbard–Stratonovich trick \cite{Hubbard:1959ub,Stratonovich1957}. This will provide a well-posed saddle point approximation for the path integral over the advanced metric components in exchange of including a set of stochastic noises. We illustrate this procedure in \App{app:mixedHS}.


\section{Background evolution}\label{Sec:Background}

The previous section provides a generic construction which exhibits a rich phenomenology (see \Sec{Sec:minimal} for a detailed discussion of the scalar sector in a minimal setup). To illustrate the dynamics of our new theory, we start by studying the background evolution of the universe implied by our modified Einstein equations. As usual, assuming homogeneity and isotropy, these reduce to two modified Friedmann equations from which one can derive a modified continuity equation. Let us consider in detail what role dissipation plays in this set of  equations.


\paragraph{Background metric.} Let us expand around an FLRW background geometry on each branch of the path integral, 
\begin{align}
   \left(g_+\right)_{\mu\nu}  &= \Bar{g}_{\mu\nu} +  \left(\delta g_+\right)_{\mu\nu} , \qquad \mathrm{and} \qquad 
   \left(g_-\right)_{\mu\nu} = \Bar{g}_{\mu\nu} +  \left(\delta g_-\right)_{\mu\nu} ,
\end{align}
where the background FLRW metric with intrinsic spatial curvature $\mathfrak{C}$ is given in spherical coordinates by
\begin{align}
    \bar{g}_{\mu\nu}&=\text{Diag}\left\{-1,\frac{a^{2}(t)}{1-\mathfrak{C}r^{2}},r^{2}a^{2}(t),r^{2}\text{sin}^{2}\left(\theta\right)\right\} \,,\\
    \bar{g}^{\mu\nu}&=\text{Diag}\left\{-1,\frac{1-\mathfrak{C}r^{2}}{a^{2}(t)},\frac{1}{a^{2}(t)r^{2}},\frac{1}{a^{2}(t)r^{2}\text{sin}^{2}\left(\theta\right)}\right\}.
\end{align}
Equivalently, in the Keldysh basis we have
\begin{align}\label{eq:background}
    g_{\mu\nu} &= \frac{\left(g_+\right)_{\mu\nu} + \left(g_-\right)_{\mu\nu}}{2} = \Bar{g}_{\mu\nu} + \delta g_{\mu\nu}, \qquad \mathrm{and} \qquad 
    a^{\mu\nu} = \left(g_+\right)^{\mu\nu} - \left(g_-\right)^{\mu\nu} = \delta a^{\mu\nu}.
\end{align}
Note that the advanced field $a^{\mu\nu}$ does not have a classical background: the advanced component starts linear in perturbations, a well-known fact of non-equilibrium QFT \cite{Calzetta:2008iqa, kamenev_2011}. Hence we simply indicate it by $a^{\mu\nu}$. It follows that the lowest order in perturbations is given by $S_1$, the noise effective functional $S_2$ starting at quadratic order in perturbations. 

In the general theory defined in \eqref{eq:S1v1}, it is enough to consider the lowest order $\ell = 0$ in the $(1 + g^{00})^\ell = (\delta g^{00})^\ell$ expansion, which had been simplified using field redefinitions in \eqref{eq:S1fin}. At last, the lowest order in perturbations is obtained by fixing the retarded metric and its derived quantities at their background value. Schematically, it amounts to consider the restriction of \Eq{eq:S1try} to 
\begin{align}\label{eq:S1tryback}
    \bar{S}_1 = \int \dd^4 x \sqrt{-g}& \, M_{\mu\nu} (\bar{R}_{\mu\nu\rho\sigma}, \bar{g}^{00}, \bar{K}_{\mu\nu}, \bar{\nabla}_\mu; t) a^{\mu\nu},  
\end{align}
where bars represent background values and we dropped the $t_a$ terms because they induce equations that are equivalent to the Einstein equations.   


\subsection{The Friedmann equations}

Let us first discuss the background Einstein equations, which then reduce to the two Friedmann equations,
\begin{align}
    \frac{\delta \bar{S}_1}{\delta a^{\mu\nu}} = 0 \qquad \Rightarrow \qquad M_{\mu\nu} (\bar{R}_{\mu\nu\rho\sigma}, \bar{g}^{00}, \bar{K}_{\mu\nu}, \bar{\nabla}_\mu; t) = 0.
\end{align}

\paragraph{The first Friedmann equation.} The first Friedmann equation consists of the $00$-component of the background Einstein equations. Considering the restriction of \Eq{eq:S1fin} to background values, we obtain
\begin{align}\label{modEE00}
   \left(\gtto{1} - \gsso{1}\right) & + \left( \gtto{2} - \gsso{2}\right) \bar{K} + \left(\gtto{3} -  \gsso{3}  \right)  \bar{K}^2 + \left( \gtto{4} - \gsso{4} \right) \bar{K}_{\alpha\beta} \bar{K}^{\alpha\beta} \nonumber \\
    +&\gamma^{ts}_{3,0}\bar{\nabla}_{\beta}K^{\beta}_{\;0}+ \left(\gtto{5} - \gtso{2} - \gsso{5} \right)\bar{\nabla}^0 \bar{K} - \gsso{6} \bar{R} - \gsso{7} \bar{R}^{00} + \frac{\Mpl^2}{2} \bar{G}_{00} = 0.
\end{align}
Injecting the background values for the retarded quantities, we obtain 
\begin{align}\label{eq:F1}
    3M_1^2H^2 = c_1 + c_2 H + c_3 \dot{H},
\end{align}
where we assumed the coefficient of $H^2$ is positive and defined
\begin{align}
    M_1^2 &= \Mpl^2 + 6 \left(\gtto{3} -  \gsso{3}  \right)  + 2 \left( \gtto{4} +\gsso{7} - \gtso{3} - \gsso{4} \right) - 8 \gsso{6} \,,
\end{align}
as well as
\begin{subequations}\label{eq:c1-3}
\begin{align}
    c_1 &= -2 \left(\gtto{1} - \gsso{1}\right)-\frac{\mathfrak{C}}{a^{2}(t)}\left(3\Mpl^{2}+12\gamma_{6,0}^{ss}\right) \,, \bigg.\\
    c_2 &= - 6  \left( \gtto{2} - \gsso{2}\right)\,,\bigg. \\
    c_3 &=  6\left(\gtto{5} - \gtso{2} - \gsso{5} \right) + 12  \gsso{6} - 6 \gsso{7}. \bigg.
\end{align}
\end{subequations}


\paragraph{The second Friedmann equation.} The second Friedmann equation consists of the trace-part of the $ij$-components of the background Einstein equations. Again, restricting ourselves to the background values of \Eq{eq:S1fin}, we obtain
\begin{align}\label{eq:tadpoleii}
    & \bar{g}_{ij} \Big( \gsso{1} + \gsso{2} \bar{K} + \gsso{3}  \bar{K^2} + \gsso{4} \bar{K_{\alpha\beta} K^{\alpha\beta}} + \gsso{5} \bar{\nabla}^0 \bar{K}  + \gsso{6} \bar{R} + \gsso{7} \bar{R}^{00} \Big) \nonumber \\
    +& \gsso{8} \bar{K}_{ij} + \gsso{9} \bar{\nabla}^0 \bar{K}_{ij} + \gsso{10} \bar{K}_{i\alpha} \bar{K}^{\alpha}{}_{j} + \gsso{11} \bar{K} \bar{K}_{ij} + \frac{\Mpl^2}{2} \bar{G}_{ij} + \gsso{13} \bar{R}_i{}^0{}_j{}^0 = 0.
\end{align}
Using the background values for the retarded quantities, we obtain 
\begin{align}\label{eq:F2}
   2 M_{2}^2 \dot{H} =  c_4 + c_5 H + c_6 H^2,
\end{align}
where we assumed the coefficient of $\dot H$ is positive and defined
\begin{align}
    M_{2}^2 &=    \Mpl^2 +  3 \gsso{5}  - 6 \gsso{6} +3 \gsso{7}  + \gsso{9} +  \gsso{13}  \,,
\end{align}
as well as
\begin{subequations}\label{eq:c4-6}
\begin{align}
    c_4 &= 2\gsso{1}+\frac{\mathfrak{C}}{a^{2}(t)}\left(\Mpl^{2}+12\gamma_{6,0}^{ss}\right) \,, \bigg.\\
    c_5 &= 6 \gsso{2} + 2 \gsso{8}\,, \bigg.  \\
    c_6 &= 18 \gsso{3} + 6  \gsso{4} + 24   \gsso{6} - 6 \gsso{7} + 2\gsso{10} + 6 \gsso{11} - 3 \Mpl^2 - 2\gsso{13}. \bigg. 
\end{align}
\end{subequations}


\paragraph{Discussion.}  The expressions for the first and second Friedmann equations \eqref{eq:F1} and \eqref{eq:F2} derived above are not of the same form as the original ones,
\begin{align}
    3 \Mpl^2 H^2 &= \rho \,,    &    2  \Mpl^2 \dot{H} &= - (\rho + p)\,,
\end{align}
where $\rho$ and $p$ represent the energy density and pressure of the fluid, as they include terms with $\dot{H}$ in the first Friedmann equation and terms with $H^{2}$ in the second one. Furthermore, they include too many parameters — for example, $M_{1,2}^{2}$ and $c_{3,6}$ are degenerate, as we can reshuffle the Friedmann equations to remove either of them. To simplify the upcoming analysis we rewrite
\begin{align}
    3\Mpl^{2}H^2 &= \alpha_{1}+\alpha_{2}H \,, \Big. \label{eq:modifiedFriedmanfinal}\\
   2\Mpl^{2}\dot{H} &= \alpha_{3}+\alpha_{4}H\,, \label{eq:modifiedFriedmanfinal2}
\end{align}
where {
\begin{subequations}\label{eq:alphac}
\begin{align}
    \alpha_{1}=& 3\Mpl^{2}\left(\frac{c_{3}c_{4}+2c_{1}M_{2}^{2}}{6M_{1}^{2}M_{2}^{2}-c_{3}c_{6}}  \right) \,, \quad\quad \alpha_{2}= 3\Mpl^{2}\left(\frac{c_{3}c_{5}+2c_{2}M_{2}^{2}}{6M_{1}^{2}M_{2}^{2}-c_{3}c_{6}} \right)  \,,\\
    \alpha_{3}=& 2\Mpl^{2}\left(\frac{c_{1}c_{6}+3c_{4}M_{1}^{2}}{6M_{1}^{2}M_{2}^{2}-c_{3}c_{6}}\right) \,,  \quad\quad \alpha_{4}= 2\Mpl^{2}\left(\frac{c_{2}c_{6}+3c_{5}M_{1}^{2}}{6M_{1}^{2}M_{2}^{2}-c_{3}c_{6}}\right) \label{eq:alphac2}  \,.
\end{align}
\end{subequations}
}
Our results for the modified Friedmann equations should be interpreted following our general discussion in the introduction around \eqref{eq:mainidea}. To make this more precise and informative, let us compare our modified equations with the standard expectation from a single fluid cosmology for a flat FLRW spacetime in a closed-system approach,
\begin{align}\label{bkgdsummary}
    \text{Open: } \left\{ \begin{array}{l}
     3\Mpl^{2}   H^2 = \alpha_{1}+\alpha_{2}H \,, \Big.\\
   2\Mpl^{2}\dot{H} = \alpha_{3}+\alpha_{4}H\,,
    \end{array} \right.
    \qquad \text{vs} \qquad
    \text{Closed: } \left\{\begin{array}{l}
        3 \Mpl^2 H^2 = \rho \,, \Big.\\
        2  \Mpl^2 \dot{H} = - (\rho + p)\,,
    \end{array}\right.
\end{align}
The effect of an open dynamics is that new terms appear that are at first sight unfamiliar from the usual closed system case. However, the emergence of terms proportional to the Hubble parameter $H$ in the Friedmann equations is often encountered in the context of non-conservative dynamics. In flat FLRW spacetime, $\mathfrak{C}=0$, let us consider a the energy-momentum tensor of a dissipative fluid
\begin{align}
    T_{\mu\nu} = \rho u_\mu u_\nu +  (p -  \zeta \nabla_\gamma u^\gamma) \left( g_{\mu\nu} + u_\mu u_\nu \right),
\end{align}
where $u_\mu$ is the four-velocity of the fluid and $\zeta$ the \textit{bulk viscosity} \cite{Weinberg:1971mx, Brevik:2017msy}. In the local comoving frame of the fluid and at the homogeneous level, $u_\mu \doteq (-1,0)$, and so we find $\nabla_\gamma u^\gamma=3H$. In this case the Friedmann equations become
\begin{align}
\text{Bulk viscosity: } \left\{\begin{array}{l}
        3 \Mpl^2 H^2 = \rho \,, \Big.\\
        2  \Mpl^2 \dot{H} = - \left[\rho + \left(p - 3 H \zeta \right)\right]\,,
    \end{array}\right.
\end{align}
A matching with the left-hand side of \Eq{bkgdsummary} can be made by setting
\begin{align}
    \alpha_1 = \rho\,, \qquad\alpha_{2}=0\,,\qquad \alpha_{3} = -(\rho + p)\,, \qquad \alpha_4 = 3 \zeta.
\end{align}
A very explicit example discussed in \cite{Lau:2024mqm} can be found by picking 
\begin{align}
    S_{\mathrm{eff}} \supset \int \dd^4x \sqrt{-g} \gsso{8} K_{\mu\nu} a^{\mu\nu}.
\end{align}
In the unitary gauge, this operator correspond to a bulk-viscosity $\zeta = 2 \gsso{8}/3$. As we shall see in \Sec{sec:tensor}, it also comes with other physical implications at the level of the perturbations, for instance dissipation in the tensor sector. 

Another example of the open dynamics presented in the left-hand side of \Eq{bkgdsummary} comes from Dvali-Gabadadze-Porrati (DGP) \textit{brane-world gravity} \cite{Dvali:2000hr}. Consider a four-dimensional braneworld embedded in a five-dimensional Minkowski spacetime, whose action is given by 
\begin{align}
    S_{(5)} = - \frac{1}{16\pi} M^3 \int \dd^5x \sqrt{-g_{(5)}} R_{(5)} - \frac{1}{16\pi} \Mpl^2 \int \dd^4x \sqrt{-g} R + \int \dd^4x \sqrt{-g} \mathcal{L}_m + S_{\mathrm{GH}},
\end{align}
with $M$ the five-dimensional Planck scale and $S_{\mathrm{GH}}$ the Gibbons-Hawking action. The first term is the Einstein-Hilbert action in $5d$. The cosmological implications of this model has been investigated in \cite{Lue:2004rj} where the authors found that the first Friedmann equation now contains a term linear in $H$,
\begin{align}\label{eq:DGP}
    3  \Mpl^2 \left(H^2 \pm \frac{H}{r_0} \right) = \rho\,, \quad \text{with} \quad r_0 \equiv \frac{\Mpl}{2M^3}.
\end{align}
Here $r_0$ is a crossover scale, which controls the distance over which metric fluctuations dissipate into the bulk \cite{Dvali:2000hr}, and the sign depends on the geometry of the brane. This modification is sometimes interpreted as leakage of gravity into extra dimensions. Again, a matching with the left hand side of \Eq{bkgdsummary} can be made by setting
\begin{align}
     \alpha_1 = \rho\,, \qquad \alpha_2 = \pm\frac{3\Mpl^2}{r_0} \,,\qquad \alpha_{3} = -(\rho + p)\,, \qquad  \alpha_{4}= 0.
\end{align}

Far from being exhaustive, these two examples simply highlight how non-trivial $\alpha_i$ coefficients can emerge whenever the background evolution exhibits dissipative and friction-like effects. Note that constraints and bounds could be placed on these coefficients using physical principles such as the notion of a second law of thermodynamics \cite{Cheung:2018cwt, Cheung:2023hkq, Aoude:2024xpx, DuasoPueyo:2024rsa, Lau:2024mqm}.


\subsection{The total continuity equation}\label{subsec:cont}

Let us now turn our attention the modified continuity equation obtained from the modified Einstein equation and the Bianchi identity,
\begin{align}
    \Mpl^2 \bar{G}_{\mu\nu}+(\text{modifications})_{\mu\nu} = \bar{T}^{(\phi)}_{\mu\nu}  \qquad \Rightarrow \qquad \bar{\nabla}^\mu  \bar{T}^{(\phi)}_{\mu \nu} =\nabla^\mu (\text{modifications})_{\mu\nu}\,.
\end{align}
Because of the isotropy of the background, the only non-trivial equations is the one with $\nu=0$. The Friedmann equations \Eqs{eq:F1} and \eqref{eq:F2} for a flat FLRW spacetime can be formally written as 
\begin{align}
   3 \Mpl^2 H^2 &= 2(\bar{M}^{tt}_{\ell=0} + \bar{M}^{ts}_{0,\ell=0} +  \bar{\bar{M}}^{ss}_{00,\ell=0} ), \\
   - \Mpl^2(3H^2 + 2 \dot{H})  &=2 \bar{\bar{M}}^{ss}_{ii, \ell=0} ,
\end{align}
where we defined $\bar{\bar{M}}^{ss}_{\mu\nu, \ell=0} \equiv \bar{M}^{ss}_{\mu\nu, \ell=0}  - (\Mpl^2/2) \bar{G}_{\mu\nu}$ to extract the familiar right-hand side from \Eq{eq:S1fin}.
Combining these two equations together, we obtain the modified continuity equation 
\begin{align}\label{eq:cont2}
 &2(\dot{\bar{M}}^{tt}_{\ell=0} + \dot{\bar{M}}^{ts}_{0,\ell=0} + \dot{\bar{\bar{M}}}^{ss}_{00,\ell=0} ) + 6H( \bar{M}^{tt}_{\ell=0} + \bar{M}^{ts}_{0,\ell=0} +  \bar{\bar{M}}^{ss}_{00,\ell=0} +  \bar{\bar{M}}^{ss}_{ii, \ell=0}  ) = 0,
\end{align}
from which one can easily recognize the total energy density and the total pressure
\begin{align}
    \rho_\phi + \rho_{\mathrm{env}}  &=   2( \bar{M}^{tt}_{\ell=0} + \bar{M}^{ts}_{0,\ell=0} +  \bar{\bar{M}}^{ss}_{00,\ell=0})  \\
    P_\phi + P_{\mathrm{env}}   &= 2  \bar{\bar{M}}^{ss}_{ii, \ell=0}.
\end{align}
Explicitly, in terms of the EFT coefficients defined above, these expressions read
\begin{align}
   \rho_\phi + \rho_{\mathrm{env}} &= \alpha_{1}+\alpha_{2}H \\
   \rho_\phi + \rho_{\mathrm{env}} + P_\phi + P_{\mathrm{env}} &= - \left(\alpha_{3}+\alpha_{4}H \right),
\end{align}
which obey the standard relation 
\begin{align}
    \dot{\rho}_\phi + \dot{\rho}_{\mathrm{env}}  +3H (\rho_\phi + \rho_{\mathrm{env}} + P_\phi + P_{\mathrm{env}}) = 0.
\end{align}
This shows explicitly what we anticipated from the very beginning: the total energy-momentum  tensor of the full closed theory is always conserved as expected. However, the effective dynamics of the energy-momentum tensor of the open system appears non-standard because of the presence of an environment, or equivalently, because gravity is modified by the environment. 

Note that considering $\phi$ as a dark fluid, a natural application of this framework would be \textit{interacting dark sectors} \cite{Guo:2004vg, Cai:2004dk, 
DiValentino:2017iww} where stress-energy tensor is not conserved separately between dark matter and dark energy components. In this case, the continuity equation of the quintessence field takes the form 
\begin{align}
    \dot{\rho}_\phi + 3H( \rho_\phi + P_\phi) = Q,
\end{align}
where $Q$ encodes the lack of conservation of the $\phi$ stress-energy tensor that depends on the properties of the coupling and the environment. Depending on the time dependence of $Q$, term proportional to $H$ and $\dot{H}$ can appear in the first Friedmann equation once solving for $\rho_\phi$ \cite{Ashmita:2024ueh}.


\paragraph{Energy conditions and the space of solutions.} In general relativity, one often supplements the theory with certain energy conditions on the matter sector to avoid pathologies. The weakest and most robust\footnote{The NEC can be violated both in classical theories \cite{Dubovsky:2005xd,Creminelli:2006xe} and by quantum effects. We find it nevertheless useful to state its implications.} of these energy conditions is the Null Energy Condition (NEC). For a homogeneous and isotropic universe this requires that the full energy density and pressure satisfy $\rho+p\geq 0$. In our open system approach to gravity, one would like to know if in the full theory, the full energy-momentum tensor of $\{\text{system} + \text{environment}\}$ satisfied a given energy condition. This may be used to restrict the possible space of open theories of gravity by imposing inequalities among the effective couplings. 

If the NEC is imposed on the full system, it implies $\dot H\leq 0$. To make the implications of this constraint explicit, we first combine the dissipative Friedmann equations in \eqref{bkgdsummary} to find the algebraic quadratic equation for $H$, which is easily solved
\begin{align}\label{Hquadratic}
  3\Mpl^{2}H^2 -\alpha_{1}-\alpha_{2}H =0 \quad \rightarrow \quad H=\frac{1}{6\Mpl^{2}}\left(\alpha_{2}\pm\sqrt{12\Mpl^{2}\alpha_{1}+\alpha_{2}^{2}}\right)\,.
\end{align}
Let $H_{1,2}$ be the two solutions of this equation. Plugging these into the second Friedmann equation and demanding $\dot H\leq 0$ one finds the condition
\begin{align}
    \text{Null Energy Condition: }\quad \alpha_{3}+\frac{\alpha_{4}}{6\Mpl^{2}}\left(\alpha_{2}\pm\sqrt{12\Mpl^{2}\alpha_{1}+\alpha_{2}^{2}}\right)\leq0,
\end{align}
Moreover, if one assumes a flat FLRW universe as we are doing here, then the reality of $H$ follows from the discriminant of \eqref{Hquadratic} and requires
\begin{align}
    12\Mpl^{2}\alpha_{1}+\alpha_{2}^{2}\geq0\,.
\end{align}
Recall that the $\alpha$'s were given in terms of the $\gamma$'s through \eqref{eq:alphac}, \eqref{eq:c1-3} and \eqref{eq:c4-6}. If this condition is satisfied one can find two solutions for $H$ purely algebraically. For comparison, from gravity in a closed system one would find $H^2\propto \rho$ and so, for a given choice of $\rho$, one can find both an expanding, $H>0$, and a contracting solution, $H<0$. Conversely, in our case, the two solutions of \eqref{Hquadratic} can a priori have arbitrary signs. In particular, it is possible that they are for example both positive. In this case, the system does not admit a spatially-flat contracting solution. 

Finally, one may want to require that the two modified Friedmann equations admit some solution. A necessary condition is that the time-dependent coefficients $\alpha_{1,...,4}(t)$ are such that the function of time obtained from the algebraic solution of \eqref{Hquadratic} also solves the second Friedmann equations, where $\dot H$ must be written as time derivative of this solution. This will lead to a non-linear constraint for $\alpha_{1,...,4}(t)$ and their time evolution akin to the continuity equation. We can derive this by taking the time derivative of the first modified Friedmann equation and comparing it with the modified second Friedmann equation,
\begin{align}
    6\Mpl^{2}H\dot{H}=\dot{\alpha}_{1}+\dot{\alpha}_{2}H+\alpha_{2}\dot{H},\\
    6\Mpl^{2}H\dot{H}=3H\alpha_{3}+3H^{2}\alpha_{4}\,,
\end{align}
such that
\begin{equation}\label{conteqalpha}
    \dot{\alpha}_{1}+\dot{\alpha}_{2}H+\frac{\alpha_{2}}{2\Mpl^{2}}\left(\alpha_{3}+H\alpha_{4}\right)=3H\alpha_{3}+\frac{\alpha_{4}}{\Mpl^{2}}\left(\alpha_{1}+H\alpha_{2}\right).
\end{equation}
If one wishes one can input the algebraic solutions for $H$ in \eqref{Hquadratic} into this expression to obtain a modified continuity equation for the $\alpha(t)$'s alone. Unlike the usual continuity equation, $\dot{\rho}+3H\left(\rho+P\right)=0$, in our case the relation is non-linear, the non-linearity being induced by $\alpha_{2,4}\neq0$. This non-linearity is therefore a signal of dynamics beyond the original Friedmann equations in a perfect fluid background. 

Note that to solve the pair of Friedmann equations in \eqref{eq:modifiedFriedmanfinal} and \eqref{eq:modifiedFriedmanfinal2} one need to specify three of the four $\alpha$'s, with the fourth being obtained by solving \eqref{conteqalpha}. More intuitively, one needs to introduce three generalized ``equations of state" that relate the different $\alpha$'s, for example
\begin{equation}
    f_{1}\left(\alpha_{1},\alpha_{2},\alpha_{3},\alpha_{4}\right)=0\quad,\quad f_{2}\left(\alpha_{1},\alpha_{2},\alpha_{3},\alpha_{4}\right)=0\quad,\quad f_{3}\left(\alpha_{1},\alpha_{2},\alpha_{3},\alpha_{4}\right)=0.
\end{equation}
These generalized equations of state could be arbitrarily complicated. A tractable scenario is one where these are approximately linear relations
\begin{equation}
    \alpha_{2}=\frac{\lambda_{2}}{\Mpl}\alpha_{1}\quad,\quad\alpha_{3}=\lambda_{3}\alpha_{1}\quad,\quad \alpha_{4}=\frac{\lambda_{4}}{\Mpl}\alpha_{1}.
\end{equation}
In this case, the modified continuity equation \eqref{conteqalpha} simplifies to
\begin{equation}
    \dot{\alpha_{1}}\left(1+\frac{\lambda_{2}H}{\Mpl}\right)-3\lambda_{3}H\alpha_{1}+\alpha^{2}_{1}\frac{\lambda_{2}\lambda_{3}\Mpl-2\lambda_{4}\Mpl-\lambda_{2}\lambda_{4}H}{2\Mpl^{4}}=0.
\end{equation}

The main point to keep in mind is that the class of solutions for all possible $\alpha_{1,\dots,4}(t)$ coincides with the class of solution obtained in gravity as a closed system because if one knew how the environment evolves one could always map our open case to the closed one. However, it may happen that some solutions are easier to find or construct in our formalism. 


\section{Tensor sector: primordial gravitational waves}\label{sec:tensor}

The central novel feature of our theory is its incorporation of dynamical gravity, which allows us to study gravitational effects, such as the generation and propagation of gravitational waves (GWs) during inflation. These phenomena are encoded in Transverse and Traceless (TT) metric perturbations. In this section, we derive the quadratic action that governs this sector, and we compute the corresponding propagators and power spectrum. Finally, we extract the tensor-to-scalar ratio, which enables us to confront the parameters of our theory with current and upcoming observational data.

\subsection{The transverse traceless sector of linear perturbations}

To derive the linear equation of motion for GWs, we expand \Eq{eq:S1v1} and \eqref{eq:S2v1} to quadratic order in perturbations and perform a scalar-vector-tensor decomposition of $\delta g_{\mu\nu}$ and $a^{\mu\nu}$ (recall that the latter is already of first order in perturbations) and restrict to the TT sector according to:
\begin{equation}
    g_{ij} = a^2(t) \left(\delta_{ij} + h_{ij}\right), \qquad a^{ij} = a^{-2}(t) h_{ij}^{a},
\end{equation}
with $h_{ij}$ and $h^a_{ij}$ transverse $\partial_i h_{ij}=0=\partial_i h^a_{ij}$ and traceless $\delta^{ij} h_{ij}=0= \delta^{ij}h^a{}_{ij}$. The index $a$ on $h^a_{ij}$ denotes that this is an advanced variable.

We begin by extracting the TT sector of $S_1$ in \Eq{eq:S1v1}. $a^{00}$ and $a^{0\mu}$ do not contain any transverse and traceless component, such that we obtain
\begin{align}
    &S_1^{(2)} = \int \dd^4 x \sqrt{-g} \,  a^{ij} \Big[ a^2 h_{ij} \Bar{M}^{ss}_{\ell=0}  + \gsso{8} \delta K_{ij} + \gsso{9} \nabla^0 \delta K_{ij} + \gsso{10} \delta \left( K_{i\alpha} K^{\alpha}{}_{j} \right) \nonumber \\
    &\quad+ \gsso{11} \Bar{K} \delta K_{ij} + \frac{\Mpl^2}{2} \delta R_{ij} + \gsso{13} \delta R_i{}^0{}_j{}^0 + \gamma^{\po}_{1,0} \epsilon_i{}^{\ell m 0} \delta \left( \nabla_\ell  K_{m j} \right) + \gamma^{\po}_{2,0} \epsilon_i{}^{\ell m 0} \delta R^0{}_j{}_{\ell m}
    \Big]  \, ,
\end{align}
where $\Bar{M}^{ss}_{\ell=0}$ denotes \Eq{eq:Mss_scalar} evaluated on the background $\Bar{g}_{\mu\nu}$. The perturbed tensors are given by
\begin{align}
    \delta R_{ij} &= -\frac{1}{2} \nabla^2 h_{ij} + \frac{a^2}{2} \left[ \ddot h_{ij} + 3H \dot h_{ij} + \left(6H^2 + 2\dot H\right) h_{ij} \right] , \label{eq:deltaRijTT} \\
    \delta R_i{}^0{}_j{}^0 &= -a^2 \left[ \frac{1}{2} \ddot{h}_{ij} + H \dot h_{ij} + \left(\dot{H}+H^2\right) h_{ij} \right] , \label{eq:deltaRi0j0TT} \\ 
    \delta R^0{}_j{}_{lm} &= \frac{a^2}{2} \left(\partial_l \dot{h}_{mj} - \partial_m \dot{h}_{lj} \right) , \label{eq:deltaRlm0jTT} \\
    \delta K_{ij} &= \frac{a^2}{2} \dot{h}_{ij} + a^2 H h_{ij} , \label{eq:deltaKijTT}
\end{align}
Plugging these into the action, along with the background values yields:
\begin{align}
    S_1^{(2)} = \frac{1}{2} \int \dd^4 x \sqrt{-g} \, &h_{ij}^{a} \bigg\{\left(\frac{\Mpl^2}{2}-\gsso{9}-\gsso{13}\right)\ddot{h}_{ij} \\
    &+\left[\gsso{8}+ \left(2\gsso{10}+3\gsso{11}+3\frac{\Mpl^2}{2}+2\gsso{13} \right)H\right] \dot{h}_{ij} \nonumber \\
    &- \frac{\Mpl^2}{2} \frac{\nabla^2}{a^2} h_{ij} + \frac{1}{a} \left(\gamma^{\po}_{1,0} + 2\gamma^{\po}_{2,0} \right) \Tilde{\epsilon}_{imn} \partial_m \dot{h}_{nj} \nonumber
    \bigg\}  \, .
\end{align}
To arrive at this result, we have used \Eq{eq:tadpoleii}, the spatial background equation, which imposes $\Bar{M}^{ss}_{\ell=0} = - \Tilde{\Bar{M}}^{ss}_{ii,\ell=0}$. Consequently, every term in $a^2 h_{ij} \Bar{M}^{ss}_{\ell=0}$ cancels with a corresponding $\sim h_{ij}$ term in the perturbed quantities in \Eqs{eq:deltaRijTT}, \eqref{eq:deltaRi0j0TT}, \eqref{eq:deltaRlm0jTT} and \eqref{eq:deltaKijTT}, such that no mass term for the graviton is produced. This is in agreement with our expectation that the graviton is massless, as long as we do not break retarded spatial diffs. Our theory only contains the two helicities of the graviton and the clock field. Note that we lowered the indices on all expressions in this equation, and that $\Tilde{\epsilon}_{imn}$ denotes the totally antisymmetric symbol, not tensor, the two being related via $\epsilon_{0ijk} = \sqrt{-g} \, \Tilde{\epsilon}_{ijk}$.

At quadratic order in perturbations, we further expect to obtain contributions from $S_2$. These contributions, being quadratic in advanced fields, encode stochastic noise terms generated by random fluctuations in the environment. The only contribution to the TT sector at this order in perturbations comes from
\begin{equation}
    S_2 \supset i \int \dd^4 x \sqrt{-g} \, \beta_{6,0} a^{\mu\nu} g_{\mu\rho} g_{\nu\sigma} a^{\rho\sigma} , 
\end{equation}
which results in
\begin{equation}
    S_2^{(2)} = i \int \dd^4 x \sqrt{-g} \, \beta_{6,0} h_{ij}^{a} h_{ij}^{a} .
\end{equation}
Note that we do not have to consider any higher action $S_{\geq 3}$ as these start at least at cubic order in perturbations.

The total quadratic action for the TT sector is therefore given by $S^{(2)} = S_1^{(2)} + S_2^{(2)}$. 
To simplify this expression we introduce
\begin{align}
    c_T^{-2} &\equiv 1 - \frac{2 \gsso{9}}{\Mpl^2} - \frac{2 \gsso{13}}{\Mpl^2} , \\
    \Gamma_T & \equiv c_T^2 \frac{2 \gsso{8}}{\Mpl^2} + c_T^2 H \left[\frac{4\gsso{10}}{\Mpl^2} + \frac{6\gsso{11}}{\Mpl^2} + \frac{4\gsso{13}}{\Mpl^2} +3\left(1-\frac{1}{c_T^2}\right)\right], \\
    \chi & \equiv c_T^2 \left(\frac{2 \gamma^{\po}_{1,0}}{\Mpl^2} + \frac{4 \gamma^{\po}_{2,0}}{\Mpl^2}\right),
\end{align}
with dimensions
\begin{equation}
    \left[c_T\right] = E^0, \qquad \left[\Gamma_T\right] = E^1, \qquad \left[\chi\right] = E^0 , \qquad \left[\beta_{6,0}\right] = E^4 . 
\end{equation}
The quadratic action becomes
\begin{align}
    S^{(2)} = \int \dd^4 x \sqrt{-g} \, &\frac{\Mpl^2}{4 c_T^2} \, h_{ij}^{a}  \bigg[\ddot{h}_{ij} - c_T^2 \frac{\nabla^2}{a^2} h_{ij} +\left(\Gamma_T + 3 H\right) \dot{h}_{ij} \label{eq:S_2_coefficients} \\
    &+ \frac{\chi}{a} \Tilde{\epsilon}_{imn} \partial_m \dot{h}_{nj}  + i c_T^2 \frac{4 \beta_{6,0}}{\Mpl^2} h_{ij}^{a}
    \bigg] . \nonumber
\end{align}
We briefly comment on the effects caused by the various operators that appear in $S^{(2)}$.

\paragraph{Noise $\beta_{6,0}$.} The coefficient $\beta_{6,0}$ characterizes stochastic fluctuations of gravitational waves sourced by the environmental sector. Upon performing the Hubbard-Stratonovich trick \cite{Hubbard:1959ub,Stratonovich1957} this quadratic term in $h_{ij}^a$ can be traded for a term linear in $h_{ij}^a$, at the expense of introducing an auxiliary field $\xi_{ij}$: 
\begin{align}
    &\text{exp}\left\{-\int \dd^{4}x\sqrt{-g}\left[ c_T^2 \frac{4 \beta_{6,0}}{\Mpl^2} h_{ij}^a  h_{ij}^a\right]\right\}  \nonumber \\
    &\qquad =\int[\mathcal{D}\xi_{ij}]\;\text{exp}\left\{-\int \dd^{4}x \sqrt{-g}\left[ \frac{\Mpl^2}{\beta_{6,0} c_T^2}\xi_{ij}\xi_{ij}+i\xi_{ij}h_{ij}^a\right]\right\}.
\end{align}
The field $\xi_{ij}$ is transverse and traceless and follows Gaussian statistics. After introducing $\xi_{ij}$, the equation of motion for $h_{ij}$ reads:
\begin{align}\label{eq:GWs_bf_eom}
    \ddot{h}_{ij} - c_T^2 \frac{\nabla^2}{a^2} h_{ij} +\left(\Gamma_T + 3 H\right) \dot{h}_{ij}  + \frac{\chi}{a} \Tilde{\epsilon}_{imn} \partial_m \dot{h}_{nj} = \xi_{ij}.
\end{align}
Note that the noise constraints discussed in \App{App:noise_constraints} do not place any constraints on the transverse traceless part of $\xi_{ij}$ at first order in advanced fields. This is because advanced diffs at this order \cite{Lau:2024mqm}
\begin{equation}
    a_{\mu\nu} \rightarrow a_{\mu\nu} - \nabla_\mu \epsilon^a_\nu - \nabla_\nu \epsilon^a_{\mu},
\end{equation}
do not contain any transverse traceless parts. We should stress that this noise term cannot be removed by a field redefinition and hence constitutes a true modification of the dynamics of tensor modes and hence of the inflationary predictions for the tensor power spectrum. This should be contrasted with the nice findings of \cite{Creminelli:2014wna}, where it was shown that in the (closed) EFT of inflation the predictions for tensor modes cannot be changed at leading order in derivatives and the first modification only arises at cubic order in derivatives. More importantly, in the open EFT of inflation, the amplitude of the tensor power spectrum will turn out to be proportional to $\beta_6$ and hence does not fix the energy scale of inflation, in stark contrast to the closed case. 


\paragraph{Speed of propagation $c_T^2$.} The presence of $c_T^2$ changes the speed of propagation for tensor modes. An operator that causes such an effect has already been identified in the closed theory as $\delta K_{\mu\nu} \delta K^{\mu\nu} - \delta K^2$ \cite{Creminelli:2014wna}. We can translate this unitary operator into the Schwinger-Keldysh basis by considering the difference of two copies of this term, one for each branch of the path integral
\begin{equation}\label{eq:K2}
    \int \dd^4x \left[ \sqrt{-g_+} \left( K[g_+]^2 - K_{\mu\nu}[g_+]^2 \right) - \sqrt{-g_-} \left( K[g_-]^2 - K_{\mu\nu}[g_-]^2 \right) \right] .
\end{equation}
We can rewrite the above using \cite{Wald:1984rg}
\begin{equation}
    R_{\mu\nu} n^\mu n^\nu = K^2 - K_{\mu\nu} K^{\mu\nu} - \nabla_\mu \left(n^\mu \nabla_\nu n^\nu \right) + \nabla_\nu \left(n^\mu \nabla_\mu n^\nu \right) , 
\end{equation}
and dropping boundary terms. Expressing \eqref{eq:K2} in the Keldysh basis $g_\pm = g \pm a/2$ and expanding to first order in $a^{\mu\nu}$ yields 
\begin{equation}\label{eq:ct2_unitary_1}
    \int \dd^4x \sqrt{-g} \left(-\frac{1}{2} a^{\mu\nu} g_{\mu\nu} R_{\alpha\beta}n^\alpha n^\beta + n^\mu n^\nu \delta_a R_{\mu\nu} - 2 \frac{a^{0\mu}}{g^{00}} n^\nu R_{\mu\nu} - \frac{a^{00}}{g^{00}} n^\mu n^\nu R_{\mu\nu} \right) , 
\end{equation}
where the first term comes the expansion of the metric determinant and the last two terms originate from the expansion of $n^\mu$. Meanwhile the variation of the Ricci-tensor $\delta_a R_{\mu\nu}$ is given by the Palatini identity in \Eq{eq:Palatini}, which can be expressed in terms of the advanced metric using the steps outlined in \App{app:ft}. Plugging this into the above expression results in terms with two covariant derivatives acting on the advanced metric $a^{\mu\nu}$. Via integration by parts the covariant derivatives act on $n^\mu$, which results in extrinsic curvature terms. Neglecting contributions from the acceleration $n^\rho \nabla_\rho n_\mu = 0$ one obtains: 
\begin{equation}\label{eq:ct2_unitary_2}
    \int \dd^4x \sqrt{-g} \, n^\mu n^\nu \delta_a R_{\mu\nu} = \int \dd^4x \sqrt{-g} \, a^{\mu\nu} \left( -n^\rho \nabla_\rho K_{\mu\nu} - K K_{\mu\nu} + \frac{1}{2}g_{\mu\nu} K^2 + \frac{1}{2} g_{\mu\nu} n^\rho \nabla_\rho K \right) .
\end{equation}
From \Eqs{eq:ct2_unitary_1} and \eqref{eq:ct2_unitary_2} one can read of all the operators of the open theory that correspond to the unitary operator $K^2 - K_{\mu\nu}K^{\mu\nu}$ of the closed theory. Indeed, the first term in \Eq{eq:ct2_unitary_2} corresponds to the operator $\gsso{9}$. For simplicity, we will set $c_T = 1$ from now on.

\paragraph{Dissipation $\Gamma_T$.} The term proportional to $\Gamma_T$ captures the dissipation of GWs into the environment. It serves as an example of an operator that arises in an open theory. In a classical (closed) action the size of the term proportional to $\dot{h}_{ij}$ would be fixed by the Hubble parameter. The operator $\gsso{8}$, which generates dissipation at lowest order in derivative expansion, has previously been identified in \cite{Lau:2024mqm}.

\paragraph{Dissipative birefringence $\chi$.} When decomposing $h_{ij}$ into a polarization basis, the term $\chi$ acquires opposite signs for each polarization state due to the presence of the Levi-Civita symbol. Similar parity-violating operators also exist in the closed formulation of the EFToI, however these operators come with at least three derivatives \cite{Creminelli:2014wna}. This implies that the operator found here must correspond to a non-unitary effect. Indeed, just as it was the case for $\Gamma_T$, one can verify that it is not possible to construct a term in a closed action that yields a contribution of the form $\sim \epsilon_{ilm}\partial_l \dot{h}_{mj}$ in the equation of motion. For instance, a term like $\epsilon_{ilm}\partial_l \dot{h}_{mj} h_{ij}$ is a total derivative and thus does not contribute to the dynamics. Consequently we call the effect found here ``dissipative birefringence'', to distinguish it from the ``unitary birefringence'' described in \cite{Creminelli:2014wna}.

The difference between of the two effects can also be studied in open electromagnetism \cite{Salcedo:2024nex}, which was summarized in \Sec{sec:OpenEFT}. In this theory, the operator $\gamma_4$ in \Eq{S1old} is responsible for both unitary and dissipative birefringence. If the EFT is local, $\gamma_4$ must be an analytic function of $\omega$ and $k^2$. At zeroth order, $\gamma_4$ is just a constant, and hence yields a birefringent term $\sim \epsilon_{ijk} \partial_j A_k$ in the equation of motion. This term was shown to be unitary in \cite{Salcedo:2024nex}. At higher order in derivatives $\gamma_4$ can be expanded in its arguments, which yields a term $\sim i \omega$ that results in $\sim i \omega \epsilon_{ijk} \partial_j A_k$ in the equation of motion. The latter is anti-Hermitian and hence dissipative. Note that, contrarily to open electromagnetism, in open gravity, dissipative birefringence starts at one \textit{lower} order in derivatives than its unitary counterpart. 

Finally, note that our construction does not yield a term $\sim \epsilon_{ilm}\partial_l h_{mj}$ in the equation of motion, as this operator is not retarded diff invariant.\\

We proceed to compute the propagators from the action in \Eq{eq:S_2_coefficients}, which will ultimately be used to determine the noise induced power spectrum of GWs and the corresponding tensor-to-scalar ratio.
The action simplifies when both $h_{ij}$ and $h^{a}_{ij}$ are expanded in a polarization basis:
\begin{align}
    h_{ij}(t,x) = \int_{\bfk} \sum_{s} e_{ij}^s(\hat{\bfk}) h^s(t,\bfk) e^{i\bfk \cdot \bfx}, \\
    h^{a}_{ij}(t,\bfk) = \int_{\bfk} \sum_{s} e_{ij}^s(\hat{\bfk}) h_a^s(t,\bfk) e^{i\bfk \cdot \bfx},
\end{align}
where the polarization tensors fulfill
\begin{align}
    e_{ii}^s(\hat{\bfk}) &= k^i e_{ij}^s(\hat{\bfk}) = 0 , &\qquad
    e_{ij}^s(\hat{\bfk}) &= e_{ji}^s(\hat{\bfk}) , &\qquad
    e_{ij}^s(\hat{\bfk}) e_{jk}^s(\hat{\bfk}) &= 0 , \Big. \\
    e_{ij}^s(\hat{\bfk}) e_{ij}^{s'}(\hat{\bfk})^* &= 2 \delta_{s s'} , &\qquad
    e_{ij}^s(\hat{\bfk})^* &= e_{ij}^s(-\hat{\bfk}) &\qquad i \Tilde{\epsilon}_{ijk} k_j e_{km}^s &= \frac{s}{2} k e_{im}^{s} . \label{epsid}
\end{align}
For the left and right circular polarization we have $s=\pm 2$. The quadratic action takes the form
\begin{align}
    S^{(2)} = &\frac{\Mpl^2}{2} \sum_{s} \int_{\bfk} \int \text{d}t \sqrt{-g} \, h_{a}^s(t, -\bfk) \bigg[ \ddot{h}^{s}(t,\bfk) +  \frac{\bfk^2}{a^2} h^{s}(t,\bfk) \label{eq:S2_proptimev1} \\ 
    &+\left(\Gamma_T+ 3 H + \frac{ks}{2a} \chi \right) \dot{h}^{s}(t,\bfk) + \frac{4 \beta_{6,0}}{\Mpl^2} h_{a}^s(t,\bfk)
     \bigg]. \nonumber 
\end{align}


\subsection{Without birefringence}

For simplicity, we first restrict to the case $\chi=0$. Changing to conformal time $\dd t = a(t) \dd \eta$ and canonically normalizing 
\begin{equation}\label{eq:gws_can_norm}
    h^{s}_{c}\equiv \frac{\Mpl}{\sqrt{2}} h^s , \qquad \mathrm{and} \qquad h_{c,a}^{s}\equiv \frac{\Mpl}{\sqrt{2}} h^s_a ,
\end{equation}
results in 
\begin{align}
    S^{(2)} = \sum_{s} &\int_{\bfk} \int \text{d}\eta \, a(\eta)^2 \, h_{c,a}^{s}(\eta, -\bfk) \bigg[ h^{s}_{c}(\eta, \bfk)'' + \bfk^2 h^{s}_{c}(\eta, \bfk) \\
    &+a(\eta)\left(\Gamma_T+ 2 H\right) h^{s}_{c}(\eta, \bfk)' 
    + i a(\eta)^2 \frac{4 \beta_{6,0}}{\Mpl^2} h_{c,a}^{s}(\eta, \bfk) \bigg] . \nonumber
\end{align}
This is just two copies of the action for the decoupled Goldstone $\pir$, which has been studied in detail in \cite{Salcedo:2024smn}. Following their approach, we can rewrite the action as bilinear in the fields
\begin{equation}
    S = \frac{1}{2} \sum_{s} \int_{\bfk} \int \dd \eta \, \left(h^s_c(\eta, -\bfk), h^s_{c,a}(\eta, -\bfk) \right) \begin{pmatrix}
    0 & \hat{D}_A \\
    \hat{D}_R & 2i\hat{D}_K 
    \end{pmatrix} 
    \begin{pmatrix}
    h^s_c(\eta, \bfk) \\
    h^s_{c,a}(\eta, \bfk)
    \end{pmatrix} , 
\end{equation}
with
\begin{align}
    \hat{D}_R &= a^2(\eta) \left[ \partial_\eta^2 + \left(2H + \Gamma_T\right) a(\eta) \partial_\eta + k^2 \right] \bigg. \,,\\
    \hat{D}_A &= a^2(\eta) \left[ \partial_\eta^2 + \left(2H - \Gamma_T \right) a(\eta) \partial_\eta + k^2 - 3a^2 H \Gamma_T \right] \,,\\
    \hat{D}_K &= a^4(\eta) \frac{4 \beta_{6,0}}{\Mpl^2} . 
\end{align}
The retarded Green's function obeys
\begin{equation}
    \hat{D}_R(\eta_1) G^R(k;\eta_1, \eta_2) = \delta(\eta_1-\eta_2) , 
\end{equation}
which results in \cite{Salcedo:2024smn}
\begin{equation}\label{eq:retprop}
    G^R(k;\eta_1,\eta_2) = \frac{\pi}{2} \frac{H^2}{k^3} \left( \frac{z_1}{z_2} \right)^{\nu_\Gamma} ( z_2)^3 \left[ Y_{\nu_\Gamma}(z_1) J_{\nu_\Gamma}(z_2)  - J_{\nu_\Gamma}(z_1) Y_{\nu_\Gamma}(z_2)\right] \theta(\eta_1 - \eta_2) ,
\end{equation}
in terms of Bessel functions of the first kind, with 
\begin{equation}
     \nu_\Gamma \equiv \frac{3}{2} + \frac{\Gamma_T}{2H} \qquad \mathrm{and} \qquad z_i \equiv -k\eta_i \, .
\end{equation}
This can also be rewritten in terms of Hankel functions as
\begin{equation}\label{eq:retpropHankelfct}
    G^R(k;\eta_1,\eta_2) = \frac{\pi}{2} H^2 (\eta_1 \eta_2)^{\tfrac{3}{2}} \left( \frac{\eta_1}{\eta_2} \right)^{\frac{\Gamma_T}{2H}} \Im \text{m} \left[ H^{(1)}_{\nu_\Gamma}\left(-k\eta_1 \right) H^{(2)}_{\nu_\Gamma}\left(-k\eta_2 \right) \right] \theta(\eta_1 - \eta_2) .
\end{equation}
The Keldysh propagator is given by 
\begin{align}
    G^K(k;\eta_1,\eta_2) = i \frac{4\beta_{6,0}}{\Mpl^2} \int \frac{\dd \eta'}{H^4 \eta'{}^4} G^R(k; \eta_1, \eta') G^R(k; \eta_2, \eta') + (\eta_1 \leftrightarrow \eta_2) .
\end{align}
The power spectrum is obtained in the coincident limit of the Keldysh propagator $P_T(k, \eta) = -i G^K(k;\eta, \eta)$.
The reduced GW power spectrum is
\begin{equation}
    \Delta_h^2(k) \equiv \frac{k^3}{2\pi^2} P_T(k), \qquad \mathrm{with} \qquad
    \langle h_{ij}(\bfk) h_{ij}(\bfk') \rangle = (2\pi)^3 \delta^3(\bfk+\bfk') P_T.
\end{equation}
In the super-Hubble regime $z \ll 1$ is thus given by 
\begin{equation}\label{eq:gws_ps_Gamma}
    \Delta_h^2(k) =  \frac{4 \beta_{6,0}}{\Mpl^4} 2^{2\nu_\Gamma} \frac{\Gamma(\nu_\Gamma-1 ) \Gamma(\nu_\Gamma)^2}{\Gamma(\nu_\Gamma-\tfrac{1}{2} )\Gamma(2\nu_\Gamma-\tfrac{1}{2})},
\end{equation}
where an additional factor of 2 accounts for both polarizations. 
\begin{figure}
    \centering
    \includegraphics[width=0.7\linewidth]{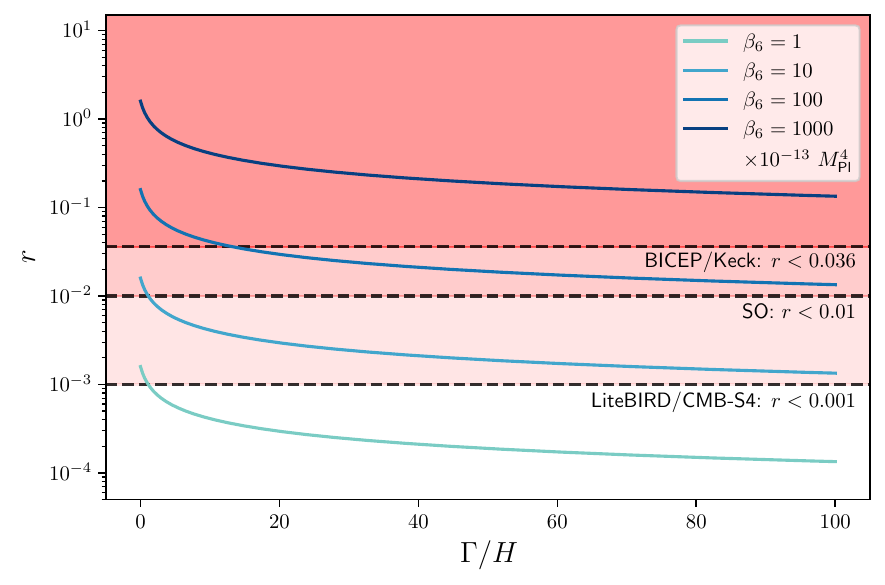}
    \caption{The tensor-to-scalar ratio $r$ as a function of $\Gamma_T/H$, for different choices of $\beta_{6,0}$. The observational constraint $r < 0.036$ \cite{BICEPKeck:2024stm} restricts the allowed region of parameters. Ongoing and future surveys, such as SO \cite{SimonsObservatory:2018koc}, CMB S4 \cite{CMB-S4:2020lpa} and LiteBIRD \cite{LiteBIRD:2022cnt} will put tighter constraint on this parameter.}
    \label{fig:gw_ps_nobf}
\end{figure}
The result is shown in \Fig{fig:gw_ps_nobf} as a function of $\Gamma_T/H$. As expected, the power spectrum is damped for large dissipation. 
Expanding in the weak $\Gamma_T \ll H$ and strong $\Gamma_T \gg H$ dissipation regime  yields: 
\begin{align}
     \Delta_h^2(k) &\propto \begin{dcases}
 \frac{\beta_{6,0}}{\Mpl^4}, &\mathrm{for} \quad \Gamma_T \ll H,\\
 \frac{\beta_{6,0}}{\Mpl^4} \sqrt{\frac{H}{\Gamma_T}} \left[ 1 + \O\left(\frac{H}{\Gamma_T}\right) \right], & \mathrm{for} \quad \Gamma_T \gg H .
\end{dcases}
\end{align}
The tensor-to-scalar ratio is given by
\begin{equation}
    r = \frac{\Delta_h^2(k)}{\Delta_\zeta^2(k)} .
\end{equation}
Using the observed $\Delta_\zeta^2 = 2.1 \times 10^{-9}$ this becomes:
\begin{equation}
    r =  \frac{\beta_{6,0}}{\Mpl^4} 2^{2\nu_\Gamma} \frac{\Gamma(\nu_\Gamma-1 ) \Gamma(\nu_\Gamma)^2}{\Gamma(\nu_\Gamma-\tfrac{1}{2} )\Gamma(2\nu_\Gamma-\tfrac{1}{2})} \times 1.9 \times10^{9} .
\end{equation}
This is plotted for different choices of $\beta_{6,0}$ against $\Gamma_T/H$ in \Fig{fig:gw_ps_nobf}. From this figure we can infer that $\beta_{6,0} \lesssim \left(10^{-3} \Mpl \right)^4 $ to be compatible with the observational bound of $r < 0.036$, for small values of $\Gamma_T/H$. For realizations near the observational bound, the noise sourcing the tensor sector is significant. It would be interesting to investigate if this can be achieved in concrete UV models scenarios (see e.g. \cite{Senatore:2011sp}). Also note that for a given UV model increasing the dissipation parameter $\Gamma_T$ might not lead to a decreased power spectrum - in realistic scenarios increasing the influence of the environment via $\Gamma_T$ also leads to an increase in the noise $\beta_{6,0}$, such that $\Delta_h$ might increase overall. This is for instance realized by the KMS condition $(4 \beta_{6,0}/\Mpl^4) = 2 \pi \Gamma_T T_{\mathrm{eq}}$ for environments that are in thermal equilibrium, if the temperature of the thermal bath $T_{\mathrm{eq}}$ remains fixed.  


\subsection{Including birefringence} 

For $\chi \neq 0$, the quadratic action is given by
\begin{align}
    S^{(2)} = &\sum_{s} \int_{\bfk} \int \text{d}t \sqrt{-g} \, h_{c,a}^s(t, -\bfk) \bigg[ \ddot{h}^{s}_c(t,\bfk) +  \frac{\bfk^2}{a^2} h^{s}_c(t,\bfk) \label{eq:S2_proptimev2} \\ 
    &+\left(\Gamma_T+ 3 H + \frac{ks}{2a} \chi \right) \dot{h}^{s}_c(t,\bfk) + \frac{4 \beta_{6,0}}{\Mpl^2} h_{c,a}^s(t,\bfk)
     \bigg] \nonumber ,
\end{align}
where we have canonically normalised the fields again according to \Eq{eq:gws_can_norm}. 
This can again be written as a bilinear of the fields 
\begin{equation}
    S = \frac{1}{2} \sum_{s} \int_{\bfk} \int \dd \eta \, \left(h^s_c(\eta, -\bfk), h^s_{c,a}(\eta, -\bfk) \right) \begin{pmatrix}
    0 & \hat{D}_A^s \\
    \hat{D}_R^s & 2i\hat{D}_K 
    \end{pmatrix} 
    \begin{pmatrix}
    h^s_c(\eta, \bfk) \\
    h^s_{c,a}(\eta, \bfk)
    \end{pmatrix} , 
\end{equation}
with
\begin{align}
    \hat{D}_R^s &= a^2(\eta) \left[ \partial_\eta^2 + \left(2H + \Gamma_T + \frac{\chi s}{2 a(\eta)} k\right) a(\eta) \partial_\eta + k^2 \right], \\
    \hat{D}_A^s &= a^2(\eta) \left[ \partial_\eta^2 + \left(2H - \Gamma_T - \frac{\chi s}{2 a(\eta)} k\right) a(\eta) \partial_\eta + k^2 - 3 a^2 H \Gamma_T - a  H\chi s k\right], \\
    \hat{D}_K &= a^4(\eta) \frac{4 \beta_{6,0}}{\Mpl^2} . \label{eq:DK}
\end{align}

\paragraph{Homogeneous solutions.}

The homogeneous solutions of $\hat{D}_R^s$ obey
\begin{equation}\label{eq:gws_bf_eom}
    \left[ \partial_\eta^2 - \left( \frac{ 2 + \tfrac{\Gamma_T}{H}}{\eta} - \frac{s \chi}{2} k \right)\partial_\eta + k^2 \right] h^s_k = 0 .
\end{equation}
Before discussing the full solution to this equation we can study the asymptotic behavior of its solutions. For the asymptotic future, $-k\eta \ll 1$, the dissipative birefringence term $\sim \chi s k$ becomes negligible. Asymptotic solutions to this equation are then given by a constant and a decaying solution as $\eta \to 0$. On the contrary, in the asymptotic past, $-k\eta \gg 1$, the dissipative birefringence term dominates, such that the resulting equation reads\footnote{Note that here the birefringent term does not have an $i$, despite having a single spatial derivative, because of the identity \eqref{epsid}.}
\begin{equation}
    \left[ \partial_\eta^2 + \frac{s \chi}{2} k \partial_\eta + k^2 \right] h^s_k = 0 ,
\end{equation}
which has plane waves $e^{-i\omega \eta}$ as solutions with dispersion relations
\begin{equation}\label{eq:gws_bf_dr}
    \omega = -\frac{1}{4} i \chi s k \pm \frac{1}{2} k \sqrt{4-\chi^2} .
\end{equation}
It is clear that the system features an instability for $\chi \neq 0$, as the mode function of the $s=+2$ polarization (since we assumed $\chi\geq 0$) experiences an exponential enhancement as $\eta \to - \infty$, whereas the $s=-2$ polarization decays. This behavior is also called ``amplitude" birefringence, which stands in contrast to ``velocity" birefringence, where the two polarizations move at different speed. Instability in one of the two polarizations also occurs in UV-models that include birefringent terms, such as Chern-Simons (CS) gravity. 
Within CS-gravity, the instability can be handled by either imposing a UV-cutoff or by coupling the $R \tilde{R}$-term to the inflaton field (or any other scalar field), and hence giving this coupling a time-dependence \cite{Alexander:2004wk, Alexander:2009tp, Creque-Sarbinowski:2023wmb, Dyda:2012rj, Alexander:2004us, Callister:2023tws}. Note that when the magnitude of $h_{ij}$ becomes comparable to the background metric, the perturbative approach breaks down and a non-perturbative treatment is necessary.  We stress, that one such modifications is necessary to tame the divergence of the power spectrum of one of the two polarizations. For the moment, we will remain agnostic about such possible model dependent modifications.

The full solution to \Eq{eq:gws_bf_eom} reads
\begin{equation}\label{eq:gws_bf_hg_sol1}
     u^{s}_k \propto e^{-\tfrac{1}{2} s k\eta \chi} \eta^{\nu_\Gamma-\tfrac{1}{2}} \, W_{\kappa_s,\nu_\Gamma}\left( 2i k\eta \lambda \right) ,
\end{equation}
and 
\begin{equation}\label{eq:gws_bf_hg_sol2}
   v^{s}_k \propto e^{- \tfrac{1}{2} s k\eta \chi} \eta^{\nu_\Gamma-\tfrac{1}{2}} \, W_{-\kappa_s,\nu_\Gamma}\left(-2i k\eta \lambda \right).
\end{equation}
Here $W_{\kappa_s,\nu_\Gamma}$ denote Whittaker W-functions and we have introduced the parameters
\begin{align}
    \nu_\Gamma &\equiv \frac{3}{2} + \frac{\Gamma_T}{2H} , \qquad   \lambda \equiv \sqrt{4-\chi^2} , \qquad
    \kappa_s \equiv -\frac{i s \chi}{2\lambda}\left(\nu_\Gamma - \frac{1}{2}\right) .
\end{align}
For $\Gamma_T\in \mathbb{R},\,\chi \in \mathbb{R}$, we have $\nu_\Gamma \in \mathbb{R}$ and  $\kappa \in i \mathbb{R}$ respectively, such that $(u_k^s)^* = v_k^s$, due to the analyticity of $W_{\kappa,\nu}(z)$. Using the asymptotic expansions of the Whittaker W-function:
\begin{align}
    W_{\kappa, \nu}(z) &\sim z^{\tfrac{1}{2}-\mu} \qquad \text{as } \qquad z \to 0 ,\\
    W_{\kappa, \nu}(z) &\sim e^{-\tfrac{1}{2}z} z^\kappa \qquad \text{as } \qquad z \to \infty ,
\end{align}
one can check that the solutions \Eqs{eq:gws_bf_hg_sol1} and \eqref{eq:gws_bf_hg_sol2} have the correct asymptotic behavior discussed above. For instance, in the far past $-k\eta \gg 1$, the two solutions become 
\begin{align}\label{eq:gw_bf_mode_fcts}
    u_k^s &\sim (2i \lambda k \eta)^{\kappa_s} \eta^{\nu_\Gamma -\tfrac{1}{2}} e^{-\tfrac{1}{2}s k\eta \chi} e^{-i\lambda k\eta} , \\
    v_k^s &\sim (2i \lambda k \eta)^{-\kappa_s} \eta^{\nu_\Gamma -\tfrac{1}{2}} e^{-\tfrac{1}{2}sk\eta \chi} e^{+i\lambda k\eta} ,
\end{align}
such that $u_k^s$ corresponds to waves of positive frequency, whereas the second solution $v_k^s$ corresponds negative frequency solutions. Depending on the sign of $s$, we indeed recover an exponential enhancement and an exponential damping. 

\paragraph{Retarded Green's function.}

The retarded Green's function $G^R_s(k;\eta_1, \eta_2)$ obeys 
\begin{equation}
    \left[ \partial_\eta^2 - \left( \frac{ 2 + \tfrac{\Gamma_T}{H}}{\eta} - \frac{s \chi}{2} k \right)\partial_\eta + k^2 \right] G^R_s(k;\eta_1, \eta_2) = H^2 \eta_1^2 \delta(\eta_1 - \eta_2) .
\end{equation}
The normalisation is fixed by the continuity of the retarded Green function at coincident time and its first derivative discontinuity is controlled by the time-dependent prefactor of the $\partial_{\eta_1}^2$ term in \Eq{eq:retprop}, that is 
\begin{align}
G^R_s(k; \eta_2,\eta_2) &= 0, \qquad \mathrm{and} \qquad
\left. \partial_{\eta_1}G^R_s(k; \eta_1,\eta_2) \right|_{\eta_2 = \eta_1} = H^2 \eta^2_2.
\end{align}
$G^R_s(k;\eta_1, \eta_2)$ can thus be build out of the homogeneous solutions as
\begin{align}
    G^R_s(k;\eta_1, \eta_2) = H^2 \eta_2^2 \frac{u_k^s(\eta_1) v_k^s(\eta_2) - u_k^s(\eta_2) v_k^s(\eta_1)}{\mathcal{W}\{u_k^s, v_k^s\}(\eta_2)} \theta(\eta_1 - \eta_2),
\end{align}
such that
\begin{align}\label{eq:ret_prop_bf}
    G^R_s(k;\eta_1, \eta_2) = &\frac{H^2}{\lambda k^3} e^{-i\pi\kappa_s} \left(\frac{z_1}{z_2} \right)^{\nu_\Gamma} (z_1 z_2)^{-\tfrac{1}{2}} (z_2)^3 e^{\tfrac{1}{2}s\chi(z_2-z_1)} \nonumber \\
    &\times \Im\mathrm{m} \Big[W_{\kappa_s,\nu_\Gamma}(-2i\lambda z_1)W_{-\kappa_s,\nu_\Gamma}(2i\lambda z_2)\Big] \theta(\eta_1 - \eta_2). 
\end{align}
To arrive at this result, we have used the Wronskian $ \mathcal{W}\{\cdot, \cdot\}$ between $u_k^s$ and $v_k^s$, which is given by
\begin{align}
    \mathcal{W}\{u_k^s, v_k^s\} = 2i \lambda k \, \eta^{2\nu_\Gamma-1} e^{- s k \eta \chi} e^{i\pi \kappa_s}, 
\end{align}
following from \cite{magnus2004formulas}
\begin{align}
     \mathcal{W}\left\{W_{\kappa,\nu}(z), W_{-\kappa,\nu}(-z)\right\} = e^{\mp i \pi \kappa}, 
\end{align}
where $+$ is used for $\Im\mathrm{m}(z)>0$ and $-$ for $\Im\mathrm{m}(z)<0$. 

Note that for $\chi=0$ the Whittaker functions reduce to Hankel functions
\begin{align}
    W_{0,\nu_\Gamma}(2i z) &= \sqrt{\frac{\pi z}{2}} e^{-i\tfrac{\pi}{4}(1+2\nu_\Gamma)} H_{\nu_\Gamma}^{(2)}(z) , \\
    W_{0,\nu_\Gamma}(-2i z) &= \sqrt{\frac{\pi z}{2}} e^{+i\tfrac{\pi}{4}(1+2\nu_\Gamma)} H_{\nu_\Gamma}^{(1)}(z) ,
\end{align}
from which we recover \Eq{eq:retpropHankelfct}. 

\paragraph{Keldysh propagator.}

The Keldysh propagator is obtained via
\begin{equation}
    G^K_s(k;\eta_1, \eta_2) = i \int \dd \eta' G^R_s(k;\eta_1, \eta') \hat{D}_K G^R_s(k;\eta_2, \eta') + ( \eta_1 \leftrightarrow \eta_2),
\end{equation}
which upon inserting \Eq{eq:ret_prop_bf} yields
\begin{align}
    &G^K_s(k;\eta_1, \eta_2) = \frac{-4 i \beta_{6,0}}{\Mpl^2} \frac{1}{\lambda^2 k^3} e^{-2\pi i \kappa_s} (z_1 z_2)^{\nu_\Gamma-\tfrac{1}{2}} e^{-\tfrac{1}{2}s\chi (z_1+z_2)}  \\
    &\quad \times\bigg( W_{1} W_{2} I_{z_2}\left[\Bar{W},\Bar{W}\right] + \Bar{W}_{1} \Bar{W}_{2} I_{z_2}\left[W,W\right] 
    - \left(W_{1} \Bar{W}_{2}+\Bar{W}_{1} W_{2}\right) I_{z_2}\left[W,\Bar{W}\right]\bigg)   + ( \eta_1 \leftrightarrow \eta_2), \nonumber
\end{align}
where we have introduced the shorthand notations
\begin{align}
    W_i \equiv W_{\kappa_s,\nu_\Gamma}(-2i\lambda z_i) , \qquad \Bar{W}_{i} \equiv W_{-\kappa_s,\nu_\Gamma}(2i\lambda z_i), 
\end{align}
and
\begin{align}
    I_{z_2}[W,\Bar{W}] \equiv \int^{\infty}_{z_2} \dd z' (z')^{1-2\nu_\Gamma} e^{s \chi z'} W_{\kappa_s,\nu_\Gamma}(-2i\lambda z') W_{-\kappa_s,\nu_\Gamma}(2i\lambda z') .
\end{align}
The power spectrum is obtained at the equal time limit $\eta_1 = \eta_2$ via $P^s_T(k,\eta) = -i G^K_s(k;\eta, \eta)$, such that
\begin{align}\label{eq:ps_pre}
    P^s_{T}(k,\eta) &= \frac{-8\beta_{6,0}}{\Mpl^2} \frac{1}{\lambda^2 k^3} e^{-2\pi i \kappa_s} (z)^{2\nu_\Gamma-1} e^{-s \chi z} \Big[ W_{\kappa_s, \nu_\Gamma}(-2i\lambda z)^2 I_z\left[\Bar{W},\Bar{W}\right] \nonumber \\
    &+ W_{-\kappa_s, \nu_\Gamma}(2i\lambda z)^2 I_z\left[W,W\right]
    - 2 W_{\kappa_s, \nu_\Gamma}(-2i\lambda z) W_{-\kappa_s, \nu_\Gamma}(2i\lambda z) I_z\left[W,\Bar{W}\right]\Big].
\end{align}

We now have to evaluate this expression. In the super-Hubble regime $z \ll 1$, we can use the small argument approximation of the Whittaker W-functions
\begin{equation}
    W_{\kappa_s, \nu_\Gamma}(z) = \frac{\Gamma(2\nu_\Gamma)}{\Gamma(\frac{1}{2}+\nu_\Gamma-\kappa_s)} z^{\tfrac{1}{2}-\nu_\Gamma} + \O\left(z^{\tfrac{3}{2}-\Re \mathrm{e}(\nu_\Gamma)}\right) ,
\end{equation}
which is valid as long as $\tfrac{1}{2}-\kappa_s \pm \nu_\Gamma \neq - n$ for $n\in \mathbb{N}$. Furthermore we can expand the Whittaker W-functions as Whittaker M-functions via
\begin{equation}
    W_{\kappa,\mu}\left(z\right)=\frac{\Gamma\left(-2\mu\right)}{\Gamma\left(\frac{1}{2}-\mu-\kappa\right)}M_{\kappa,\mu}\left(z\right)+\frac{\Gamma\left(2\mu\right)}{\Gamma\left(\frac{1}{2}+\mu-\kappa\right)}M_{\kappa,-\mu}\left(z\right),
\end{equation}
which is valid as long as $2 \mu \neq n$ for $n\in \mathbb{N}$, and rotate the Whittaker M-functions with negative arguments via 
\begin{equation}
    M_{\kappa,\mu}\left(z e^{\pm\pi i}\right) = \pm i e^{\pm\mu\pi i} M_{-\kappa,\mu}\left(z\right) .
\end{equation}
Here, we have to choose the ``+" sign in order to avoid crossing the branch cut of $ M_{\kappa,\mu}(z)$ on the negative real axis. Following these steps and taking the $\eta \rightarrow 0$ limit, one arrives at 
\begin{align}\label{eq:psbfpre}
    P^s_T(k) &= \frac{8\pi^2 \beta_{6,0}}{\Mpl^2} \frac{1}{\nu_\Gamma^2 k^3} (2 i \lambda)^{-1-2\nu_\Gamma} \int^{\infty}_{0} \dd z' (z')^{1-2\nu_\Gamma} e^{s \chi z'} M_{-\kappa_s,\nu_\Gamma}(2i\lambda z')^2 .
\end{align}
For $\chi = 0$ one can use
\begin{equation}
    M_{0,\nu_\Gamma}(2 i z) = 2^{\tfrac{1}{2}+2\nu_\Gamma} i^{\tfrac{1}{2}+\nu_\Gamma} z^{\tfrac{1}{2}} \Gamma(1+\nu_\Gamma) J_{\nu_\Gamma}(z) ,
\end{equation}
to recover 
\begin{align}
    P^s_T(k) &= \frac{8\pi^2\beta_{6,0}}{\Mpl^2} \frac{1}{k^3} 2^{2\nu_\Gamma} \Gamma(\nu_\Gamma)^2 \int^{\infty}_{0} \dd z' (z')^{2-2\nu_\Gamma} J_{\nu_\Gamma}(z')^2 \nonumber \\
    &= \frac{2\pi^2\beta_{6,0}}{\Mpl^2} \frac{1}{k^3} 2^{2\nu_\Gamma} \frac{\Gamma(\nu_\Gamma)^2 \Gamma(\nu_\Gamma-1)}{\Gamma(\nu_\Gamma-\tfrac{1}{2}) \Gamma(2\nu_\Gamma-\tfrac{1}{2})} 
\end{align}
which obtains an additional factor of $2/\Mpl^2$ upon going back to the unnormalized field, and is thus consistent with the previous result in \Eq{eq:gws_ps_Gamma}.

\paragraph{Damped polarization $s=-2$.} Within our framework, we may compute the power spectrum of the damped polarization $s=-2$ analytically. For $\chi \neq 0$, the $s=-2$ polarization experiences damping under the dissipative birefringence operator and hence converges. To compute its power spectrum, we can use the integral \cite{Gradshteyn:1943cpj}
\begin{align}\label{eq:gw_integral}
    &\int_0^\infty \dd x \, x^{\nu-1} e^{-bx} M_{\lambda_1, \delta_1-\tfrac{1}{2}}(a_1 x) M_{\lambda_2, \delta_2-\tfrac{1}{2}}(a_2 x) = \bigg. \\
    &\quad =a_1^{\delta_1} a_2^{\delta_2} (b+A)^{-\nu-M} \Gamma(\nu+M) \, F_2\left(\nu+M; \delta_1-\lambda_1, \delta_2-\lambda_2;2\delta_1, 2\delta_2; \tfrac{a_1}{b+A}, \tfrac{a_2}{b+A}\right) ,\nonumber 
\end{align}
with $M = \delta_1+\delta_2$ and $A = (a_1+a_2)/2$, where $F_2$ the second Appell function. From this expression, we obtain 
\begin{align}\label{eq:gw_ps_v1}
     P^{-2}_T(k) &= \frac{2\pi^2\beta_{6,0}}{\Mpl^2} \frac{1}{\nu_\Gamma^2 k^3} \left(\frac{s \chi}{2} +i\lambda\right)^{-3} \nonumber \\ &\times F_2\left(3;\tfrac{1}{2}+\nu_\Gamma+\kappa_s, \tfrac{1}{2}+\nu_\Gamma+\kappa_s; 2\nu_\Gamma+1, 2\nu_\Gamma+1; \tfrac{1}{1-\tfrac{is \chi}{2 \lambda}}, \tfrac{1}{1-\tfrac{i s \chi}{2 \lambda}}\right),
\end{align}
with $F_2$ the second Appell function. 
Via an Euler transformation \cite{Ananthanarayan:2021bqz}, \Eq{eq:gw_ps_v1} can be rewritten as
\begin{align}
     P^{-2}_T(k) 
    = \frac{2\pi^2\beta_{6,0}}{\Mpl^2} \frac{1}{\nu_\Gamma^2 \chi^{3} k^3}  \nonumber F_2\bigg(&3;\tfrac{1}{2}+\nu_\Gamma-\kappa_{-2}, \tfrac{1}{2}+\nu_\Gamma+\kappa_{-2}; \\
    &\quad 2\nu_\Gamma+1, 2\nu_\Gamma+1; i \sqrt{\tfrac{4}{\chi^2}-1}, -i \sqrt{\tfrac{4}{\chi^2}-1} \bigg),
\end{align}
such that the reduced power spectrum $ [\Delta_h^{s}(k)]^2\equiv\frac{k^3}{2\pi^2} P_s(k)$ for the unnormalised field is given by
\begin{align}\label{eq:ps_euler_trafo}
    \left[\Delta_h^{-2}(k)\right]^2  = \frac{\beta_{6,0}}{\Mpl^4} \frac{1}{\nu_\Gamma^2 \chi^{3} } F_2\bigg(&3;\tfrac{1}{2}+\nu_\Gamma-\kappa_{-2}, \tfrac{1}{2}+\nu_\Gamma+\kappa_{-2}; \nonumber \\
    &\quad 2\nu_\Gamma+1, 2\nu_\Gamma+1; \sqrt{1-\tfrac{4}{\chi^2}}, -\sqrt{1-\tfrac{4}{\chi^2}} \bigg) \, ,
\end{align}
which is scale-invariant. 

\paragraph{Enhanced polarization $s=+2$.} If $\chi \neq 0$, the power spectrum of the $s=+2$ polarization diverges, as can be seen from \Eq{eq:psbfpre}. This divergence originates from the exponential enhancement of this polarization at early times, which can also be seen in the dispersion relation in \Eq{eq:gws_bf_dr}. This divergence may be expected, as we are attempting to compute the power spectrum of a GW that travels through an enhancing medium. As previously noted, this approximation cannot remain valid indefinitely: when the amplitude of GWs becomes comparable to that of the background metric, the perturbative expansion ceases to be reliable. In this work we restrict ourselves to the perturbative regime of the theory. It would be interesting to investigate if this effect can explain the observed signal for pulsar timing arrays (PTA).


\section{Scalar sector: mixing with gravity and decoupling}\label{sec:Stuck}

Time diffeomorphism invariance in both the plus and minus branches independently is restored by performing both a \textit{retarded} and an \textit{advanced} St\"uckelberg trick\footnote{A more accurate name would be the Baron Ernst Carl Gerlach St\"uckelberg von Breidenbach zu Breidenstein und Melsbach trick, but we opt for simplicity.}. While the former consists in performing the same infinitesimal time diffeomorphism $t \rightarrow t + \pi_r(x)$ in each branch of the path integral, the latter consists in transforming the two branches in opposite directions, that is the $+$ branch by $t \rightarrow t + \pi_a(x)/2$ and the $-$ branch by $t \rightarrow t - \pi_a(x)/2$. This changes the effective functional according to 
\begin{align}
    S_{\text{3d-diff}_r} [g_{\mu\nu}, a_{\mu\nu}] \rightarrow S_{\text{4d-diff}_r \times \text{t-diff}_a} [g_{\mu\nu}, a_{\mu\nu}, \pir, \pia] \,.
\end{align}
While these re-writing of our theory naively appear to make it more complicated, they actually lead a remarkably eventual simplification: the theory with $\pia$ and $\pir$ admits a simple decoupling limit in which the retarded metric is simply set to its unperturbed background and the advanced metric vanishes
\begin{align}
    \text{decoupling: }\quad S_{\text{4d-diff}_r \times \text{t-diff}_a} [g_{\mu\nu}, a_{\mu\nu}, \pir, \pia] \longrightarrow S[\pir,\pia]\,.
\end{align}
We will show in Subsection \ref{ss:decoupling} that this matches our previous construction in \cite{Salcedo:2024nex}. At last, we will assess the regime of validity of this approximation in Subsection \ref{Sec:minimal}.


\subsection{St\"uckelberg tricks}\label{subsec:stuck}

Let us start by describing the two different types of St\"uckelberg fields we can introduce in a theory that breaks both retarded and advanced time diffeomorphisms.

\subsubsection{Retarded St\"uckelberg trick}\label{subsec:retstuck}

We can explicitly reintroduce the scalar field $\pir$ by performing an infinitesimal \textit{time} diff $t \rightarrow t + \pir(x)$.
This transformation, which acts similarly on both branches of the path integral, belongs to the diagonal subgroup of diffs$_+$ $\times$ diffs$_-$\cite{Akyuz:2023lsm}. Consequently, retarded and advanced fields transform as known from EFTs that break time diffs: $4d$ diff invariant operators simply transform as tensors and hence do not introduce any $\pir$ terms \cite{Lau:2024mqm}. Since advanced and retarded tensors are just linear combinations of tensors in the plus and minus branches, which in turn transform in the same covariant way, they also transform as ordinary tensors under retarded diffs:
\begin{align}
    \text{Retarded 4-diffs: } \quad g_{\mu\nu}^\pm\,, g_{\mu\nu}\,, a^{\mu\nu},R^\mu_{\,\, \nu\rho\sigma},\nabla_\mu \quad \sim \quad \text{Covariant tensors}\,.
\end{align}
More in details, under $x^\mu\to x'^{\mu} = x^\mu+\e^\mu$ one has for example
\begin{align}
    g^{\mu\nu}_{\pm}(x) &\rightarrow  g'^{\mu'\nu'}_{\pm}(x')= \frac{\partial x'^{\mu'} }{\partial x^\alpha} \frac{\partial x'^{\nu'} }{\partial x^\beta} g^{\alpha \beta}(x), 
\end{align}
where on the right-hand side $x=x(x')$. As usual in the EFToI, it is convenient to also perform a change of coordinates in the open functional $x\to x'$, which leaves the functional invariant, in such a way that the spacetime argument in all transformed fields is always $x$ and the shift to $x'$ only appears in the unknown Wilsonian coefficients.

Operators that are only covariant under spatial diffs generate terms with derivatives on $\pir$. This is for instance the case for
    \begin{align}
    g^{00}(x) &\rightarrow  \frac{\partial ( t + \pir) }{\partial x^\alpha} \frac{\partial ( t + \pir) }{\partial x^\beta} g^{\alpha \beta}(x), \\
    a^{00}(x) &\rightarrow  \frac{\partial ( t + \pir) }{\partial x^\alpha} \frac{\partial ( t + \pir) }{\partial x^\beta} a^{\alpha \beta}(x),
    \end{align}
which lead to
	\begin{align}
		g^{00} &\rightarrow g^{00} + 2 g^{0\mu} \partial_\mu \pir + g^{\mu \nu} \partial_\mu \pir \partial_\nu \pir, \label{eq:g00stuck}\\
		a^{00} &\rightarrow a^{00} + 2 a^{0\mu} \partial_\mu \pir + a^{\mu \nu} \partial_\mu \pir \partial_\nu \pir.\label{eq:a00stuck}
	\end{align}
Similarly, 
    \begin{align}
    g^{0i}(x) &\rightarrow  \frac{\partial ( t + \pir) }{\partial x^\alpha} g^{\alpha i}(x), \qquad  a^{0i}(x) \rightarrow  \frac{\partial ( t + \pir) }{\partial x^\alpha}  a^{\alpha i}(x),
    \end{align}
    generates
     \begin{align}
		g^{0i} &\rightarrow g^{0i} +  g^{i\nu} \partial_\nu \pir, \qquad 
		a^{0i} \rightarrow  a^{0i} +  a^{i\nu} \partial_\nu \pir, \label{eq:g0mustuck}
    \end{align}
Likewise 
\begin{align}
    R^{00} &\rightarrow R^{00} + 2 R^{0\mu} \partial_\mu \pir + R^{\mu\nu} \partial_\mu \pir \partial_\nu \pir \,, \\
    a^{\mu\nu} R_\mu{}^0{}_\nu{}^0 &\rightarrow a^{\mu\nu} R_\mu{}^0{}_\nu{}^0 + 2 a^{\mu\nu} R_\mu{}^\rho{}_\nu{}^0 \partial_\rho \pir + a^{\mu\nu} R_\mu{}^\rho{}_\nu{}^\sigma \partial_\rho \pir \partial_\sigma \pir ,
\end{align}
where the contravariant transformation of $a^{\mu\nu}$ cancels with the covariant transformation of the two spacetime indices on $R_\mu{}^0{}_\nu{}^0$. At last, the zero component $v^0$ of a covariant vector $v^{\mu}$, such as for example $a^{0\mu}\partial_\mu \pir$, transforms as
\begin{align}
    v^0 \to v^0 +v^\mu \partial_\mu \pir\,.
\end{align}

Note that $K_{\mu\nu}$, being defined with reference to the temporal foliation of spacetime by $n_\mu$, does not transform as a four tensor and so its change upon performing the St\"uckelberg trick is more complicated. The spatial components of the
extrinsic curvature orthogonal to the constant time hypersurface are given by \cite{Piazza:2013coa} 
\begin{align}
    K_{ij} = \frac{1}{2} \sqrt{-g^{00}}(\partial_0 g_{ij} - \partial_i g_{0j} - \partial_j g_{0i})
\end{align}
By making use of \Eqs{eq:g00stuck} and \eqref{eq:g0mustuck}, one can derive the transformation rule of $K_{ij}$ at a given order in $\pir$.
Finally, note that in the usual discussion of the EFToI \cite{Cheung:2007st} one performs a change of coordinates in the integral that defines the action \textit{after} having performed a diff on all the fields. As a consequence of this, all time-dependent coefficients, such as $c(t)$, also generate terms in which $\pir$ appears with no derivatives. For example, 
\begin{align}\label{eq:cexp}
	c(t) \rightarrow c(t + \pir) = c(t) + \dot{c}(t) \pir + \frac{1}{2} \Ddot{c}(t) \pir^2 + \cdots
\end{align}

We first consider the part of the effective functional linear in the advanced fields discussed in \Eq{eq:S1try}.
The full 4-diff invariant operators inside $M_{\mu\nu} (R_{\mu\nu\rho\sigma}, g^{00}, K_{\mu\nu}, \nabla_\mu; t)$ and $M (R_{\mu\nu\rho\sigma}, g^{00}, K_{\mu\nu}, \nabla_\mu; t)$
may only receive $\pir$ contributions through the time dependence of their EFT coefficients, as in \Eq{eq:cexp}. Since these contributions are controlled by time derivatives, they are generally slow-roll suppressed in the inflationary context \cite{Cheung:2007st} when one further assumes that there is a shift symmetry acting on an inflaton with an approximately linear time dependence \cite{Finelli:2018upr}. Therefore, if one wants to recover the Open EFToI results \cite{Salcedo:2024smn}, one can safely neglect these terms. Conversely, in the late-time cosmology context, keeping these terms may have important consequences, as emphasized in \cite{Gubitosi:2012hu}. For instance, adding a $f(t) R$ operator in the theory without further modification is enough to introduce a massless scalar propagating degree of freedom.

Before closing this section, let us discuss the transformation of the second order effective functional $S_2$ given in \Eq{eq:S2try}. Since these contributions are already quadratic in perturbations, reintroducing the $\pir$ field through the time diff $t \rightarrow t + \pir(x)$ only generates cubic and higher order operators. 
At the level of the equations of motion, this class of operators has been identified in \cite{Creminelli:2023aly, Salcedo:2024smn} as a coupling between the Gaussian noise and an operator of the EFT. In this work, we choose to focus on the linear dynamics for simplicity, since the non-linear terms that survive in the decoupling limit were already studied in \cite{Salcedo:2024nex}. Therefore, we will not further investigate these terms which may be the subject of future work.


\subsubsection{Advanced St\"uckelberg trick}\label{subsec:advstuck}


We now aim at reintroducing manifest covariance under advanced diffs transforming the two branches of the path integral in opposite directions. 


\subsubsection*{Passive and active transformations: Method I vs Method II}

A new subtlety arises with advanced diffeomorphisms that transform fields belonging to different branches of the path integral in opposite directions. The approach presented in \Sec{subsec:retstuck} performs a change coordinates in the integral that defines the action. This is very useful in the standard unitary case because it allows one to write how the St\"uckelberg field appears to all orders, for example in $g^{00}$. However, this procedure is much more complicated in the presence of terms mixing the two branches of the path integral, as we see below. The reason is that there is not a single change of coordinates such that all the transformed fields in a given interaction term are evaluated at the same spacetime point. This leaves the advanced St\"uckelberg field $\pi_a$ inside the spacetime argument of other fields. Because of this we will prefer to avoid this additional change of coordinates. We explain the two possible methods and their difference in the following. The reader comfortable with active versus passive transformations can jump straight to the transformation rules below. 

\paragraph{Scalar fields.} Let us consider the coordinate transformation $x^\mu \rightarrow x'^\mu = x^\mu + \epsilon^\mu$. We first discuss the case of a scalar $\phi$ that transforms as 
\begin{align}
   \phi(x) \quad \rightarrow \quad \phi'(x + \epsilon) = \phi(x) \quad \leftrightarrow \quad \phi'(x)=\phi(x-\epsilon)\,.
\end{align}
There are two ways to deal with the gauge transformation. To illustrate this, let us consider the flat spacetime action 
\begin{align}
    S = \int \dd^4 x \mathcal{L}[\phi(x);x] ,
\end{align}
where $\mathcal{L}[\phi(x);x]$ is a Lagrangian density that depends on both the scalar field $\phi(x)$ and the spacetime coordinate $x$. Under coordinate transform, the action becomes
\begin{align}
    S = \int \dd^4 x \mathcal{L}[\phi(x);x]  \rightarrow \int \dd^4 x \mathcal{L}[\phi'(x);x] &= \overbrace{\int \dd^4 x \mathcal{L}[\phi(x- \epsilon);x]}^{\text{Method I}} 
    = \overbrace{\int \dd^4 \widetilde{x} \mathcal{L}[\phi(\widetilde{x});\widetilde{x} + \epsilon]}^{\text{Method II}}. 
\end{align}
The last step, from Method I to Method II is a change of coordinates in the integral, as opposed to a transformation of the fields.

Let us now consider the advanced diff $x^\mu \rightarrow x^\mu + \epsilon^\mu_a$, which transform the two branches of the path integral in opposite directions,
\begin{align}
   \phi_+(x) \rightarrow \phi_+'(x + \epsilon_a) = \phi_+(x) \quad \leftrightarrow \quad \phi_+'(x ) = \phi_+(x-\epsilon_a)\,, \\
    \phi_-(x) \rightarrow \phi_+'(x - \epsilon_a) = \phi_-(x)  \quad \leftrightarrow \quad \phi_-'(x ) = \phi_-(x+\epsilon_a) \,.
\end{align}
In this case, the effective functional becomes
\begin{align}
    S = \int \dd^4 x \mathcal{L}[\phi_+(x), \phi_-(x);x]  &\rightarrow \overbrace{\int \dd^4 x \mathcal{L}[\phi_+(x- \epsilon_a), \phi_-(x+ \epsilon_a);x]}^{\text{Method I}},
\end{align}
One might try to implement something analogous to Method II, but now there is no transformation such that the $\epsilon_a$ dependence disappears from the fields and is moved exclusively to the time-dependent couplings. One can still proceed with Method II, however, this is much less convenient in the open case than it is in the standard unitary case. Conversely, while Method I is less convenient in the standard unitary case, it becomes the method of choice for us. 


\paragraph{Metric.} When considering the metric transformation corresponding to the change of coordinates $x^\mu \to x'^\mu=x^\mu+\epsilon^\mu$, we have
\begin{align}\label{eq:gtransfov0}
    g_{\mu\nu}(x) \quad \rightarrow \quad g'_{\mu\nu}(x')= \frac{\partial x^{\alpha} }{\partial x^{\prime \mu}} \frac{\partial x^{\beta} }{\partial x^{\prime\nu}} g_{\alpha \beta}(x)\,.
\end{align}
To make this expression absolutely clear, it is useful to re-write it as (relabeling $x'$ as $x$)
\begin{align}\label{eq:gtransfo}
    g_{\mu\nu}(x) \quad \rightarrow \quad g'_{\mu\nu}(x)= \frac{\partial (x^{\alpha}-\epsilon^\alpha )}{\partial x^{ \mu}} \frac{\partial (x^{\beta}-\epsilon^\beta) }{\partial x^{\nu}} g_{\alpha \beta}(x-\epsilon)\,.
\end{align}
The difference between Method I and Method II is that, in one case, one transforms the argument of the metric. It leads to 
\begin{align}
    S = \int \dd^4 x \mathcal{L}[g_{\mu\nu}(x);x]  &\rightarrow \int \dd^4 x \mathcal{L}[g'_{\mu\nu}(x);x] \\
    &= \underbrace{\int \dd^4 x \mathcal{L}\left[\frac{\partial (x^{\alpha}-\epsilon^\alpha )}{\partial x^{ \mu}} \frac{\partial (x^{\beta}-\epsilon^\beta) }{\partial x^{\nu}} g_{\alpha \beta}(x-\epsilon);x\right]}_{\text{Method I}}.
\end{align}
If one further uses \Eq{eq:gtransfo} expended at linear order in $\epsilon$, one obtains 
\begin{align}
    S &\rightarrow \int \dd^4 x \mathcal{L}\left[g_{\mu\nu}(x) - g_{\lambda \mu}(x) \partial_\nu \epsilon^\lambda - g_{\lambda \nu}(x) \partial_\mu \epsilon^\lambda  - \epsilon^\lambda \partial_\lambda g_{\mu\nu}(x) ;x\right] \\
    &= \int \dd^4 x \mathcal{L}\left[g_{\mu\nu}(x) - 2 \nabla_{(\mu} \epsilon_{\nu)};x\right]
\end{align}
Here one recognizes that the change of the metric is nothing but the familiar Lie derivative in the $\epsilon^\mu$ direction. Note that, since $g^{\mu\nu}g_{\mu\nu}=4$ is clearly invariant under diffs, the inverse metric transforms with the opposite sign to the metric:
\begin{align}\label{plusminus}
     \boxed{ \text{Method I: } \qquad \Delta g_{\mu\nu}=- 2 \nabla_{(\mu} \epsilon_{\nu)} \,,  \qquad \Delta g^{\mu\nu}=+ 2 \nabla^{(\mu} \epsilon^{\nu)} \Big.} \,. 
\end{align}
Method II is then nothing but Method I plus a change of coordinate such that
\begin{align}
    S  = \int \dd^4 x \mathcal{L}[g_{\mu\nu}(x);x]  &\rightarrow \int \dd^4 x \mathcal{L}[g'_{\mu\nu}(x);x] \equiv \int \dd^4 x \sqrt{g'(x)} \tilde{\mathcal{L}}[g'_{\mu\nu}(x);x]  \\
    &= \underbrace{\int \dd^4 \widetilde{x} \sqrt{g(\widetilde{x})} \tilde{\mathcal{L}}\left[\frac{\partial (\widetilde{x}^{\alpha}+\epsilon^\alpha )}{\partial \widetilde{x}^{ \mu}} \frac{\partial (\widetilde{x}^{\beta}+\epsilon^\beta) }{\partial \widetilde{x}^{\nu}} g_{\alpha \beta}(\widetilde{x}); \widetilde{x} + \epsilon \right]}_{\text{Method II}}.
\end{align}
According to Method II, the transformation of the metric to linear order in $\e$ is
\begin{align}
    \boxed{ \text{Method II: } \qquad \Delta g_{\mu\nu}=- g_{\mu\alpha} \partial_{\nu} \epsilon_{\alpha}- g_{\nu\alpha} \partial_{\nu} \epsilon_{\alpha} \,,  \qquad \Delta g^{\mu\nu}=+ 2 \partial^{(\mu} \epsilon^{\nu)} \Big.}\,.
\end{align}
What is particularly convenient about Method II is that the transformation of the inverse metric $g^{\mu\nu}$ to all orders in $\epsilon^\mu$ contains only terms that are linear and quadratic in $\epsilon^\mu$. All higher order terms arise exclusively from the explicit spacetime dependence of the effective couplings. This is in contrast to Method I, where the transformation of the metric itself contains terms to all orders in $\epsilon$, and the Lie derivative transformation in \eqref{plusminus} is only the leading term. However, as we will see shortly, the need to perform a change of coordinates in Method II becomes a nuisance in the presence of cross-branch couplings, which are the defining property of open systems. Because of this, we will find it more convenient to employ Method I when studying advanced gauge transformations in open systems. 


We briefly illustrate that Method I and Method II lead to the same results by considering the cosmological constant term
\begin{align}
    S =- \int \dd^4 x \sqrt{-g(x)} \Lambda(t). 
\end{align}
Under Method II, the transformation under time diffeomorphisms is trivial and yields the familiar result
\begin{align}\label{eq:cctest}
     S \rightarrow -\int \dd^4 x \sqrt{-g(x)} \Lambda\left[t + \pi(x)\right] = -\int \dd^4 x \sqrt{-g(x)} \left[\Lambda(t) + \dot{\Lambda}(t) \pi(x) 
     + \cdots \right]\,,
\end{align}
An equivalent result can be reached from Method I order by order in perturbation theory, using
\begin{align}
    \sqrt{-g(x)} \rightarrow  \sqrt{-g(x)} \left( 1 -  \nabla_\mu \epsilon^\mu+ \cdots \right) \,,
\end{align}
such that the action transforms as
\begin{align}
    \Delta S = \int \dd^4 x \sqrt{-g} \Lambda(t)  \nabla_\mu \epsilon^\mu  = \int \dd^4 x  \partial_\mu \left( \sqrt{-g} \epsilon^\mu \right) \Lambda(t)\,.
\end{align}
Upon integrating by part and restricting ourselves to a time diff, $\epsilon^\mu=(\pi,0,0,0)$, we recover the right hand side of \Eq{eq:cctest}.  


\subsubsection*{Transformation rules}

The coordinate on the $+$ branch transforms as $x^\mu \rightarrow x^\mu + \epsilon^\mu_a/2$ and on the $-$ branch by $x^\mu \rightarrow x^\mu - \epsilon_a^\mu/2$. At linear order, this leads to the transformation of the metric appearing in the functional by (using \eqref{plusminus})
\begin{align}
    g^{\pm}_{\mu\nu} \rightarrow g^{\pm}_{\mu\nu} \mp  \nabla_{(\mu} \epsilon_{a\nu)}, \qquad \text{and} \qquad g_{\pm}^{\mu\nu} \rightarrow g_{\pm}^{\mu\nu} \pm \nabla^{(\mu} \epsilon_a^{\nu)}\,.
\end{align}
It follows that
\begin{align}
    g_{\mu\nu} \rightarrow g_{\mu\nu}, \qquad a^{\mu\nu} \rightarrow a^{\mu\nu} + 2 \nabla^{(\mu} \epsilon_a^{\nu)}.
\end{align}
One can equivalently expand the expression for the advanced metric components to
\begin{align}
    a^{00} &\rightarrow \mathcal{A}^{00} \equiv  a^{00} + 2g^{0\mu} \partial_\mu \pia  + 2 g^{0\mu} \Gamma^{0}_{~\mu 0} \pia, \label{eq:ta1}\\
    a^{0i} &\rightarrow \mathcal{A}^{0i} \equiv   a^{0i} + g^{i\mu} \partial_\mu \pia +  g^{i\mu}_{~} \Gamma^{0}_{~\mu 0} \pia+g^{0\mu}\Gamma^{i}_{\,\mu 0} \pia, \label{eq:ta2}\\
    a^{ij} &\rightarrow \mathcal{A}^{ij} \equiv   a^{ij} +  g^{i\mu}_{~} \Gamma^{j}_{~\mu 0} \pia +  g^{j\mu}_{~} \Gamma^{i}_{~\mu 0} \pia, \label{eq:ta3}
\end{align}
where we defined $\mathcal{A}^{\alpha\beta} \equiv a^{\alpha\beta} +  2 \nabla^{(\alpha} \epsilon_a^{\beta)}$ for convenience and restricting ourselves to $\epsilon^\mu_a = \pia \delta^\mu_{~0}$
Compared to the retarded St\"uckelberg trick which relies on Method II and the implicit coordinate redefinition in the integrals, the advanced St\"uckelberg trick features some Christoffel symbols that follow from Method I and the lack of coordinate redefinition. 

We now derive the transformation of the scalar clock under advanced time diff. Using Method I, the linear transformation of a scalar under diff $\epsilon^\mu$ is 
\begin{align}
    \Delta \phi(t) = - \epsilon^\mu \partial_\mu \phi.
\end{align}
Applying this rule to the advanced time diff on the Schwinger-Keldysh contour, we obtain on each branch
\begin{align}
    \Delta \phi_\pm(t) = \mp \frac{\epsilon_a^\mu}{2} \partial_\mu \phi_\pm\,.
\end{align}
Changing variables to work with the $t_\pm$, we have 
\begin{align}
    \phi_\pm(t,\bfx) = \bar{\phi}[t_\pm(t,\bfx)].
\end{align}
We deduce the transformation of $t_\pm$ by matching the transformation of $\phi_\pm$. Explicitly, 
\begin{align}
    \Delta\phi_\pm(t) = \dot{\bar{\phi}} \Delta t_\pm\,.
\end{align}
Restricting $\epsilon^\mu_a = \pia \delta^\mu_{~0}$, we find $ \Delta t_\pm = \mp \pia/2$, leading to 
\begin{align}
    t_a \rightarrow t_a - \pi_a , \qquad t_r \rightarrow t_r\,,
\end{align}
at linear order in $\pia$. We conclude that if we stick to the linear theory, it is then enough to consider the transformation of the advanced metric and advanced time under advanced time diff, while leaving the retarded metric and retarded time invariant.

Deriving the transformation rules of $K_{ij}$ and $R^{00}$ under advanced time diff is indeed more involved. For the scope of this work where we focus on the linear dynamics, it is enough to observe that the transformation of $K_{ij}$ and $R^{00}$ generates terms that are at least quadratic in the advanced fields as $\delta g^{00}\sim\mathcal{O}\left(a^{\mu\nu}\pia\right)$. 
Given that these terms always multiply operators linear in $a^{\mu\nu}$ in $S_1$, their contributions are always postponed to cubic operators in advanced fields and higher, $S_{2p+1\geq3}$. Note that there is no risk of ``contamination'' of noise operators belonging to $S_2$ from the advanced time diff transformation of operators in $S_1$. This can be demonstrated to all orders using the constraint \eqref{eq:herm}. 


\subsection{The decoupling limit}\label{ss:decoupling}

During inflation, the \textit{decoupling limit} \cite{Cheung:2007st, Green:2024hbw} is often invoked as a regime in which the mixing between the metric perturbations and $\pi$ is suppressed compared to self-interactions. This limit can be taken after performing the \stuck tricks, by replacing the retarded tensors by their background values, while dropping contributions involving $a^{\mu\nu}$, the latter having no background and therefore only contributing to the mixing with gravity. To ensure that no term is missed it is essential that the decoupling limit is taken only \textit{after} both the retarded and advanced St\"uckelberg tricks have been performed.
The main reason why we introduced the advanced St\"uckelberg field $\pi_a$ is to get the correct decoupling limit. If we limited ourselves to study the full theory including dynamical gravity there would not be much advantage in having $\pi_a$ around. 
Indeed, it ends up being equivalent to describe the physics in terms of 
\begin{align}
    \int \mathcal{D}[g_{\mu\nu}]  \int \mathcal{D}[a^{\mu\nu}]  \int \mathcal{D}[\phi_r] \int \mathcal{D}[\phi_a] \ee^{iS_{\mathrm{eff}}[g_{\mu\nu},\,a^{\mu\nu};\, \phi_r,\,\phi_a]},
\end{align}
or 
\begin{align}
    \int \mathcal{D}[g_{\mu\nu}]  \int \mathcal{D}[a^{\mu\nu}]  \int \mathcal{D}[t_r] \int \mathcal{D}[t_a] \ee^{iS_{\mathrm{eff}}[g_{\mu\nu},\,a^{\mu\nu};\, t_r,\,t_a]},
\end{align}
or 
\begin{align}\label{eq:last}
    \int \mathcal{D}[g_{\mu\nu}]  \int \mathcal{D}[a^{\mu\nu}]  \int \mathcal{D}[\pi_r] \int \mathcal{D}[\pi_a] \ee^{iS_{\mathrm{eff}}[g_{\mu\nu},\,a^{\mu\nu};\, \pi_r,\,\pi_a]}.
\end{align}
It turns out \eqref{eq:last} is particularly convenient when the mixing with gravity can be neglected. 

In the following we report the contributions coming from different metric components after performing the advanced and retarded St\"ucklerberg trick in the decoupling limit. Note that the commutativity of the \stuck tricks, despite not obvious at first sight, can be checked explicitly. To simplify the derivation, it is useful to consider 4d diff $\epsilon_r^\mu$ and $\epsilon_a^\mu$ and to restrict to $\epsilon^\mu_r = \pir \delta^\mu_{~0}$ and $\epsilon^\mu_a = \pia \delta^\mu_{~0}$ at the really end. On the one hand, 
\begin{align}
 	a^{\mu\nu} &\xrightarrow{(\text{t-diff}_a)}   a^{\mu\nu} +  2 \nabla^{(\mu} \epsilon_a^{\nu)} \xrightarrow{(\text{t-diff}_r)} \left[  a^{\alpha\beta} +  2 \nabla^{(\alpha} \epsilon_a^{\beta)}\right]  \Big[ \partial_\alpha\left(x^\mu + \epsilon_r^\mu\right) \partial_\beta \left(x^\nu + \epsilon_r^\nu\right)\Big]\,,
\end{align}
where in the last expression, we used the fact that $ a^{\mu\nu}$ and $\nabla^{(\mu} \epsilon_a^{\nu)}$ transform as rank $2$ tensors under retarded time diff. Similarly, if we first perform the retarded St\"uckelberg trick then the advanced one, we obtain
\begin{align}
 	a^{\mu\nu} &\xrightarrow{(\text{t-diff}_r)}  a^{\alpha\beta}   \Big[ \partial_\alpha\left(x^\mu + \epsilon_r^\mu\right) \partial_\beta \left(x^\nu + \epsilon_r^\nu\right)\Big]  \xrightarrow{(\text{t-diff}_a)} \left[  a^{\alpha\beta} +  2 \nabla^{(\alpha} \epsilon_a^{\beta)}\right]  \Big[ \partial_\alpha\left(x^\mu + \epsilon_r^\mu\right) \partial_\beta \left(x^\nu + \epsilon_r^\nu\right)\Big]\,,
\end{align}
which is indeed equivalent. The same goes for the retarded metric that does not transform under advanced time diff. 

Explicitly, we find that 
\begin{align}
    a^{00} &\xrightarrow[(\text{t-diff}_a)]{(\text{t-diff}_r)}\mathcal{A}^{00} + 2 \mathcal{A}^{0\mu} \partial_\mu\pir +  \mathcal{A}^{\mu\nu}\partial_\mu\pir\partial_\nu\pir\,, \\
    a^{0i} &\xrightarrow[(\text{t-diff}_a)]{(\text{t-diff}_r)} \mathcal{A}^{0i} + \mathcal{A}^{i\mu} \partial_\mu\pir\,, \\
    a^{ij} &\xrightarrow[(\text{t-diff}_a)]{(\text{t-diff}_r)} \mathcal{A}^{ij},
\end{align}
where the $\mathcal{A}^{\mu\nu}$ elements are all given between \eqref{eq:ta1} and \eqref{eq:ta3}.
At last, we take the decoupling limit by replacing the retarded metric and Christoffel symbols by their background values, while dropping contributions involving $a^{\mu\nu}$. In this limit, \eqref{eq:ta1} to \eqref{eq:ta3} become
\begin{align}
    \mathcal{A}^{00} &\xrightarrow{(\text{decoupl.})}  - 2\dpia, \qquad 
    \mathcal{A}^{0i} \xrightarrow{(\text{decoupl.})}  \bar{g}^{ij}\partial_j\pia, \qquad
    \mathcal{A}^{ij} \xrightarrow{(\text{decoupl.})}  2 H \bar{g}^{ij} \pia,
\end{align}
from which we deduce
\begin{align}
     a^{00} &\xrightarrow{(\text{decoupl.})} 2 \left(-\dpia  - 2 \dpia   \dpir +  \bar{g}^{ij}\partial_i\pir\partial_j\pia \right) \bigg.  \\
     &\qquad \quad  - 2 \dpia \dot{\pi}^2_r + 2 \dpir  \bar{g}^{ij}\partial_i\pir\partial_j\pia  + 2 H  \bar{g}^{ij}\partial_i\pir\partial_j\pir  \pia, \bigg. \nonumber  \label{eq:kintermv2} \\
    a^{0i} &\xrightarrow{(\text{decoupl.})}  (1 + \dpir)  \bar{g}^{ij}\partial_j\pia + 2 H  \bar{g}^{ij}\partial_j\pir  \pia, \bigg. \\
    a^{ij} &\xrightarrow{(\text{decoupl.})} 2 H \bar{g}^{ij} \pia. \bigg.
\end{align}
For the retarded metric, $g^{00}$ and $g^{0i}$ are invariant at linear order under advanced time diffeomorphism. Consequently, one finds
\begin{align}
 	g^{00} &\xrightarrow{(\text{t-diff}_r)}   g^{00}+2g^{0\mu}\partial_\mu \pir +g^{\mu\nu}\partial_\mu \pir \partial_\nu \pir  \xrightarrow{(\text{decoupl.})} - 1 -2\dpir - \dpir^2 +\bar{g}^{ij}\partial_i\pir\partial_j\pir  \bigg. \,, \\
    g^{0i} &\xrightarrow{(\text{t-diff}_r)}  g^{0i}+g^{i\mu}\partial_\mu \pir \xrightarrow{(\text{decoupl.})}  \bar{g}^{ij}\partial_j\pir\bigg. \,, \\
    g^{ij} &\xrightarrow{(\text{t-diff}_r)}  g^{ij}\xrightarrow{(\text{decoupl.})}   \bar{g}^{ij} \bigg. \,,
\end{align}
with $\bar{g}^{ij} = \delta^{ij}/a^2(t)$. 


\paragraph{Recovering the Open EFToI.}

Let us now discuss the decoupling limit of the theory. We will study in the next section in what regime this decoupling limit is a good description of the theory. In \cite{Salcedo:2024smn}, three of the authors wrote down a dissipative and stochastic theory of inflation fluctuations in an unperturbed inflationary background. This is the theory of a dissipative shift symmetric scalar in de Sitter spacetime. By enforcing the retarded shift symmetry but spontaneously breaking the advanced one, we recover and extent the usual EFToI results supplemented by local dissipation and noise. There, the construction relied on the building blocks
$-\dot{\pi}_{a} + \partial^\mu \pi_{r} \partial_\mu \pi_{a}$ and $\pi_a$ multiplied by powers of $  -2 \dot{\pi}_{r} + ( \partial_\mu \pi_{r})^2$.
Following \Sec{subsec:stuck}, these operators can be recovered by combining\footnote{While \eqref{eq:Stuck1} and \eqref{eq:Stuck3} are straightforward, \eqref{eq:Stuck2} requires more thinking. One first need to gather all $\pir$ and $\pia$ through a relatively long and cumbersome computation, leading to 
\begin{align}
    c(t) \left(-a^{00} + \frac{g^{00}}{2} g_{\mu\nu} a^{\mu\nu}\right)  &\rightarrow \bigg[2c(t) \dpia + \dot{c}(t)\pia\bigg] + \bigg[2 c(t) \dpia \dpir - 2 c(t) \bar{g}^{ij}\partial_i \pir \partial_j \pia   \nonumber \\
     +&2 \dot{c}(t) \left(\dot{\pi}_{r}\pia+\pir\dot{\pi}_{a}\right) + \ddot{c}(t) \pir \pia \bigg]  + \bigg[ \dot{c}(t)\dpir^2 \pia - \dot{c}(t) \bar{g}^{ij}\partial_i \pir \partial_j \pir  \pia\bigg],
\end{align}
where the first bracket gathers the tadpole contribution, the second contribution the quadratic operators (which contains a total derivative that vanishes at the boundary of the path integral) and the third contribution the cubic operators. Neglecting derivatives of $c(t)$ which are slow-roll suppressed in the decoupling limit, we reach \eqref{eq:Stuck2}.}  
\begin{align}
    - \frac{1}{2} \left(- a^{00} + \frac{g^{00}}{2} g_{\mu\nu} a^{\mu\nu} \right)\quad &\xrightarrow{(\text{decoupl.})} \quad -\dot{\pi}_{a} + \partial^\mu \pi_{r} \partial_\mu \pi_{a}\,, \bigg. \label{eq:Stuck2}\\
    g_{\mu\nu} a^{\mu\nu} &\xrightarrow{(\text{decoupl.})} 2 \dpia + 6H\pia\,,\bigg.\label{eq:Stuck1} \\
     1 + g^{00} \quad &\xrightarrow{(\text{decoupl.})} \quad -2 \dot{\pi}_{r} + (\partial_\mu \pi_{r})^2, \bigg. \label{eq:Stuck3}
\end{align}
where the right-hand side holds in the decoupling limit, where fluctuations of the metric are discarded. Based on these building blocks, one can reconstruct the theory investigated in \cite{Salcedo:2024smn} from $S_1$ and $S_2$ found in \Eqs{eq:S1v1} and \eqref{eq:S2v1}. 

In \App{app:ope}, we present specific operators which play a particular role in the scalar dynamics. For the remaining of this work, we will focus on the minimal subset controlling the linear dynamics in the presence of local dissipation and noise, 
\begin{align}\label{eq:th2}
	S_{\rm eff}=\int \dd^4x \sqrt{-g}\bigg[&\frac{\Mpl^{2}}{2}G_{\mu\nu}a^{\mu\nu}+\frac{\Lambda(t)}{2}g_{\mu\nu}a^{\mu\nu}-c(t)a^{00}+\frac{c(t)}{2}g^{00}g_{\mu\nu}a^{\mu\nu}\nonumber \\
	 -&\frac{M_{2}(t)}{4}\left(1+g^{00}\right)^{2}g_{\mu\nu}a^{\mu\nu}+M_{2}(t)\left(1+g^{00}\right)a^{00} \bigg.\nonumber\\
     +& \frac{\Gamma(t)}{3H}\left(1 + g^{00} \right)\left( a^{00} + g_{\mu\nu} a^{\mu\nu}\right)-\xi_{\mu\nu}a^{\mu\nu}\bigg], 
\end{align}
Each of these contributions are separately discussed in \App{app:ope}. $S_{\mathrm{eff}}$ encompasses the universal part of the EFToI controlled by the EFT coefficients $\Lambda(t)$ and $c(t)$, as we show in \App{app:univ}. It also contains a speed of sound controlled by $M_{2}(t)$ and a simple non-unitary extension made of the dissipative coefficient $\Gamma(t)$ and the noise contribution controlled by $\xi_{\mu\nu}$. These operators are explicitly constructed in \App{app:minimal}.  

Note that there exists an equivalent formulation of the theory where $t_a$ terms are kept. For the theory to admit solution, the $t_a$ contributions have to follow form the Einstein equations, as discussed in \Eq{singleclockstructure}. Explicitly, 
\begin{align}\label{eq:th1}
   S_{\rm eff} \rightarrow  S_{\rm eff} + \int \dd^4x \sqrt{-g}\bigg\{&-\dot{\Lambda}(t)t_{a} -2c(t)g^{0\mu}\partial_{\mu}t_{a}-\dot{c}(t)g^{00}t_{a} \nonumber \\
    +& 2M_{2}(t) (1+g^{00})g^{0\mu}\partial_{\mu}t_{a} + \frac{\dot{M}_{2}(t)}{2}\left(1+g^{00}\right)^{2}t_{a} \bigg. \nonumber \\
    +& 2\Gamma(t) (1 + g^{00}) t_a - 2 \xi_{00} \dot{t}_a + 2 \bar{g}^{ij} \xi_{0i} \partial_j t_a + 2H \bar{g}^{ij} \xi_{ij} t_a\bigg\},
\end{align}
in which case $t_a$ perfectly supplants the role of $\pia$. In particular, the effective functional is  invariant under advanced time diff and equations of motion for $\pir$ are obtained by varying with respect to $t_a$. We will not further discuss this case in this work and consider \Eq{eq:th2} to assess mixing between gravity and the dynamical scalar.


\subsection{Mixing with gravity}\label{Sec:minimal}

The formulation of the Open EFToI \cite{Salcedo:2024smn} in terms of geometric quantities provides a natural framework to study the mixing between the scalar field and the metric perturbations. In particular, it enables a systematic evaluation of the regime of validity of the results in \cite{Salcedo:2024smn}, where the decoupling of $\pi_r$ was assumed. The main conclusion of this section is that the theory developed in \cite{Salcedo:2024smn} can indeed be interpreted as the decoupling limit of a gravitational Open EFT, formulated according to the rules presented in \Sec{sec:unitgauge}.

In the standard EFToI \cite{Cheung:2007st}, it was shown that scalar perturbations decouple from gravity at energy scales $E \gg \sqrt{\epsilon}\, H$, where $E \sim H$ is the typical perturbation energy, $H$ is the Hubble parameter, and $\epsilon \equiv -\dot{H}/H^2$ is the first slow-roll parameter. In a first study of dissipative effects within the EFToI framework \cite{LopezNacir:2011kk}, it was found that dissipation introduces a modified decoupling scale $E \gg \sqrt{\epsilon H \Gamma}$, where $\Gamma$ controls dissipation in the scalar sector. Here, we revisit these two decoupling scales in the context of the current formalism and extend the analysis to include the effects of noise.

\paragraph{Einstein's equations.}

Let us first consider the metric effective functional given in \Eq{eq:th2}.
We first derive the background Friedmann equations to fix the EFT parameters $c(t)$ and $\Lambda(t)$ in terms of the Hubble parameter $H$ and the slow-roll parameter $\epsilon$. At the background level, \Eq{eq:th2} reads
\begin{equation}
    S_{\rm eff}=\int \dd^4x \sqrt{-g}\left[\frac{\Mpl^{2}}{2}\bar{G}_{\mu\nu}+\frac{\Lambda(t)-c(t)}{2}\bar{g}_{\mu\nu}a^{\mu\nu}-c(t)a^{00}\right],
\end{equation}
from which we obtain the background Einstein's equations by varying with respect to $a^{\mu\nu}$, 
\begin{equation}
    \frac{\Mpl^{2}}{2}\bar{G}_{\mu\nu}+\frac{\Lambda(t)-c(t)}{2}\bar{g}_{\mu\nu}-c(t)\delta_{\mu}^{0}\delta_{\nu}^{0}=0.
\end{equation}
We recover the usual Friedmann equations
\begin{align}\label{eq:Fried}
    3\Mpl^{2}H^{2}=&\Lambda(t)+c(t),\qquad
    2\Mpl^{2}\dot{H}=-2c(t),
\end{align}
which allow us to fix the Wilson coefficients in terms of 
\begin{equation}
    c(t)=\epsilon H^{2}\Mpl^{2},\qquad\Lambda(t)=\left(3-\epsilon\right)H^{2}\Mpl^{2}.
\end{equation}

We now derive the perturbed Einstein's equations from which the mixing between $\pir$ and $\delta g^{00}$ emerges. Following \Sec{subsec:retstuck}, we first reintroduce $\pi_r$ by performing a retarded St\"uckelberg field to $S_{\mathrm{eff}}$ given in \Eq{eq:th2}. The operators from $S_{\mathrm{eff}}$ breaking retarded time-diffeomorphism generate terms that are linear in $\pi_r$. Computations being slightly cumbersome, we gather the details in \App{app:mixing} and here state the main result which are the perturbed $00$ and $0i$ Einstein's equations, 
\begin{align}
    \frac{\Mpl^{2}}{2}\delta G_{00}+\frac{\Lambda(t)-2M_{2}(t)}{2}\delta g_{00}+3Hc(t)\pir-\left[c(t)+2M_{2}(t)\right]\dot{\pi}_r&=\xi_{00}\,,\label{eq:E1v2} \\
    \frac{\Mpl^{2}}{2}\delta G_{0i}+\frac{\Lambda(t)-c(t)}{2}\delta g_{0i}-c(t)\partial_{i}\pir&=\xi_{0i}. \label{eq:E2v2}
\end{align}
\Eqs{eq:E1v2} and \eqref{eq:E2v2} are the manifestly covariant Einstein's equations, once $\pi_r$ has been reintroduced. We are now in a position to solve the constraints and evaluate the mixing with gravity.
The introduction of the Goldstone mode allows us to work in the flat gauge
\begin{equation}
    \delta g_{ij}^{\rm scalar}=0\quad,\quad\delta g_{00}=-2\phi=-\delta g^{00}\quad,\quad \delta g_{0i}=a(t)\partial_{i}F=\delta g^{0i}.
\end{equation}
To assess the mixing between gravity and the Goldstone mode, we need to solve the constraints to find $\phi$ and $F$ as a function of $\pir$ and the noise \cite{Maldacena:2002vr}. The solution for $\phi$ is obtained by solving the $0i$ component, leading to 
\begin{equation}\label{eq:c1}
    \phi=\frac{1}{H\Mpl^{2}}\left[c(t)\pir+\xi_{||}\right],
\end{equation}
where we defined $\xi_{0i} = \partial_i \xi_{\parallel} $, while the solution for $F$ is given by the $00$ component, reading
\begin{equation}\label{eq:c2}
    \nabla^{2}F=a(t)\left[\frac{\epsilon^{2}H^{2}}{c_{s}^{2}}\pir - \frac{\epsilon H}{c_{s}^{2}}\dot{\pi}_r - \frac{\xi_{00}}{\Mpl^{2}H} - \left(\frac{3c_{s}^{2}-\epsilon}{c_{s}^{2}\Mpl^{2}}\right) \xi_{||}\right].
\end{equation}
In the above expression, we introduced the speed of sound $c_{s}^{2}$ through
\begin{equation}
    \frac{1}{c_{s}^{2}} \equiv 1+\frac{2M_{2}(t)}{c(t)}.
\end{equation}
Importantly, so far, the only difference with the usual EFToI is through the appearance of the noise terms $\xi_{00}$ and $\xi_{\parallel}$. This comes from the fact that the dissipative operator $2\Gamma(t)(1 + g^{00})(a^{00} + g_{\mu\nu}a^{\mu\nu})$ does not affect the $00$ or $0i$ part of the Einstein equations which control the constraints. Nevertheless, this operator still contributes to the equation of motion for the Goldstone mode and to the trace component of the Einstein equations.\footnote{Indeed, if $2\Gamma(t)(1 + g^{00})(a^{00} + g_{\mu\nu}a^{\mu\nu})$ did not contribute to the Einstein equations, then it could not consistently appear in the equation of motion for $\pi_r$.} This goes on the same line as the inclusion of $\xi_{ii}$ on the RHS for the equation of motion for $\pir$. 

\paragraph{Equation of motion for $\pi_r$.}

Once the constraints are solved, we can focus on the terms that control the $\pi_r$ equation of motion, which ultimately determines the amplitude of the primordial power spectrum. We here need to reintroduce $\pia$ by performing an advanced \stuck trick. This step is explicitly performed in \App{app:minimal} and leads to
\begin{align}\label{eq:Ssca}
    S_{\pi}=\int \dd^4x \sqrt{-g}\bigg\{&-\dot{\Lambda}(t)\pia -2c(t)g^{0\mu}\partial_{\mu}\pia-\dot{c}(t)g^{00}\pia \nonumber \\
    +& 2M_{2}(t) (1+g^{00})g^{0\mu}\partial_{\mu}\pia + \frac{\dot{M}_{2}(t)}{2}\left(1+g^{00}\right)^{2}\pia \nonumber  \bigg.  \\
    +& 2H\Gamma(t) (1 + g^{00}) \pia - 2 \xi_{00} \dot{\pi}_a + 2 \bar{g}^{ij} \xi_{0i} \partial_j \pia + 2H \bar{g}^{ij} \xi_{ij} \pia\bigg\}.
\end{align}
By performing a retarded St\"uckelberg trick and varying with respect to $\pia$, we obtain the equations of motion for $\pir$. 
At leading order, we recover the usual background continuity equation
\begin{equation}
    \dot{\Lambda}(t)+\dot{c}(t)+6Hc(t)=0, 
\end{equation}
consistently with the Friedmann equations found in \Eq{eq:Fried}. Note that in this particular implementation, the dissipative operator proportional to $\Gamma$ does not affect the background equations of motion, neither for $\pir$ or the Einstein equations. This does not need to be the case in general, but it is a consistent choice of model we can make. 
We can then expand at second order in perturbations. This long but straightforward computation detailed in  \App{app:mixing} leads to the equation of motion for $\pir$
\begin{equation}\label{eq:pireomexp}
    \ddot{\pi}_r+\Upsilon\dot{\pi}_r - \cs \partial_{i}^{2}\pir+3\epsilon c_{s}^{2}H^{2}\pir+\tilde{G}_{0}\delta g^{00}+\tilde{G}_{1}\partial_{0}\delta g^{00}+\tilde{G}_{2}\partial_{i}\delta g^{i0}=-\frac{c_{s}^{2}}{\epsilon H^{2}\Mpl^{2}}\xi_{\pi},
\end{equation}
which features dissipation, noise, and mixing with gravity. The self dynamics of the scalar is controlled by 
\begin{align}
    \Upsilon\equiv&\left(3+\eta-2\epsilon\right)H+\frac{2\Gamma(t)c_{s}^{2}}{\epsilon H^{2}\Mpl^{2}}-\frac{2\dot{c}_{s}}{c_{s}},
\end{align}
with the second slow-roll parameter $\eta \equiv \dot{\epsilon}/(H \epsilon)$, while the mixing with gravity follows from
\begin{align}
    \tilde{G}_{0}\equiv&-\frac{3H}{2}\left(1+c_{s}^{2}\right)+\frac{\dot{c}_{s}}{c_{s}}+\frac{\Gamma c_{s}^{2}}{\epsilon H^{2}\Mpl^{2}}+\left(\frac{2\epsilon-\eta}{2}\right)H; \qquad  \tilde{G}_{1}\equiv-\frac{1}{2}; \qquad  \tilde{G}_{2}\equiv-c_{s}^{2}.
\end{align}
At last, the noise in the scalar sector $\xi_\pi$ can be deduced from $\xi_{\mu\nu}$, 
\begin{align}\label{eq:noisepi}
    \xi_\pi &= 2( \dot{\xi}_{00} + 3H \xi_{00} ) - \frac{2}{a^2(t)} \left(\nabla^2 \xi_{\parallel}  - 3H \Xi_{\parallel}\right)\,,
\end{align}
where we further decomposed $\xi_{\mu\nu}$ into scalar, vector and tensor, with the scalar sector reading
\begin{align}\label{eq:noiseSVTv3}
    \xi_{00}, \qquad  \xi_{0i} =  \partial_i \xi_{\parallel}, \qquad \xi_{ij} = \Xi_{\parallel} \delta_{ij} + \left(\partial_i \partial_i - \frac{1}{3}  \delta_{ij} \partial^2 \right)  \Xi^{\mathrm{TT}}.
\end{align}
Armed with this we are able to study the decoupling limit in several regimes.

\paragraph{Decoupling regimes.}

We first recover the usual decoupling limit of the EFToI \cite{Cheung:2007st} by setting the dissipation $\Gamma$ and the noises $\xi_{00}$ and $\xi_{||}$ to zero. In this case, we find the constraints to be
\begin{align}
    \phi=\epsilon H\pir\quad,\quad \nabla^{2}F=-\frac{\epsilon H}{c_{s}^{2}}\left[\dot{\pi}_r-\epsilon H\pir\right],
\end{align}
To study decoupling, we have to compare the kinetic term $\ddot{\pi}_r \sim E^2 \pir$ against the mixing with gravity controlled by $\tilde{G}_{0,1,2}$. Injecting 
$\delta g^{00} = 2\phi$ and $\delta g^{0i} = a(t)\partial_{i}F$ into \Eq{eq:pireomexp}, we obtain 
\begin{align}
   &\tilde{G}_{0}~ \text{mixing (no dissipation):} \qquad E\gg \sqrt{\epsilon}  H, \\
   &\tilde{G}_{1}~ \text{mixing (no dissipation):} \qquad  E\gg \sqrt{\epsilon} H, \\
   &\tilde{G}_{2}~ \text{mixing (no dissipation):} \qquad   E\gg \epsilon H, 
\end{align}
where we neglected numerical prefactors and worked at leading order in slow-roll. We recover the usual estimate that decoupling holds as long as $E \gg \sqrt{\epsilon} H$ \cite{Cheung:2007st, Cheung:2007sv} assuming a standard slow-roll hierarchy.

We now consider how these results are modified in the presence of dissipation $\Gamma$. We first study the low dissipation regime where $E \sim H \gg \gamma$, where we defined the rescaled dissipation coefficient
\begin{equation}\label{eq:smalldiss}
     \gamma \equiv \frac{\Gamma c_{s}^{2}}{\epsilon H^{2}\Mpl^{2}}.
\end{equation}
In this regime, the kinetic term of the field is still the dominant linear contribution \cite{Salcedo:2024smn}. The only term that is modified by $\Gamma$ is $\tilde{G}_{0}$, leading to
\begin{equation}
   \tilde{G}_{0}~ \text{mixing (low dissipation):} \qquad   E\gg\sqrt{\gamma \epsilon H}.
\end{equation}
This bound is trivially satisfied at low dissipation when the other bounds hold, from which we conclude that small dissipation does not affect the decoupling limit. The large dissipation regime is reached when $\gamma \gg H \sim E$. In this regime, the operator that dominates is the dissipation operator $\Upsilon \dpir \sim  \gamma E \pir$ \cite{Salcedo:2024smn}. Therefore, we need to redo the analysis for $\tilde{G}_{0,1,2}$, leading to
\begin{align}
	&\tilde{G}_{0}~ \text{mixing (large dissipation):} \qquad	 E \gg \epsilon H  , \\
	&\tilde{G}_{1}~ \text{mixing (large dissipation):} \qquad   \gamma E \gg  \epsilon H^2, \\
	&\tilde{G}_{2}~ \text{mixing (large dissipation):} \qquad  	\gamma \gg \epsilon H. 
\end{align}
Here, we recover the results from \cite{LopezNacir:2011kk} that these bounds are all easily verified as long as the slow-roll hierarchy holds. We conclude that as long as dissipation is concerned, the slow-roll hierarchy is enough to guarantee decoupling of $\pir$ from the perturbations of the metric. 

\begin{table}
    \begin{center}
    \begin{tabular}{ | P{2.8cm}|| P{5.5cm} | P{5.5cm} |  }
    \hline 
    \bigg. & $\gamma \ll H$ & $H \gg \gamma$ \\
    \hline
    \bigg.$\tilde{G}_{0}~ \text{mixing:}$ & $\epsilon \ll 1$ & $ \epsilon \ll 1$ \\
    \hline 
    \bigg.$\tilde{G}_{1}~ \text{mixing:}$ & $\epsilon \ll 1$ & $\epsilon H \ll \gamma$\\
    \hline
    \bigg.$\tilde{G}_{2}~ \text{mixing:}$ & $\epsilon \ll 1$  & $\cs\epsilon H \ll \gamma$ \\
     \hline
    \end{tabular}
    \end{center}
\caption{Hierarchy of scales among the noise coefficients to reach decoupling of the $\pi$ sector in the presence of noise in the gravity sector.}
\label{tab:noisedecoupv0}
\end{table}

At last, we include the noise contributions appearing in the constraints \eqref{eq:c1} and \eqref{eq:c2}, that is
\begin{align}
    \phi \supset \frac{\xi_{||}}{H\Mpl^{2}},\quad \mathrm{and}\quad
    \nabla^{2}F \supset a(t)\left[ - \frac{\xi_{00}}{\Mpl^{2}H} - \left(\frac{3c_{s}^{2}-\epsilon}{c_{s}^{2}\Mpl^{2}}\right)\xi_{||}\right].
\end{align}
By demanding that $\xi_{\pi}$ dominates over the stochastic terms coming from the constraints, we obtain a set of requirements. Making use of \Eq{eq:noisepi}, we further ask that the $\xi_{00}$ contributions from $\xi_{\pi}$ dominates from the $\xi_{00}$ contributions from the constraints, and similarly for $\xi_{\parallel}$.\footnote{One may need the heuristic estimate $\cs k^2 \sim (a H)^2$ at low dissipation and $\cs k^2 \sim  (\gamma/H) \times (a H)^2$ at large dissipation \cite{Salcedo:2024smn}.} This is a conservative requirement, as the aggregate contributions of $\xi_{\pi}$ could collectively dominate over the stochastic contributions of the constraints. We obtain the results stated in Table \ref{tab:noisedecoupv0} for decoupling to hold. Again, these bounds are easy to satisfied when the slow-roll hierarchy holds. 
We conclude that the quadratic dynamics of \cite{Salcedo:2024smn},
\begin{align}\label{eq:Sdecoup}
    S_{\pi}&\xrightarrow{(\text{decoupl.})} \int \dd^4x \sqrt{-g}\left\{- \dpir \dpia + \cs \bar{g}^{ij} \partial_i\pir \partial_j \pia + 2 \gamma \dpir \pia +  \frac{c_{s}^{2}}{\epsilon H^{2}\Mpl^{2}}\xi_{\pi} \pia\right\}, 
\end{align}
can indeed be understood as the decoupling limit of the theory of open gravity.


\section{Conclusion}\label{sec:conclu}

In this work, we develop an open effective field theory describing gravity in a medium, wherein the two graviton helicities and a scalar clock field interact with an unspecified environment. In contrast to the original EFT of Inflation \cite{Cheung:2007st} and EFT of Dark Energy \cite{Gubitosi:2012hu}, our framework systematically incorporates local dissipation and noise. These effects, which stem from a coarse graining of the system, are commonly encountered in real-world systems.\\

Inspired by the framework of open electromagnetism \cite{Salcedo:2024nex} {highlighted} in \Sec{sec:OpenEFT}, we construct an analogue open EFT for gravity in a medium within the Schwinger–Keldysh formalism. In \Sec{sec:unitgauge}, we define a notion of unitary gauge in the Schwinger-Keldysh formalism from which one can write down the most generic open functional for a clock field and dynamical gravity. At second order in derivatives acting on the retarded metric, this was given explicitly in \Eq{eq:S1v1}. The noise can be included in a systematic manner, following the same approach, which leads to \Eq{eq:S2v1}. 

Our construction naturally incorporates local dissipation and noise, leading to modifications of the background cosmology, as explored in \Sec{Sec:Background}. The Friedmann equations \eqref{eq:F1} and \eqref{eq:F2} receive additional contributions that can accommodate, for example, fluids with bulk viscosity \cite{Weinberg:1971mx, Brevik:2017msy}, interactions within the dark sector \cite{Guo:2004vg, Cai:2004dk, DiValentino:2017iww}, or scenarios such as brane-world cosmology \cite{Dvali:2000hr, Lue:2004rj}. 

The phenomenology of this class of model is further explored in \Sec{sec:tensor} where we derive predictions for primordial gravitational waves. Tensor modes exhibit modified equations of motion featuring non-trivial speed of propagation, dissipation, birefringence and noise. Remarkably, the leading-order gravitational birefringence is dissipative in nature, whereas conservative birefringence appears only at a higher derivative order\footnote{The counting of derivative is most precise in flat spacetime, where there are no time-dependent functions that can be hit by a derivative hence reducing the total number.} — opposite to the electromagnetic case. These effects alter predictions such as the tensor-to-scalar ratio, which we recomputed for this theory. Since tensors are sourced by noise, their power spectrum does not fix the scale of inflation. 
Current observational bounds can be used to constrain some of the EFT parameters affecting the tensor sector of the theory, as done in \Fig{fig:gw_ps_nobf}.

Finally, the presence a dynamical scalar can be made explicit by performing two St\"uckelberg tricks. In \Sec{sec:Stuck}, we discuss how one can introduce $\pi_r$ and $\pi_a$, the retarded and advanced St\"ueckelberg fields respectively. Neglecting the mixing with the fluctuations of the metric, we recover the decoupling limit of the Open Effective Field Theory of Inflation (Open EFToI) \cite{Salcedo:2024smn}. With the current framework, we are now able to go beyond this assumption and assess the mixing with gravity. For the linear dynamics, we find that decoupling holds in the slow-roll regime of inflation, independently of the inclusion of dissipation effects, consistently with the findings in \cite{LopezNacir:2011kk}. This confirms that the open EFT of inflation \cite{Salcedo:2024smn} can be understood as the decoupling limit of a theory of open gravity. \\

Our work provides a systematic framework to study dissipative and stochastic effects on gravitational waves and dynamical dark energy. There is a rich symmetry structure yet to be explored. In open electromagnetism, one is able to fix both the retarded and advanced gauges because a \textit{deformed advanced gauge} automatically follows from retarded gauge invariance at the linear order \cite{Salcedo:2024nex}. This is true also for gravity. However, we only fixed the retarded unitary gauge in this work. Instead, we used a field redefinition to reabsorb the $t_a$ contributions and reintroduced the $\pi_a$ through an advanced St\"uckelberg trick. In \Sec{ss:decoupling}, we observed that all operators in the Goldstone sector can be reintroduced without the need of an advanced clock, see \Eqs{eq:Stuck2}, \eqref{eq:Stuck1} and \eqref{eq:Stuck3}. This might be peculiar to single-clock settings and future work may elucidate when the $t_a$ contributions bring new physical effects. 

Apart from the peculiar symmetry structure of the theory, our construction needs to be further developed. Future work may assess the cutoff of our open EFT for gravity, following the methodology of perturbative unitary bounds in curved spacetime, see e.g. \cite{Pueyo:2024twm}. Applications to late-time cosmology may also need to address the frequency dependence of EFT coefficients away from the slow-roll limit. At last, non-linear aspects of retarded and advanced diffeomorphisms may be crucial to place requirements such as the noise constraints we found in open electromagnetism in the gravity setting, as discussed in \App{App:noise_constraints}.

Further work on open systems in cosmology is fundamental to make progress in the study of loops in cosmology. This is because the changing size of the horizon has to be incorporated in cutoff regularization to preserve scale invariance \cite{Senatore:2009cf}, which is akin to having an open system for the modes inside the horizon. A better understanding of this apparent openness may shed light on recent discussions of the analytic structure of loop corrections to cosmological correlators \cite{Weinberg:2005vy,Weinberg:2006ac,Senatore:2009cf,Salcedo:2022aal,Lee:2023kno,AguiSalcedo:2023nds,Qin:2023nhv,Baumann:2024mvm,Benincasa:2024ptf,Qin:2024gtr,Bhowmick:2025mxh,Cespedes:2025dnq}. 

Ultimately, finding explicit matching with UV models remains essential to confirm that our construction truly captures consistent modelts. Notably, interactions between gravitational waves and relativistic neutrinos are known to induce damping and rescattering effects \cite{Weinberg:2003ur, Flauger:2007es}, providing a concrete example of a cosmological environment where the techniques developed here may be applicable — though the locality assumptions underlying the present construction may require revision in such cases. 

In the scalar sector, cosmologies with viscous and stochastic corrections beyond the perfect fluid approximation have been the subject of extensive investigation over several decades  \cite{Weinberg:1971mx, Garcia-Bellido:2021idr, Espinosa-Portales:2021cac, Arjona:2021uxs, Garcia-Bellido:2024qau, Lapi:2023plb}. Our formalism of open gravity may provide a systematic framework to discuss gravity in a medium \cite{Modrekiladze:2024htc} such as gravitational aether \cite{Afshordi:2008xu, Donnelly:2010cr, Aslanbeigi:2011si, Blas:2014aca}. It would be worthwhile to explore whether the present framework can reproduce or shed light on certain features of these earlier constructions.

Interacting dark sector models have recently attracted considerable attention in the context of cosmological tensions \cite{Tudes:2024jpg, Wolf:2025jlc, Grimm:2025fwc, Silva:2025hxw, Aoki:2025bmj, Khoury:2025txd}. 
These models often incorporate dissipative and stochastic effects, capturing interactions between spacetime geometry, dark matter, and dark energy sectors. It would be worthwhile to explore the extent to which the present framework overlaps with this class of models. In particular, establishing a connection between our approach and recent developments in effective field theories of coupled dark energy \cite{Aoki:2025bmj} as well as of dark matter and chiral gravitational waves \cite{Aoki:2025uwz} would be especially valuable. 

At last, the present approach bears structural similarities to bigravity and massive gravity theories \cite{Isham:1971gm, Dubovsky:2004sg, Dubovsky:2009xk, deRham:2010kj, DAmico:2011eto, deRham:2014zqa, Bello-Morales:2023btf}, where two copies of diffeomorphism invariance are broken down to a diagonal subgroup. A closer examination of the symmetry breaking patterns in these theories and in our construction may offer further insight, and could potentially inform the search for partial UV completions, for example via dimensional deconstruction \cite{Dvali:2000hr, Arkani-Hamed:2001kyx}. \\

In this context, a natural extension of the present work would be its application to dark energy phenomenology in the late universe, in the spirit of the EFT of dark energy \cite{Gubitosi:2012hu}. Data analysis can be pursued at multiple levels to constrain this class of models. The background evolution presented in \Sec{Sec:Background} could be tested using observations of the cosmic microwave background, baryon acoustic oscillations and type Ia supernovae \cite{Hu:2013twa, Raveri:2014cka, DESI:2024mwx, DESI:2025zgx, DESI:2025kuo}. Upon including a matter sector, we generically find modifications to scalar perturbations at the level of the Poisson, continuity, and Euler equations, which affect the growth and clustering of matter. These effects can be probed using redshift-space distortions from galaxy surveys (BOSS, DESI, EUCLID, LSST) \cite{Colas:2019ret, Taule:2024bot, DESI:2024yrg, Euclid:2025tpw}, as well as through weak lensing measurements (KiDS, DES) \cite{Silvestri:2013ne, Pogosian:2016pwr, Sailer:2025rks, Castello:2024lhl, Grimm:2025fwc}.

Furthermore, dissipative and stochastic effects in the gravitational wave sector could be explored via observations of binary merger events \cite{Baker:2017hug, LIGOScientific:2021sio, Baker:2022eiz, Chen:2023wpj, LISACosmologyWorkingGroup:2022kbp, Domenech:2025bvr}, as well as data from pulsar timing arrays \cite{Weltman:2018zrl, Jenkins:2019uzp, Jaraba:2023djs, EPTA:2023xxk}. 
By systematizing the inclusion of dissipative and stochastic effects that naturally arise from the coarse-grained modeling of environmental degrees of freedom, the framework developed here offers physically motivated templates for data analysis in gravitation and cosmology. We hope this tool will assist the community in extracting meaningful physical insights from the wealth of forthcoming observational data. 


\paragraph{Acknowledgements.} We thank Niayesh Afshordi, Austin Joyce, Toshifumi Noumi, Riccardo Penco, Harvey Reall, Claudia de Rham, Leonardo Senatore, David Stefanyszyn, Tony Padilla, Andrew Tolley, Gianmassimo Tasinato, Xi Tong and Gonzalo Villa for the useful discussions. T.C. thanks the  organizers and participants of the 2025 Peyresq Spacetime Meeting, in particular Eugenio Bianchi, Laurent Freidel, Bei-Lok Hu and Albert Roura, together with the Julian Schwinger Foundation for financial support of the workshop. This work has been supported by STFC consolidated grant ST/X001113/1, ST/T000694/1, ST/X000664/1, ST/Y509127/1 and EP/V048422/1. S.A.S. is supported by a Harding Distinguished Postgraduate Scholarship. S.A.S. is specially thankful to Austin Joyce and the Kavli Institute for Cosmological Physics at the University of Chicago for their hospitality during part of the writing of this work.


\appendix


\section{Operators in open gravity}\label{app:ope}

In this appendix, we discuss specific operators to build intuition about the structure of the terms appearing in $S_1$ (\Eq{eq:S1v1}) and $S_2$ (\Eq{eq:S2v1}).


\subsection{Universal part}\label{app:univ}

We begin by expressing the universal part of the EFToI  \cite{Cheung:2007st} in the Keldysh basis. The universal part \eqref{eq:universal} governs the background dynamics and contributes to the quadratic action. The theory being unitary, the effective functional separates into
\begin{align}
S_{\mathrm{univ}}[g_+,g_-] = S_{\mathrm{univ}}[g_+] - S_{\mathrm{univ}}[g_-]\,.
\end{align}
If the only gauge fixing in this theory is the retarded unitary clock gauge $t_r=t$, see \eqref{tr=t}, then one would expect this theory to be invariant under 3d retarded diffs \textit{and} 4d advanced diffs. This is indeed what one finds including all $t_a$ terms. However, as we discussed in the main text, in a single clock cosmology $t_a$ can be removed by a field redefinition of $a^{\mu\nu}$. This procedure does not change the predictions of the theory, but breaks invariance under advanced time diffs. In the following we remedy this by performing an advanced St\"uckelberg trick, which introduces $\pia$. This in effect brings back all the $t_a$ terms we had removed, since if we had included the $t_a$ term the theory would have been 4d advanced diff invariant and the St\"uckelberg trick would have been a trivial operation. Therefore we proceed with
\begin{align}\label{eq:unitpmv2}
&S_{\mathrm{univ}}[g]= \int \dd^4 x \sqrt{-g} \left[ \frac{M^2_{\mathrm{\mathrm{Pl}}}}{2}R - \Lambda(t) - c(t )g^{00}\right].
\end{align}
We need to re-write this expression in the Keldysh basis, and then expand at linear order in the advanced fields. A useful expression is
\begin{align}
\sqrt{-g_\pm} &= \sqrt{- \mathrm{det}\left( g \pm \frac{a}{2}\right)}=\sqrt{-g}\left[1 \mp \frac{1}{4} g_{\mu\nu} a^{\mu\nu} + \frac{1}{16} g_{\mu\nu} a^{\nu\sigma} g_{\sigma\rho} a^{\rho\mu}  + \frac{1}{32} \left(g_{\mu\nu} a^{\mu\nu}\right)^2 + \cdots \right]. \label{eq:sqrt1v2}
\end{align}
Perturbing the Ricci scalar \cite{Weinberg:2008zzc}, we obtain
\begin{align}\label{eq:Sunivv2}
 S_{\mathrm{univ}} = \int \dd^4 x \sqrt{-g} \Bigg\{& \left[\frac{M^2_{\mathrm{\mathrm{Pl}}}}{2} G_{\mu\nu}a^{\mu \nu}  + \frac{M^2_{\mathrm{\mathrm{Pl}}}}{2} g^{\mu\nu} \delta_{a}R_{\mu\nu} \right] + \left[\frac{\Lambda(t)}{2} g_{\mu\nu} a^{\mu\nu} \right]\\
+& \left[\frac{c(t)}{2} \gr g_{\mu\nu} a^{\mu\nu} - c(t) \ga   \right]\Bigg\} + \cdots \nonumber 
\end{align}
where geometric objects such as the Einstein tensor $G_{\mu\nu}$ are constructed from the retarded metric and $\delta_{a}R_{\mu\nu}$ stands for the perturbation of the Ricci tensor induced by the advanced metric, that is
\begin{align}\label{eq:Rmunuv2}
\delta_{a}R_{\mu\nu} \equiv \delta R_{\mu \nu}\left[g + \frac{a}{2}\right] - \delta R_{\mu \nu}\left[g - \frac{a}{2}\right].
\end{align}
The contribution from this term is explicitly computed in \App{app:ft} and vanishes by integration by part (as long as we do not consider an $f(t)R$ extension of general relativity). We conclude that, at linear order in the advanced field, the unitary effective functional reads\footnote{Note that the gravitational side of the effective functional can equally be recovered from the deterministic equations of motion — for instance, as derived in \cite{Gubitosi:2012hu}:
\begin{align}
    M_{\mathrm{Pl}}^2G_{\mu\nu} + \Lambda(t)  g_{\mu\nu} + c(t) g^{00} g_{\mu\nu} - c(t) \delta^0_\mu \delta^0_\nu = 0,
\end{align}
from which we deduce that the universal part of \cite{Cheung:2007st, Gubitosi:2012hu} generates \Eq{eq:univfinv2}. It indeed fulfills the construction rules found above. One can then reintroduce the $\pir$ and $\pia$ contributions by performing a retarded and advanced \stuck trick.}
\begin{align}\label{eq:univfinv2}
    S_{\mathrm{univ}}=\int d^{4}x\sqrt{-g}&\bigg[\frac{\Mpl^{2}}{2}G_{\mu\nu}a^{\mu\nu}+\frac{\Lambda(t)}{2} g_{\mu\nu}a^{\mu\nu}+\frac{c(t)}{2}g^{00}g_{\mu\nu}a^{\mu\nu}-c(t)a^{00} \bigg].
\end{align}
Explicitly, in the language of \Eq{eq:decomp}, it amounts to consider the EFT operators controlled by the coefficients
\begin{align}
  \gtto{1} = - c(t), \quad \gssi{1} =\frac{c(t)}{2}, \quad \gsso{1} = \frac{\Lambda(t)}{2} - \frac{c(t)}{2}.
\end{align}
\Eq{eq:univfinv2} indeed fulfills the structure discussed above, namely the constraints given in \Eqs{eq:norm}, \eqref{eq:herm} and \eqref{eq:pos} are satisfied, all the terms are at least linear in an advanced component and the effective functional is invariant under retarded spatial diffeomorphisms. 

Upon performing the advanced St\"uckelberg trick (\Sec{subsec:advstuck}), we obtain the $\pi_a$ contributions,
\begin{align}
   S_{\mathrm{univ}} \rightarrow S_{\mathrm{univ}} + \int \dd^4 x \sqrt{-g} \left[ - \dot{\Lambda}(t) \pia  - 2 c(t) g^{0\mu}\partial_\mu \pia  - \dot{c}(t) \gr \pia \right], 
\end{align}
While the second term can be easily deduced from the advanced St\"uckelberg of $-c(t) a^{00}$,
the other two terms come with no derivatives acting on $\pia$, originate from the integration by part of 
\begin{align}
     \frac{\Lambda(t)}{2} g_{\mu\nu}a^{\mu\nu} \quad&\xrightarrow{(\text{t-diff}_a)} \quad\Lambda(t) \left(\dpia + 3H\pia \right)\quad \xrightarrow{(\text{IBP})}\quad -\dot{\Lambda} \pia. 
\end{align}
and similarly for $- \dot{c}(t) \gr \pia $ (up to higher order operators that can eventually be reabsorbed into non-universal operators).
At the background level, these contributions generate the tadpole equation
\begin{align}\label{eq:tadv1v2}
   \frac{\delta \bar{S}_{\mathrm{eff}}}{\delta \pia} = 0 \qquad \Rightarrow \qquad \dot{\Lambda}(t)+\dot{c}(t)+6Hc(t)=0.
\end{align}
Equivalently, using the first and second Friedmann equations, 
\begin{align}\label{eq:tadv2v2}
   3 H^{2}\Mpl^{2}&=\Lambda(t)+c(t)\quad,\quad2\Mpl^{2}\dot{H}=-2c(t),
\end{align}
we recover the exact same equation. Indeed, there is no new information in the continuity equation compared to the Einstein equations.  

Note that one can explicitly check that once the $\pia$ is reintroduced, the theory is indeed invariant under advanced diffs $\epsilon_a^\mu$. First, the Einstein-Hilbert term is invariant\footnote{In checking diff-invariance we systematically drop all boundary terms.} thanks to the Bianchi identity
\begin{equation}
    \Delta S_{\rm EH}=\Mpl^{2}\int \dd^4x \sqrt{-g}\left[G_{\mu\nu}\nabla^{\mu}\epsilon^{\nu}_{a}\right]=\Mpl^{2}\int \dd^4x \sqrt{-g}\nabla^{\mu}\left[G_{\mu\nu}\epsilon^{\nu}_{a}\right]=0.
\end{equation}
Second, the $\Lambda(t)$ term transform as
\begin{equation}
    \Delta S_{\Lambda(t)}=\int \dd^4x \sqrt{-g}\left[\Lambda(t)\nabla_{\mu}\epsilon^{\mu}_{a}+\dot{\Lambda}(t)\epsilon^{0}_{a}\right]=\int \dd^4x \sqrt{-g}\nabla_{\mu}\left[\Lambda(t)\epsilon^{\mu}_{a}\right]=0,
\end{equation}
which is also invariant. At last, the invariance under advanced diffs of the $c(t)$ terms can be checked through
\begin{align}
    \Delta S_{c(t)}=&\int \dd^4x \sqrt{-g}\left[-2c(t)\nabla^{0}\epsilon^{0}_{a}+c(t)g^{00}\nabla_{\mu}\epsilon^{\mu}_{a}+2c(t)g^{0\mu}\partial_{\mu}\epsilon_{a}^{0}+\dot{c}(t)g^{00}\epsilon_{a}^{0}\right]\\
    =&\int \dd^4x \sqrt{-g}\left\{\left[-2c(t)\nabla^{0}\epsilon^{0}_{a}+2c(t)g^{0\mu}\partial_{\mu}\epsilon_{a}^{0}\right]+\nabla_{\mu}\left[c(t)g^{00}\epsilon_{a}^{\mu}\right]\right\}=0,
\end{align}
where we have used that $\partial=\nabla$ when acting on a scalar. Note the cancellation here takes places in two different ways, as there is a piece that sums up to a boundary term and a piece that cancels identically.

\subsection{Minimal extension}\label{app:minimal}

In \cite{Salcedo:2024smn}, we constructed an open EFT for inflation in which the inflaton field experiences dissipation and noise from its interactions with an unknown environment. We considered the linear functional for the canonically normalized scalar field
\begin{align}\label{eq:target}
    S_{\mathrm{eff}} = \int \dd^4x \sqrt{-g} \bigg\{\dpir \dpia - c_s^2 \partial_i \pir \partial_i \pia - \gamma \dpir \pia + i \beta_\pi  \pia^2  \bigg\},
\end{align}
where we neglect derivative noises.
While the standard kinetic term can be extracted from the $c(t)$ terms of $S_{\mathrm{univ}}$, as discussed around \Eq{eq:Stuck2}, the remaining contributions require additional ingredients.

\paragraph{Speed of sound.} In addition to the standard kinetic term for the $\pi$ sector, breaking Lorentz invariance allows one to modify the speed of sound for the Goldstone mode \cite{Cheung:2007st, Piazza:2013coa}. In the Schwinger-Keldysh contour, this can be achieved by considering $S_{\cs}[g_+] - S_{\cs}[g_-]$ with 
\begin{align}\label{eq:onebranchv0v2}
    S_{\cs}[g] = \frac{M_2}{2}\int \dd^4 x \sqrt{-g} (\delta g^{00} )^2,
\end{align}
where we introduced the notation $\delta g^{00} = 1 + g^{00}$.
In the Keldysh basis, at linear order in the advanced metric, it leads to
\begin{align}\label{eq:addcs}
    S_{\cs} = \frac{M_2}{2}\int \dd^4 x \sqrt{-g} \bigg[ - \frac{1}{2} (\delta g^{00})^2 g_{\mu\nu} a^{\mu\nu} + 2 \delta g^{00} a^{00} \bigg],
\end{align}
that is $\gssii{1} = M_2 /4$ and $\gtti{1} = M_2$.
One can reintroduce the $\pia$ by performing an advanced \stuck trick, leading to 
\begin{align}\label{eq:csfin}
    S_{\cs} \rightarrow S_{\cs} + \int \dd^4 x\sqrt{-g}\left[2M_{2}(t)(1+g^{00})g^{0\mu} \partial_{\mu}\pi_{a} +\frac{\dot{M}_{2}(t)}{2}\left(1+g^{00}\right)^{2}\pi_{a}\right]\,.
\end{align}
At last, upon reintroducing $\pir$, the first term in \Eq{eq:csfin} generates a quadratic operator $ 4 M_2 \dpir \dpia$, which, once added to the standard kinetic term of the universal part, generates a non-trivial speed of sound $c^2_s$.


\paragraph{Dissipation.}
Comparing the dissipation term $- \gamma \dpir \pia$ with the building blocks identified above, the former can be captured by 
\begin{align}\label{eq:dissiprefv2}
    S_{\mathrm{dissip}} &= \int \dd^4 x \sqrt{ - g}  \frac{\Gamma(t)}{3H}\left( 1 + g^{00}\right) \left( g_{\mu\nu} a^{\mu\nu} + a^{00}\right) .
\end{align}
Indeed, reintroducing the $\pia$ by performing an advanced \stuck trick, we obtain
\begin{align}
     S_{\mathrm{dissip}} & \rightarrow S_{\mathrm{dissip}} + \int \dd^4 x \sqrt{ - g}  2\Gamma(t)\left( 1 + g^{00}\right) \pia,
\end{align}
which generates a dissipative operator $\dpir \pia$ once the retarded \stuck trick is performed.  

\paragraph{Noise.}
At last, the noise term appearing in \Eq{eq:target} is easily embedded in our open gravity EFT once a Hubbard-Stratonovich trick \cite{Hubbard:1959ub,Stratonovich1957} 
is performed, leading to
\begin{align}\label{eq:noisev2}
    S_{\mathrm{noise}} 
\equiv -  \int d^{4}x\sqrt{-g}&\xi_{\mu\nu} a^{\mu\nu}.
\end{align}
Reintroducing the $\pia$ by performing an advanced \stuck trick, we obtain a noise for the $\pi$ sector, that is 
\begin{align}
     S_{\mathrm{noise}} & \rightarrow S_{\mathrm{noise}} + \int \dd^4 x \sqrt{ - g}  \left(- 2 \xi_{00} \dot{\pi}_a + 2 \bar{g}^{ij} \xi_{0i} \partial_j \pia + 2H \bar{g}^{ij} \xi_{ij} \pia\right).
\end{align}


\subsection{Modified Einstein-Hilbert term}\label{app:ft}

At last, we discuss a well known modification of the Einstein-Hilbert term known as $f(t)$ gravity \cite{Gubitosi:2012hu}. Let us consider $S_{f(t)}[g_+,g_-] = S_{f(t)}[g_+] - S_{f(t)}[g_-] $ with
\begin{align}
    S_{f(t)}[g] = \int \dd^4 x \sqrt{-g} \frac{M^2_*}{2} f(t) R,
\end{align}
where we introduced a time dependent function $f(t)$. In the Keldysh basis, after performing the advanced \stuck trick, we obtain
\begin{equation}
    S_{f(t)} =\int\dd^4 x\sqrt{-{g}}\frac{M^{2}_{*}}{2}\left[f(t){G}_{\mu\nu}a^{\mu\nu}+f(t)\delta_{a}R_{\mu\nu}{g}^{\mu\nu} - {R}\dot{f}(t)\pia\right],
\end{equation}
where the variation of the Ricci tensor is given by the Palatini's identity
\begin{equation}\label{eq:Palatini}        \delta_{a}R_{\mu\nu}=\nabla_{\rho}\delta_{a}\Gamma^{\rho}_{\mu\nu}-\nabla_{\nu}\delta_{a}\Gamma^{\rho}_{\rho\mu}\,,
\end{equation}
in which
\begin{equation}
    \delta_{a}\Gamma^{\rho}_{\mu\nu}=\frac{1}{2}{g}^{\rho\alpha}\left(\nabla_{\mu}\delta_{a}g_{\alpha\nu}+\nabla_{\nu}\delta_{a}g_{\alpha\mu}-\nabla_{\alpha}\delta_{a}g_{\mu\nu}\right).
\end{equation}
Using the relation between inverse and direct variations of the metric
\begin{equation}
    \delta_{a}g_{\mu\nu}=-{g}_{\mu\mu'}{g}_{\nu\nu'}a^{\mu'\nu'}\,,
\end{equation}
one finds
\begin{equation}
    \delta_{a}\Gamma^{\rho}_{\mu\nu}=-\frac{1}{2}{g}^{\rho\alpha}\nabla_{\mu}{g}_{\alpha\alpha'}{g}_{\nu\nu'}a^{\alpha'\nu'}-\frac{1}{2}{g}^{\rho\alpha}\nabla_{\nu}\delta_{a}{g}_{\mu\mu'}{g}_{\alpha\alpha'}a^{\mu'\alpha'}+\frac{1}{2}{g}^{\rho\alpha}\nabla_{\alpha}{g}_{\mu\mu'}{g}_{\nu\nu'}a^{\mu'\nu'}.
\end{equation}
At last, using that the connection is metric affine, we obtain 
\begin{equation}
    \delta_{a}\Gamma^{\rho}_{\mu\nu}=-\frac{1}{2}{g}_{\nu\nu'}\nabla_{\mu}a^{\rho\nu'}-\frac{1}{2}{g}_{\mu\mu'}\nabla_{\nu}a^{\rho\mu'}+\frac{1}{2}{g}^{\rho\alpha}{g}_{\mu\mu'}{g}_{\nu\nu'}\nabla_{\alpha}a^{\mu'\nu'},
\end{equation}
and
\begin{equation}
    \delta_{a}\Gamma^{\rho}_{\rho\nu}=-\frac{1}{2}{g}_{\rho\mu'}\nabla_{\nu}a^{\rho\mu'}.
\end{equation}
It leaves a term of the form
\begin{equation}
    \int \dd^4 x \sqrt{-{g}}\frac{M^{2}_{*}}{2}\left[f(t)\delta_{a}R_{\mu\nu}{g}^{\mu\nu}\right]=\int \dd^4 x \sqrt{-{g}}\frac{M^{2}_{*}}{2}f(t)\left({g}_{\mu\nu}\Box-\nabla_{\mu}\nabla_{\nu}\right)a^{\mu\nu}\,,
\end{equation}
where
\begin{align}
    \Box f(t) \equiv \frac{1}{\sqrt{- {g}}} \partial_\mu \left[\sqrt{-{g}} {g}^{\mu\nu}\partial_\nu f(t)\right]\,.
\end{align}
This can be integrated by parts into
\begin{equation}
    \int \dd^4 x\sqrt{-{g}}\frac{M^{2}_{*}}{2}\left[f(t)\delta_{a}R_{\mu\nu}{g}^{\mu\nu}\right]=\int \dd^4 x \sqrt{-{g}}\frac{M^{2}_{*}}{2}\left[\left({g}_{\mu\nu}\Box-\nabla_{\mu}\nabla_{\nu}\right)f(t)\right]a^{\mu\nu}.
\end{equation}
In the end, the full modified Einstein-Hilbert term reads
\begin{align}
     S_{f(t)} &= \int \dd^4x \sqrt{-{g}}  
    \frac{M^2_*}{2} \bigg\{\Big[ \left(  {G}_{\mu \nu} +{g}_{\mu\nu}\Box -\nabla_{\mu}\nabla_{\nu} \right)f(t) \Big] a^{\mu\nu} - \dot{f}(t) {R} \pia  \bigg\}.         
      \label{eq:ft1}
\end{align}
Whenever $f(t)$ is time independent, only the $G_{\mu\nu}$ contribution survives in \Eq{eq:ft1}. This particular modification of the theory will be relevant when we will use our theory of open gravity to describe dark energy as an open system.


\section{Noise sector}

This appendix gathers a series of technical results on the noise part of the effective functional $S_2$ given in \eqref{eq:S2v1}. In \App{app:S2_second_order}, we extend the main text construction up to second order in derivatives. In \App{App:noise_constraints}, we derive so-called noise constraints that limit the number of independent advanced variables. At last, in \App{app:mixedHS}, we discuss the mixing between noises in the gravity sector and scalar sector.

\subsection{Noise functional to second order in derivatives}\label{app:S2_second_order}

To construct $S_2$ up to second order in derivatives in a systematic manner, we follow a similar approach to the construction of $S_1$ in \Sec{subsec:functional}. We begin again by expanding each term in $S_2$ in powers of $(1+g^{00})^\ell$ 
\begin{align}
    N &= \sum_{\ell=0}\left(g^{00} + 1 \right)^\ell N_{\ell}(R_{\mu\nu\rho\sigma}, K_{\mu\nu}, n_\mu, \nabla_\mu; t)  \, ,\\
    N_{\mu\nu} &= \sum_{\ell=0}\left(g^{00} + 1 \right)^\ell N_{\mu\nu, \ell}(R_{\mu\nu\rho\sigma}, K_{\mu\nu}, n_\mu, \nabla_\mu; t)   \, , \\
    N_{\mu\nu\rho\sigma} &= \sum_{\ell=0}\left(g^{00} + 1 \right)^\ell N_{\mu\nu\rho\sigma, \ell}(R_{\mu\nu\rho\sigma}, K_{\mu\nu}, n_\mu, \nabla_\mu; t)  \, .
\end{align}
$N_\ell$, $N_{\mu\nu, \ell}$ and $N_{\mu\nu\rho\sigma}$ are constructed with the same building blocks $M_\ell$, $M^{ts}_{\mu, \ell}$ and $M^{ss}_{\mu\nu, \ell}$ that were used to construct $S_1$ in \Sec{subsec:functional}. A major difference to the construction of $S_1$ is additional covariant derivatives that can appear when constructing $S_2$. When constructing $S_1$, we did not have to worry about covariant derivatives acting on the advanced fields $t_a$ or $a^{\mu\nu}$, as these can always be integrated by part to yield only terms already considered in \Eq{eq:S1v1}. Contrarily, we encounter terms of the form $\sim t_a \nabla_\mu t_a$ in $S_2$, which must be considered explicitly. To abbreviate the construction we introduce the notation $\{ M_\ell \}_n$, which denotes a building block, which is made out of the same operators as $M_\ell$, up to $n$th order in derivatives. Each operator in $\{ M_\ell \}_n$ comes with a new EFT coefficient. For instance
\begin{align}
    \{ M_\ell \}_1 &= \beta_{1,\ell} K + \beta_{2,\ell} \, , \bigg. \\
    \{ M_\ell \}_0 &= \beta_{3,\ell} \, .
\end{align}
Note that $\{ M_\ell \}_0 \subseteq \{ M_\ell \}_1$, as $\{ M_\ell \}_1$ also includes all operators at zeroth order in derivatives. Using this notation, we obtain
\begin{align}\label{eq:noise_scalar_bb}
    N_\ell &= \{ M_\ell \}_2 + \{ M_\ell \}_1 n^\alpha \nabla_\alpha + \{ M_\ell \}_0 n^\alpha n^\beta \nabla_\alpha \nabla_\beta + \{ M_\ell \}_0 \Box \\
    &= \btt{1} + \btt{2} K + \btt{3}  K^2 + \btt{4} K_{\alpha\beta} K^{\alpha\beta} + \btt{5} \nabla^0 K + \btt{6} R + \btt{7} R^{00} \nonumber \bigg. \\
    +& \btt{8} \nabla^0 + \btt{9} K \nabla^0 + \btt{10} \nabla^0 \nabla^0 + \btt{11} \Box 
    \, .
\end{align}
In this expression, covariant derivatives are understood to act on one of the two $t_a$'s that appear in $S_2 \supset \int \dd^4 x \sqrt{-g} \, N t_a^2$.
$N_{\mu\nu, \ell} (R_{\mu\nu\rho\sigma}, g^{00}, K_{\mu\nu}, \nabla_\mu; t)$ can be split along the foliation just as $M_{\mu\nu, \ell}$ in \Eq{eq:decomp} and \Eq{eq:decomp_Mmunu},
\begin{equation}
    N_{\mu\nu, \ell} = n_\mu n_\nu N^{tt}_\ell + g_{\mu\nu} N^{ss}_\ell + n_{(\mu} N^{ts}_{\nu),\ell} + \tilde{N}^{ss}_{\mu \nu, \ell}\,.
\end{equation}
Taking the new derivative terms into account, we find: 
\begin{align}
    N^{tt}_\ell &= \{M_\ell\}_2 + \{ M_\ell \}_1 n^\alpha \nabla_\alpha + \{ M_\ell \}_0 n^\alpha n^\beta \nabla_\alpha \nabla_\beta + \{ M_\ell \}_0 \Box \bigg.\,, \\
    N^{ss}_\ell &= \{M_\ell\}_2 + \{ M_\ell \}_1 n^\alpha \nabla_\alpha + \{ M_\ell \}_0 n^\alpha n^\beta \nabla_\alpha \nabla_\beta + \{ M_\ell \}_0 \Box \bigg. \,,\\
    N^{ts}_{\mu,\ell} &= \{M^{ts}_{\mu,\ell}\}_2 + \{M^{ts}_{\mu,\ell}\}_1 n^\alpha \nabla_\alpha + \{M^{ts}_{\mu,\ell}\}_0 n^\alpha n^\beta \nabla_\alpha \nabla_\beta \bigg. \nonumber  \\
    +& \{M^{ts}_{\mu,\ell}\}_0 \Box + \{M_\ell\}_1 \nabla_\mu + \{M_\ell\}_0 n^\alpha \nabla_\alpha \nabla_\mu \nonumber \bigg. \\
    &= \bts{1} R^{0}{}_\mu + \bts{2} \nabla_\mu K + \bts{3} \nabla_\beta K^{\beta}{}_\mu + \left( \bts{4}K + \bts{5} \right) \nabla_\mu +\bts{6} \nabla^0 \nabla_\mu  \bigg. \,,\label{eq:noise_vector_bb}\\
    \tilde{N}^{ss}_{\mu\nu,\ell} &= \{\tilde{M}^{ss}_{\mu\nu,\ell}\}_2 + \{\tilde{M}^{ss}_{\mu\nu,\ell}\}_1 n^\alpha \nabla_\alpha + \{\tilde{M}^{ss}_{\mu\nu,\ell}\}_0 n^\alpha n^\beta \nabla_\alpha \nabla_\beta \bigg.  \nonumber\\
    +& \{\tilde{M}^{ss}_{\mu\nu,\ell}\}_0 \Box + \{M^{ts}_{(\mu,\ell}\}_1 \nabla_{\nu)} + \{M^{ts}_{(\mu,\ell}\}_0 n^\alpha \nabla_{\nu)} \nabla_\alpha + \{M_\ell\}_0 \nabla_{(\mu} \nabla_{\nu)} \nonumber \bigg. \\
    &= \bss{12} R_{\mu\nu} + \bss{13} R_\mu{}^0{}_\nu{}^0 + \bss{14} K_{\mu\nu} + \bss{15} \nabla^0 K_{\mu\nu} \nonumber \bigg. \\
    +&  \bss{16} K_{\mu\alpha} K^{\alpha}{}_{\nu} + \bss{17} K K_{\mu\nu} + \bss{18} K_{\mu\nu} \nabla^0 + \bss{19} \nabla_{(\mu} \nabla_{\nu)} \, .\label{eq:noise_rank2tensor_bb}
\end{align}
Note that $\{M^{ts}_{\mu,\ell}\}_1 = \{M^{ts}_{\mu,\ell}\}_0 = \{\tilde{M}^{ss}_{\mu\nu,\ell}\}_0 = 0$.

Lastly, we also decompose $N_{\mu\nu\rho\sigma, \ell}$ along the foliation. As this tensor is contracted with two advanced metrics in \Eq{eq:S2try}, we decide to make its symmetries manifest. In particular it is symmetric in $\mu \leftrightarrow \nu$ and $\rho \leftrightarrow \sigma$, as well as in $(\mu\nu) \leftrightarrow (\rho\sigma)$. 
\begin{align}\label{eq:noise4_decomp}
    N_{\mu\nu\rho\sigma, \ell} &= n_\mu n_\nu n_\rho n_\sigma N^1_l + g_{\mu\nu} g_{\rho\sigma} N^2_l + g_{\mu(\rho} g_{\sigma)\nu} N^3_l + \left(g_{\mu\nu}n_\rho n_\sigma + n_\mu n_\nu g_{\rho\sigma} \right) N^4_l \bigg. \\
    +& \left(g_{\mu(\rho} n_{\sigma)} n_\nu + g_{\nu(\rho} n_{\sigma)} n_\mu \right) N^5_l + \left( n_\nu n_\rho n_\sigma  N^1_{\mu,\ell} + \textrm{3 perm's} \right) \nonumber \bigg. \\
    +& \left( g_{\nu\rho} n_\sigma  N^2_{\mu,\ell} + \textrm{7 perm's} \right) + \left( g_{\rho\sigma} n_\nu  N^3_{\mu,\ell} + \textrm{7 perm's} \right) \nonumber \bigg. \\
    +& \left( N^1_{\mu\nu,\ell} g_{\rho\sigma} + g_{\mu\nu} N^1_{\rho\sigma,\ell} \right)
    + \left(N^2_{\mu(\rho,\ell} g_{\sigma)\nu} + N^2_{\nu(\rho,\ell} g_{\sigma)\mu} \right) \nonumber \bigg. \\
    +& \left(N^3_{\mu\nu,\ell} n_\rho n_\sigma + N^3_{\rho\sigma,\ell} n_\mu n_\nu\right) + \left(N^4_{\mu(\rho,\ell} n_{\sigma)} n_\nu +N^4_{\nu(\rho,\ell} n_{\sigma)} n_\mu \right) \nonumber \bigg. \\
    +& \left( N_{\mu\nu\rho,\ell} n_\sigma + \textrm{7 perm's} \right) + \left( N_{\mu\nu\rho\sigma,\ell} + \textrm{7 perm's} \right) \nonumber  \bigg. \, ,
\end{align}
where the various permutations, abbreviated with $\text{perm's}$, are such that every term is individually symmetric in $\mu \leftrightarrow \nu$ and $\rho \leftrightarrow \sigma$, as well as in $(\mu\nu) \leftrightarrow (\rho\sigma)$.
The various $N$-scalars, vectors and tensors are build completely analogous to before, such that for instance all five $N^i{}_\ell$ feature the same operators, namely the ones in \Eq{eq:noise_scalar_bb}, but they may in general all come with different EFT coefficients. Likewise, the three $N^i{}_{\alpha, \ell}$ are build similarly to \Eq{eq:noise_vector_bb} and the four $N^i{}_{\alpha\beta, \ell}$ are build similar to \Eq{eq:noise_rank2tensor_bb}. Finally, we have to construct $N_{\alpha\beta\gamma,\ell}$ and $N_{\alpha\beta\gamma\delta,\ell}$ as the most generic rank-3 and rank-4 tensors constructed out of $R_{\mu\nu\rho\sigma}$, $K_{\mu\nu}$, $g_{\mu\nu}$, $n_\mu$ and $\nabla_\mu$ as before. As before, any term that has free indices on $n_\mu$ or $g_{\mu\nu}$ is redundant with the other terms in our decomposition in \Eq{eq:noise4_decomp}. Conversely leaving free indices on $\nabla_\mu$ is not, as explained previously. We arrive at   
\begin{align}
    N_{\alpha\beta\gamma,\ell} &= \beta^3_{10,\ell} n^\delta R_{\alpha\gamma\beta\delta} + \{M^{ts}_{\alpha, \ell}\}_0 \nabla_\beta \nabla_\gamma + \{M^{ss}_{\alpha \beta, \ell}\}_1 \nabla_\gamma + \{M^{ts}_{\alpha, \ell}\}_1 K_{\beta\gamma} \bigg.\nonumber \\
    +& \{M_{\alpha\beta,\ell}^{ss}\}_0 \nabla^{\mu}K_{\mu\gamma} + \{M_{\alpha\beta,\ell}^{ss}\}_0 \nabla_{\gamma}K \nonumber \bigg. \\
    &= \beta^3_{10,\ell} n^\delta R_{\alpha\gamma\beta\delta} + \beta^3_{11,\ell} K_{\alpha \beta} \nabla_\gamma \bigg. \,,\\
    N_{\alpha\beta\gamma\delta,\ell} &= \beta^4_{31,\ell} R_{\alpha\beta\gamma\delta} + \{\tilde{M}_{\alpha\beta}\}_1 K_{\gamma\delta} + \{\tilde{M}^{ss}_{\alpha\beta}\}_0 \nabla_\gamma \nabla_\delta + \{N_{\alpha\beta\gamma,\ell}\}_1 \nabla_\delta \bigg. \nonumber\\
    &= \beta^4_{31,\ell} R_{\alpha\beta\gamma\delta} + \beta^4_{32,\ell} K_{\alpha\beta}K_{\gamma\delta} \bigg. \,.
\end{align}
Note that $\{M^{ts}_{\alpha, \ell}\}_0 = \{\tilde{M}^{ss}_{\alpha\beta}\}_0 = \{N_{\alpha\beta\gamma,\ell}\}_1 = 0$.


\subsection{Noise constraints in open gravity}\label{App:noise_constraints}

This Appendix aims at the generalization to gravity of the noise constraint derived in \cite{Salcedo:2024nex}. Noise constraints are present to ensure the correct number of degrees of freedom in the noise, such that it matches the number of independent degrees of freedom. To illustrate this, we first present the gravitational case where the noise is the only non-unitary extension. Then, we extend the discussion to account for the interplay between dissipation and noise in open gravity. 

\subsubsection{Gravitational noise constraints without dissipation} 

Physically, the noise constraints originate from the the modified conservation laws in the presence of an environment \cite{2016RPPh...79i6001S}. To appreciate this fact, let us consider the equations of motion for the metric which can be written to all orders in perturbation theory as
\begin{equation}
    G_{\mu\nu}=T_{\mu\nu}+\xi_{\mu\nu}\quad,\quad
\end{equation}
where $T_{\mu\nu}$ is a set of deterministic currents and $\xi_{\mu\nu}$ is the stochastic noise that is induced by the environment. The Bianchi-identity imposes that
\begin{equation}
    \nabla^{\mu}G_{\mu\nu}=0,
\end{equation}
such that the RHS of the Einstein equations is conserved 
\begin{equation}
    \nabla^{\mu}T_{\mu\nu}+\nabla^{\mu}\xi_{\mu\nu}=0.
\end{equation}
This equation yields a set of constraints that the noise variables have to fulfill.

\paragraph{Advanced diffeomorphisms.}

As we shall now see, these equations can be derived from the variation of the functional under advanced diffs. Consider the Hubbard-Stratonovich transformed action \cite{Hubbard:1959ub,Stratonovich1957}
\begin{equation}
	S_{2}= \int \dd^4x \sqrt{-g} \left(a^{\mu \nu}\xi_{\mu\nu}+\xi_{t_a}t_{a}\right).
\end{equation}
Consider the transformation rules for the advanced metric and the advanced time under a generic advanced diff $\epsilon_a^\mu$,
\begin{equation}
    a^{\mu\nu} \rightarrow a^{\mu\nu} + 2 \nabla^{(\mu} \epsilon_a^{\nu)}\,,\quad t_{a}\to t_{a}-\epsilon_{a}^{0}.
\end{equation}
The transformation of the noise functional is
\begin{equation}
    \Delta S_{2} =\int \dd^4x \sqrt{-g}\left[2\xi_{\mu\nu}\nabla^{\mu}\epsilon^{\nu}_{a}-\xi_{t_{a}}\epsilon_{a}^{0}\right],
\end{equation}
which, upon integrating by part, leads to
\begin{equation}
    \Delta S_{2} = \int \dd^4x \sqrt{-g}\left[-2\epsilon^{\nu}_{a}\nabla^{\mu}\xi_{\mu\nu}-\xi_{t_{a}}\epsilon_{a}^{0}\right],
\end{equation}
where we were able to drop the boundary thanks to the vanishing of advanced diffeomorphisms on the boundary. In the case of open electromagnetism asking for $\Delta S_{2} = 0$ yields the constraints discussed above using the equations of motion \cite{Salcedo:2024nex}. Let us consider the form of the noise constraints in this case depending on whether we consider spatial or time diffeomorphisms. \\

\textit{Advanced spatial diffeomorphisms.} First we focus on spatial diffeomorphism. Consider the case $\epsilon^{0}_{a}=0$ such that the variation simplifies to
\begin{equation}
    \Delta S_{2}=\int \dd^4x \sqrt{-g}\left[-2\epsilon^{i}_{a}\nabla^{\mu}\xi_{\mu i}\right].
\end{equation}
We can then expand the divergence of the noise tensor,
\begin{equation}
    \nabla^{\mu}\xi_{\mu i}=g^{\mu\beta}\nabla_{\beta}\xi_{\mu i}=g^{\mu\beta}\partial_{\beta}\xi_{\mu i}-g^{\mu\beta}\Gamma_{\beta\mu}^{\sigma}\xi_{\sigma i}-g^{\mu\beta}\Gamma_{\beta i}^{\sigma}\xi_{\mu \sigma}.
\end{equation}
In an FLRW spacetime, these terms reduce to
\begin{align}
    g^{\mu\beta}\partial_{\beta}\xi_{\mu i}&=-\dot{\xi}_{0i}+\frac{1}{a^{2}(t)}\partial_{j}\xi_{ji}\,,\\
    -g^{\mu\beta}\Gamma_{\beta\mu}^{\sigma}\xi_{\sigma i}&=-3H\xi_{0i}\,, \bigg. \\
    -g^{\mu\beta}\Gamma_{\beta i}^{\sigma}\xi_{\mu \sigma}&=\Gamma_{0i}^{k}\xi_{0k}-\frac{\delta^{mn}}{a^{2}(t)}\Gamma_{n i}^{0}\xi_{m 0}=H\xi_{0i}-H\xi_{0i}=0.
\end{align}
It leaves us a first contraint of the form
\begin{tcolorbox}
    \begin{equation}\label{eq:const1}
        \dot{\xi}_{0i}+3H\xi_{0i}=\frac{1}{a^{2}(t)}\partial_{j}\xi_{ji}.
    \end{equation}
\end{tcolorbox}
\noindent This equation is imposing three constraints, one on the scalar sector and two on the vector sector. In the scalar sector, it implies the existence of a relation between the stochastic noise variables, reducing the number of independent noises by one. In the vector sector, this constraint is removing two independent noises, which matches the number of vector degrees of freedom that cannot be removed by gauge transformations. \\

\textit{Advanced time diffeomorphisms.} Now we can restrict to time diffeomorphisms, where only $\epsilon_a^{0}=\pia$ is non-vanishing. The variation of the noise functional is
\begin{equation}
    \Delta S_{2}=\int \dd^4x \sqrt{-g}\left[-2\epsilon^{0}_{a}\nabla^{\mu}\xi_{\mu0}-\xi_{t_{a}}\epsilon_{a}^{0}\right].
\end{equation}
We can expand the divergence,
\begin{align}
    \nabla^{\mu}\xi_{\mu0}=g^{\alpha\beta}\partial_{\alpha}\xi_{\beta 0}-g^{\alpha\beta}\Gamma_{\alpha\beta}^{\sigma}\xi_{\sigma 0}-g^{\alpha\beta}\Gamma^{\sigma}_{\alpha 0}\xi_{\beta \sigma},
\end{align}
which leads to
\begin{align}
    g^{\alpha\beta}\partial_{\alpha}\xi_{\beta 0}&=-\dot{\xi}_{00}+\frac{1}{a^{2}(t)}\partial_{i}\xi_{i0},\\
    -g^{\alpha\beta}\Gamma_{\alpha\beta}^{\sigma}\xi_{\sigma 0}&=-\frac{\delta^{ij}}{a^{2}(t)}H a^{2}(t)\delta_{ij}\xi_{00}=-3H\xi_{00},\\
    -g^{\alpha\beta}\Gamma^{\sigma}_{\alpha 0}\xi_{\beta \sigma}&=-\frac{\delta^{ik}}{a^{2}(t)}H \delta_{i}^{j}\xi_{k j}=-\frac{H}{a^{2}(t)}\delta^{ij}\xi_{ij}.
\end{align}
At last, we obtain
\begin{tcolorbox}
    \begin{equation}\label{eq:const2}
        \frac{1}{a^{2}(t)}\partial_{i}\xi_{i0}=\dot{\xi}_{00}+3H\xi_{00}+\frac{H}{a^{2}(t)}\delta^{ij}\xi_{ij}+\frac{\xi_{t_{a}}}{2}.
    \end{equation}
\end{tcolorbox}
\noindent At the level of the scalar sector this is imposing a constraint between the scalar noises variables. Therefore, it is again reducing the number of independent noises in one.

\paragraph{Flat space example I.} Let us illustrate the above findings with a simple scenario given by considering the scalar sector of GR in flat space. Here, we derive the noise constraints for the noises that couple to the scalar sector of GR in the absence of deterministic currents. Consider the perturbed Einstein's equations in flat space
\begin{equation}
	G_{\mu\nu}=\xi_{\mu\nu}.
\end{equation}
Perturbing around a FLRW background, we consider the scalar sector of the metric fluctuations\footnote{If one wants to convert into the notations of \cite{Flauger:2009uta}, we have used $E = 2\phi$, $A = - 2 \psi - (2/3) \partial^2 \beta$ and $B = 2 \beta$.} 
\begin{align}
	\delta g_{00} = - 2 \phi,\qquad \delta g_{0i} = a (t)\partial_i F, \qquad \delta g_{ij} = 2a^2(t)\left(- \psi \delta_{ij} + \chi_{ij}\right)
\end{align}
with the traceless part $\chi_{ij} \equiv (\partial_i \partial_j - \frac{1}{3} \delta_{ij} \partial^2) \beta$. The perturbations of the Einstein tensor read \cite{Weinberg:2008zzc}
 \begin{align}
	\delta G_{00} &=2 \left[ \frac{\partial^2 \psi}{a^2(t)}  - 3 H \dot{\psi} \right] + \frac{2}{3} \frac{\partial^2 \partial^2 \beta }{a^2(t)} - \frac{2}{a(t)} H \partial^2 F, \Bigg. \label{eq:G00}\\
	\delta G_{0i} &= \partial_i \left[ 2H\phi + 2\dot{\psi} + \frac{2}{3} \partial^2 \dot{\beta} - \left(2 \dot{H} + 3H^2 \right) a(t) F\right], \Bigg. \label{eq:G0i} \\
	\delta G_{ij} &= 2a^2(t) \delta_{ij} \left\{ \left[  H \dot{\phi} + \left(2 \dot{H} + 3H^2 \right) \phi\right] + \left[\Ddot{\psi} + 3 H\dot{\psi} +  \left(2 \dot{H} + 3H^2 \right) \psi \right]\right\}\Bigg. \nonumber \\
	-& \left(\partial_i \partial_j - \delta_{ij} \partial^2 \right)  \left[\left(\phi - \psi - \frac{\partial^2}{3} \beta\right) + \left(2 a(t) H F + a(t)\dot{F} \right)  \right] \Bigg.\nonumber \\
	+& a^2(t) \left(\partial_i \partial_j - \frac{1}{3 }\delta_{ij} \partial^2 \right) \left[ \Ddot{\beta} + 3 H\dot{\beta}  - 2 \left(2 \dot{H} + 3H^2 \right) \beta \right] . \label{eq:Gij}
\end{align}
At last, the flat space results are obtained by setting $a\rightarrow1$ and $H, \dot{H} \rightarrow 0$.
It follows that 
\begin{align}
	2\nabla^{2}\psi+\frac{2}{3}\nabla^{4}\beta&=\xi_{00},\\
	\partial_{i}\left(2\dot{\psi}+\frac{2}{3}\nabla^{2}\dot{\beta}\right)&=\xi_{0i}.
\end{align}
and
\begin{align}
	&\frac{2}{3}\delta_{ij}\left(\nabla^{2}\phi+\nabla^{2}\dot{F}-\frac{1}{3}\nabla^{4}\beta-\nabla^{2}\psi+3\ddot{\psi}\right) \nonumber \\
	& \qquad+\left(\partial_{i}\partial_{j}-\frac{1}{3}\nabla^{2}\delta_{ij}\right)\left(\psi-\phi-\dot{F}+\Ddot{\beta}+\frac{1}{3}\nabla^{2}\beta\right)=\xi_{ij}.
\end{align}
It is trivial to combine the first and second equation to find
\begin{equation}
	\dot{\xi}_{00}=\partial_{i}\xi_{0i},
\end{equation}
which agrees with the expectation set by time diffeomorphisms in \eqref{eq:const2}. Similarly, one can combine the last two equations to find
\begin{equation}
	\partial_{i}\dot{\xi}_{0i}=\partial_{i}\partial_{j}\xi_{ij},
\end{equation}
which again agrees with the naive expectation from the noise constraints from spatial scalar diffeomorphisms in \eqref{eq:const1}.

\paragraph{Flat space example II.} One way to have a consistent Minkowski background and a non-trivial Goldstone mode is by allowing for a time dependent Planck mass $\Mpl^2 \rightarrow f(t) M_*^2$ (à la EFTofDE \cite{Gubitosi:2012hu})
\begin{equation}
	S_{\mathrm{eff}}=\int \dd^4 x \sqrt{-g}\bigg\{\frac{M_{*}^{2}}{2}\left[f(t) G_{\mu\nu}+g_{\mu\nu}\Box f(t)-\nabla_{\mu}\nabla_{\nu}f(t)\right]a^{\mu\nu} - \xi_{\mu\nu} a^{\mu\nu} + \dot{f}(t)R t_{a} - \xi_{t_a} t_a \bigg\}.
\end{equation}
This way, we will be able to study the constraint that arises from $t_{a}$. 
The equations of motion are given by
\begin{equation}
    f(t)G_{\mu\nu}+g_{\mu\nu}\Box f(t)-\nabla_{\mu}\nabla_{\nu}f(t)=\xi_{\mu\nu},
\end{equation}
from which we observe that flat space is only a solution if $\ddot{f}(t)$ vanishes exactly. Now we can study the fluctuations
\begin{equation}
    f(t)\delta G_{\mu\nu}+\bar{g}_{\mu\nu}\delta\Box f(t)-\delta\left(\nabla_{\mu}\nabla_{\nu}\right)f(t)=\xi_{\mu\nu},
\end{equation}
where now the main object of study is the variation of the connection and the Box operator. For the connection, we find
\begin{align}
    \delta\left(\nabla_{\mu}\nabla_{\nu}\right)f(t)&=-\delta\Gamma_{\mu\nu}^{0}\dot{f}(t)\bigg.\nonumber\\
    &=-\frac{1}{2}\bar{g}^{0\sigma}\left[\partial_{\mu}\delta g_{\sigma \nu}+\partial_{\nu}\delta g_{\sigma \mu}-\partial_{\sigma}\delta g_{\mu\nu}\right]\nonumber\\
    &=\frac{1}{2}\left(\partial_{\mu}\delta g_{0 \nu}+\partial_{\nu}\delta g_{0 \mu}-\delta \dot{g}_{\mu\nu}\right),
\end{align}
and for the Box operator
\begin{align}
    \delta \Box f(t)&=\delta\left\{\frac{1}{\sqrt{-g}}\partial_{\mu}\left[\sqrt{-g}g^{\mu0}\dot{f}(t)\right]\right\}\nonumber\\
    &=\delta\left[g^{\mu0}\partial_{\mu}\text{Log}\left(\sqrt{-g}\right)+\partial_{\mu}g^{\mu 0}\right]\dot{f}(t) \bigg.\nonumber\\
    &=\dot{f}(t)\left[\frac{1}{2}\partial_{0}\left(\delta g_{00}-\delta^{ij}\delta g_{ij}\right)+\partial_{\mu}\delta g^{\mu0}\right]\,.
\end{align}

The equation from varying with respect to $t_{a}$, that is $\dot{f}(t)\delta R=\xi_{t_{a}}$, then becomes
\begin{equation}
    \dot{f}(t)\left[- 2 \left(3 \Ddot{\psi} - 2 \partial^2 \psi +\partial^2 \phi + \partial^2 \dot{F} \right) + \frac{4}{3} \partial^2 \partial^2 \beta\right]=\xi_{t_{a}},
\end{equation}
while the Einstein equations expand to
\begin{align}
    f(t)\left(2\nabla^{2}\psi+\frac{2}{3}\nabla^{4}\beta\right)-\dot{f}(t)\left[\frac{1}{2}\partial_{0}\left(\delta g_{00}-\delta^{ij}\delta g_{ij}\right)+\partial_{\mu}\delta g^{\mu0}\right]-\frac{\dot{f}(t)}{2}\delta\dot{g}_{00}&=\xi_{00},\\
    f(t)\partial_{i}\left(2\dot{\psi}+\frac{2}{3}\nabla^{2}\dot{\beta}\right)-\frac{\dot{f}(t)}{2}\partial_{i}\delta g_{00}&=\xi_{0i},
\end{align}
and 
\begin{align}
    &\frac{2 f(t)}{3}\delta_{ij}\left(\nabla^{2}\phi+\nabla^{2}\dot{F}-\frac{1}{3}\nabla^{4}\beta-\nabla^{2}\psi+3\ddot{\psi}\right) +f(t)\left(\partial_{i}\partial_{j}-\frac{1}{3}\nabla^{2}\delta_{ij}\right)\left(\psi-\phi-\dot{F}+\Ddot{\beta}+\frac{1}{3}\nabla^{2}\beta\right)\nonumber\\
     &  \qquad +\delta_{ij}\dot{f}(t)\left[\frac{1}{2}\partial_{0}\left(\delta g_{00}-\delta^{ij}\delta g_{ij}\right)+\partial_{\mu}\delta g^{\mu0}\right]-\frac{\dot{f}(t)}{2}\left(\partial_{i}\delta g_{0j}+\partial_{j}\delta g_{0i}-\delta \dot{g}_{ij}\right)=\xi_{ij}.
\end{align}
Again, an investigation of the equations of motion allows us to conclude that the constraints \Eqs{eq:const1} and \eqref{eq:const2} hold.

\subsubsection{Deformed gravitational noise constraints with dissipation} 

In \cite{Salcedo:2024nex}, some of the authors found in open electromagnetism a constraint for the noise of the form given in \Eq{eq:noiseconstEM}, that we repeat here for clarity
\begin{align}
    \partial^\mu (j_\mu + \xi_\mu) = \Gamma (j_0 + \xi_0)\,.
\end{align}
In this expression, $j_\mu$ is the system's current, $\xi_\mu$ the environmental noise and $\Gamma$ the Wilson coefficient of the dissipation operator. The presence of the later apparently breaks the advanced gauge invariance. If this breaking was effective, there would be no such constraint. Rather, the authors of \cite{Salcedo:2024nex} found that dissipation operators deform the advanced gauge symmetry. This makes the number of independent stochastic noises match the number of degrees of freedom in the theory. The deformation of the symmetries follows
\begin{equation}
	\begin{split}
		& \; A^{\mu} \xrightarrow{(\text{ret.})} A^{\mu}+\partial^{\mu}\lambda\\
		&\; a^{\mu}  \xrightarrow{(\text{adv.})} a^{\mu}+\partial^{\mu}\lambda
	\end{split} \quad \xrightarrow{(\text{dissip.})} \quad \begin{split}
		&\; A^{\mu} \xrightarrow{(\text{ret.})} A^{\mu}+\partial^{\mu}\lambda\\
		& \; a^{\mu} \xrightarrow{(\text{def. adv.})} a^{\mu}+\partial^{\mu}\lambda+\Gamma \delta^{\mu}_{~0}.
	\end{split}
\end{equation}
We here explore if an analogous phenomenon takes place in open gravity by considering two simple examples. 

\paragraph{Dissipative flat space example I.} 

To this end, we introduce operators that deform the advanced time diff. Consider the addition of a term of the form $Ra^{00}$ to the Einstein's equations,
\begin{equation}
    G_{\mu\nu}+\lambda R\delta_{\mu}^{0}\delta_{\nu}^{0}=\xi_{\mu\nu}.
\end{equation}
It is trivial to see that the equations allow for a Minkowski background such that we can comfortably write
\begin{equation}
    \delta G_{\mu\nu}+\lambda \delta R\delta_{\mu}^{0}\delta_{\nu}^{0}=\xi_{\mu\nu}.
\end{equation}
The equations of motion then take the form of
\begin{align}
    2\nabla^{2}\psi+\frac{2}{3}\nabla^{4}\beta+\lambda\left[\frac{4}{3}\nabla^{4}\beta-2\left(\nabla^{2}\phi+\nabla^{2}\dot{F}-2\nabla^{2}\psi+3\Ddot{\psi}\right)\right]&=\xi_{00},\\
    \partial_{i}\left(2\dot{\psi}+\frac{2}{3}\nabla^{2}\dot{\beta}\right)&=\xi_{0i},
\end{align}
with 
\begin{align}
    &\frac{2}{3}\delta_{ij}\left(\nabla^{2}\phi+\nabla^{2}\dot{F}-\frac{1}{3}\nabla^{4}\beta-\nabla^{2}\psi+3\ddot{\psi}\right)\\
    &\qquad+\left(\partial_{i}\partial_{j}-\frac{1}{3}\nabla^{2}\delta_{ij}\right)\left(\psi-\phi-\dot{F}+\Ddot{\beta}+\frac{1}{3}\nabla^{2}\beta\right)=\xi_{ij}.
\end{align}
The noise constraint associated with advanced spatial diffs is unchanged
\begin{equation}
    \partial_{0}\partial_{i}\xi_{0i}-\partial_{i}\partial_{j}\xi_{ij}=0,
\end{equation}
while the noise constraint from advanced time diff is modified by the presence of $\lambda$, leading to
\begin{equation}
    \partial_{0}\xi_{00}-(1+\lambda)\partial_{i}\xi_{0i}+\lambda \delta_{ij}\dot{\xi}_{ij}=0.
\end{equation}
We conclude that just as in open electromagnetism, dissipative effects deform the noise constraints in open gravity. In would be interesting to investigate if this observation suggests the existence of a deformed advanced diff under which the effective functional is invariant. We leave it for future work.

\paragraph{Dissipative flat space example II.}  Now consider the addition of a term of the form $Rg_{ij}a^{ij}$, with the Einstein equations reading
\begin{equation}
    G_{\mu\nu}+\lambda R \left(g_{\mu\nu}+n_{\mu}n_{\nu}\right)=\xi_{\mu\nu}.
\end{equation}
Again, these equations allow for a Minkowski background. In this case, the perturbed Einstein's equations reads
\begin{equation}
    \delta G_{\mu\nu}+\lambda\delta R\left(\bar{g}_{\mu\nu}+\bar{n}_{\mu}\bar{n}_{\nu}\right)=\xi_{\mu\nu},
\end{equation}
which expand to:
\begin{align}
    &2\nabla^{2}\psi+\frac{2}{3}\nabla^{4}\beta=\xi_{00},\\
    &\partial_{i}\left(2\dot{\psi}+\frac{2}{3}\nabla^{2}\dot{\beta}\right)=\xi_{0i},
\end{align}
and
\begin{align}
    &\frac{2}{3}\delta_{ij}\left(\nabla^{2}\phi+\nabla^{2}\dot{F}-\frac{1}{3}\nabla^{4}\beta-\nabla^{2}\psi+3\ddot{\psi}\right)+\lambda\delta_{ij}\left[\frac{4}{3}\nabla^{4}\beta-2\left(\nabla^{2}\phi+\nabla^{2}\dot{F}-2\nabla^{2}\psi+3\Ddot{\psi}\right)\right]\nonumber\\
    &\qquad +\left(\partial_{i}\partial_{j}-\frac{1}{3}\nabla^{2}\delta_{ij}\right)\left(\psi-\phi-\dot{F}+\Ddot{\beta}+\frac{1}{3}\nabla^{2}\beta\right)=\xi_{ij}.
\end{align}
The constraint equation from advanced time diff is unchanged
\begin{equation}
    \partial_{0}\xi_{00}=\partial_{i}\xi_{0i}.
\end{equation}
Meanwhile, the constrained from advanced spatial diffs is deformed into
\begin{equation}
   - \partial_{i}\dot{\xi}_{0i}+\partial_{i}\partial_{j}\xi_{ij}-\lambda \nabla^{2}\xi_{00}+\frac{\lambda}{1-3\lambda}\nabla^{2}\left(\delta_{ij}\xi_{ij}\right)=0.
\end{equation}
Again, we observe that dissipative effects deform the noise constraint in open gravity.


\subsection{Noise mixing between gravity and the scalar}\label{app:mixedHS}

In general, the quadratic action includes mixing between the advanced metric $a^{\mu\nu}$ and the advanced clock $t_{a}$ at the level of the noise operators, 
\begin{equation}
    S_2 = i \int \dd^4 x \sqrt{-g} \Big[N_{\mu\nu\rho\sigma}  a^{\mu\nu}  a^{\rho\sigma}  +  N_{\mu\nu} a^{\mu\nu} t_a  + N t_a^2 \Big]. 
\end{equation}
It is instrumental to assess how strong this mixing is at the level of the stochastic noises. To do this, we have to generalize the Hubbard-Stratonovich \cite{Hubbard:1959ub,Stratonovich1957} trick to include various tensors of different rank. The simplest way consists in considering the SVT decomposition of the advanced metric
\begin{equation}
    a^{\mu\nu}\equiv \begin{pmatrix}
       2\phi_{a} & a^{-1}(t)\partial_{i}F_{a}+C^{a}_{i}\\
       a^{-1}(t)\partial_{i}F_{a}+C^{a}_{i} & 2\psi_{a}a^{-2}(t)\delta_{ij}-2a^{-2}(t)\left(\partial_{i}\partial_{j}-\frac{1}{3}\nabla^{2}\delta_{ij}\right)\beta_{a}+2\partial_{(i}D_{j)}+\gamma_{ij}^{TT,a}
    \end{pmatrix},
\end{equation}
where there are 4 scalar variables, 4 components corresponding to two transverse vectors and 2 components corresponding to the two transverse and traceless polarizations. This means we can rearrange the variables into a list $\mathfrak{A}^{c}$, such that
\begin{equation}
    \mathfrak{A}^{c} \equiv \left(\phia,\psi_{a},F_{a},\beta_{a},C^{a}_{i},D^{a}_{i},\gamma^{TT,a}_{ij},t_{a}\right),
\end{equation}
which allows to rewrite the quadratic action as
\begin{equation}
    S_{2}=i\int \dd^4x \sqrt{-g}\mathfrak{A}^{c}\tilde{\mathfrak{N}}_{cd}\mathfrak{A}^{d}\quad,\quad\tilde{\mathfrak{N}}_{cd}=-i\frac{\delta^{2}S_{2}}{\delta\mathfrak{A}^{c}\delta\mathfrak{A}^{d}}.
\end{equation}
Under this decomposition, the first $10\times10$ block of $\mathfrak{N}_{cd}$ corresponds to $N_{\mu\nu\rho\sigma}$, while the last column (row) corresponds to $N_{\mu\nu}$ and $N$:
\begin{equation}
    \tilde{\mathfrak{N}}_{cd}=\begin{pmatrix}
        \frac{\partial^{2}}{\partial\mathfrak{A}^{c}\partial\mathfrak{A}^{d}}N_{\mu\nu\rho\sigma}a^{\mu\nu}a^{\rho\sigma} & \frac{\partial}{\partial\mathfrak{A}^{c}} N_{\mu\nu}a^{\mu\nu}\\
        \frac{\partial}{\partial\mathfrak{A}^{d}} N_{\mu\nu}a^{\mu\nu} & N
    \end{pmatrix}=\begin{pmatrix}
        \mathfrak{N}_{cd} & \mathfrak{N}_{c}\\
        \mathfrak{N}_{d} & \mathfrak{N}^{2}+\mathfrak{N}^{ab}\mathfrak{N}_{a}\mathfrak{N}_{b}
    \end{pmatrix}.
\end{equation}

We can then use the results for the noise in open electromagnetism \cite{Salcedo:2024nex} to rewrite $S_{2}$ in terms of these new variables:
\begin{equation}
    \text{exp}\left(-S_{2}\left[\mathfrak{A}^{c}\right]\right)=\int\mathcal{D}\mathfrak{E}_{c}\text{exp}\left\{\int \dd^4 x \sqrt{-g}\left[-\frac{1}{4}\left(\tilde{\mathfrak{N}}^{-1}\right)^{cd}\mathfrak{E}_{c}\mathfrak{E}_{d}+i\mathfrak{E}_{c}\mathfrak{A}^{c}\right]\right\},
\end{equation}
with
\begin{equation}
    \left(\tilde{\mathfrak{N}}^{-1}\right)^{cd}=\begin{pmatrix}
      \left(\mathfrak{N}^{-1}\right)^{cd}+\frac{\mathfrak{N}^{c}\mathfrak{N}^{d}}{\mathfrak{N}^{2}}& -\frac{\mathfrak{N}^{c}}{\mathfrak{N}^{2}} \\
      -\frac{\mathfrak{N}^{d}}{\mathfrak{N}^{2}}  & \frac{1}{\mathfrak{N}^{2}}
    \end{pmatrix}
\end{equation}
such that the correlation function of the stochastic noises $\mathfrak{E}_{c}$ is given by: 

\begin{equation}
    \langle\mathfrak{E}_{c}(t,\bfx)\mathfrak{E}_{d}(t',\bfx')\rangle=2\tilde{\mathfrak{N}}_{cd}\delta(t-t')\delta(\bfx-\bfx').
\end{equation}

We are particularly interested in the mixing between the \stuck field noise $\xi_{\pi}$ and the scalar sector of the advanced metric, composed of the variables $(\phi_a,\psi_a,F_a,\beta_a)$\footnote{Note that in this case the noises associated to advanced components $F_a, \beta_a$ are not simply $\xi_{0i}$ and $\xi_{ij}$, but rather:
\begin{align}
    \frac{\delta}{\delta F_{a}}\int \dd^4x \sqrt{-g} 2a^{0i}\xi_{0i}&=-2\frac{1}{a(t)}\partial_{i}\xi_{0i}=-2\frac{1}{a(t)}\partial^{2}\xi^{||}=\xi_{F}\\
    \frac{\delta}{\delta \beta_{a}}\int \dd^4x \sqrt{-g} a^{ij}\xi_{ij}&=2\frac{1}{a^{2}(t)}\left(\partial_{i}\partial_{j}-\frac{1}{3}\delta_{ij}\partial^{2}\right)\xi_{ij}=\frac{2 k^{4}\Xi^{TT}}{3a^{2}(t)}=\xi_{\beta}
\end{align}
}. Therefore, we expand \eqref{eq:S2v1} in terms of these variables and the advanced time $t_{a}$. In this paper we focus on the linear dynamics, and hence it is enough to truncate at $\ell=0$:
\begin{align}
    S_{2}=i\int \dd^4x \sqrt{-g}&\bigg[\beta_{1,0}t_{a}^{2}+2\beta_{2,0}\phi_{a}t_{a}-2\beta_{3,0}\left(\phi_{a}-3\psi_{a}\right)t_{a}+4\beta_{4,0}\phi_{a}^{2}+4\beta_{5,0}\left(\phi_{a}-3\psi_{a}\right)^{2} \nonumber \\
    &+\beta_{6,0}\left(4\phia^{2}+2F_{a}\partial^{2}F_{a}+12\psi_{a}^{2}+\frac{8}{3}\partial^{2}\partial^{2}\beta_{a}^{2}\right) \nonumber\\
    &-4\beta_{7,0}\phia\left(\phia-3\psi_{a}\right)+\beta_{8,0}\left(2\phia^{2}-F_{a}\partial^{2}F_{a}\right)\bigg],
\end{align}
which can be rewritten as
\begin{equation}
    S_{2}=i\int \dd^4x \sqrt{-g}\begin{pmatrix}
        \phia & \psi_{a} & F_{a} & \beta_{a} & t_{a}
    \end{pmatrix}^c \mathfrak{N}_{cd} \begin{pmatrix}
        \phia \\ \psi_{a} \\ F_{a} \\ \beta_{a} \\ t_{a}
    \end{pmatrix}^d.
\end{equation}
with 
\begin{align}
	\mathfrak{N}_{cd} = \begin{pmatrix}
	 	8 (\beta_{4}+\beta_{5}+\beta_{6}-\beta_{7}+\beta_{8}) & 12 (\beta_{7}-2 \beta_{5}) & 0 & 0 & 2 (\beta_{2}-\beta_{3}) \\
	 	12 (\beta_{7}-2 \beta_{5}) & 24 (3 \beta_{5}+\beta_{6}) & 0 & 0 & 6 \beta_{3} \\
	 	0 & 0 & 2 k^2 (\beta_{8}-2 \beta_{6}) & 0 & 0 \\
	 	0 & 0 & 0 & \frac{16}{3}\beta_{6,0}k^{4} & 0 \\
	 	2 (\beta_{2}-\beta_{3}) & 6 \beta_{3} & 0 & 0 & 2 \beta_{1} \\
	 \end{pmatrix}.
\end{align}
This matrix must verify the Silvester's criteria for positivity of a matrix, coming from the constraints \eqref{eq:pos}. It implies the bounds
\begin{align}
    &\beta_{4}+\beta_{5}+\beta_{6}-\beta_{7}+\beta_{8}>0, \qquad \quad ~~ \beta_{8}-2 \beta_{6}>0, \qquad \quad  ~~ \beta_{6}>0, \bigg.  \\
    &4 \beta_{4} (3 \beta_{5}+\beta_{6})+4 \beta_{8} (3 \beta_{5}+\beta_{6})+16 \beta_{5} \beta_{6}+4 \beta_{6}^2-4 \beta_{6} \beta_{7}-3 \beta_{7}^2>0, \bigg.\\
    &\beta_{1} \left[4 \beta_{4} \left(3 \beta_{5}+\beta_{6}\right)+4 \beta_{8} \left((3 \beta_{5}+\beta_{6}\right)+16 \beta_{5} \beta_{6}+4 \beta_{6}^2-4 \beta_{6} \beta_{7}-3 \beta_{7}^2\right] \Big. \nonumber \\
    &\quad -\beta_{2}^2 (3 \beta_{5}+\beta_{6})+\beta_{2} \beta_{3} (2 \beta_{6}+3 \beta_{7})-\beta_{3}^2 (3 \beta_{4}+4 \beta_{6}+3 \beta_{8})>0. 
\end{align}

We close this Appendix by noticing that the advanced \stuck mixes the noises $\xi_{t_{a}},\xi_{00}$ and $\xi_{0i}$ to form the noise $\xi_{\pi}$
\begin{align}
    \int \dd^4x \sqrt{-g}\left(\xi_{t_{a}}t_{a}+\xi_{\mu\nu}a^{\mu\nu}\right)&\xrightarrow{(\text{t-diff}_a)} \int \dd^4x \sqrt{-g}\Bigg\{\left(\xi_{t_{a}}t_{a}+\xi_{\mu\nu}a^{\mu\nu} \right) \nonumber \\
    +&\pia\left[\xi_{t_{a}}+2\left(\dot{\xi}_{00}+3H\xi_{00}\right)-\frac{2}{a^{2}(t)}\partial_{i}\xi_{0i}+\frac{2H}{a^{2}(t)}\delta^{ij}\xi_{ij}\right]\Bigg\},
\end{align}
such that
\begin{equation}
    \xi_{\pi}=\xi_{t_{a}}+2\left(\dot{\xi}_{00}+3H\xi_{00}\right)-\frac{2}{a^{2}(t)}\partial_{i}\xi_{0i}+\frac{2H}{a^{2}(t)}\delta^{ij}\xi_{ij}.
\end{equation}
In particular, this provides a mechanism to generate noise in the $\pi$ sector even if $t_a = 0$. 


\section{Technical derivations for mixing with gravity}\label{app:mixing}

In this appendix, we gather the technical details relative to the estimate of the mixing with gravity studied in \Sec{Sec:minimal}.

\subsection*{Perturbed Einstein's equations} 

We start from \Eq{eq:th2}, which we reproduce here for clarity,
\begin{align}\label{eq:th2app}
	S_{\rm eff}=\int \dd^4x \sqrt{-g}\bigg[&\frac{\Mpl^{2}}{2}G_{\mu\nu}a^{\mu\nu}+\frac{\Lambda(t)}{2}g_{\mu\nu}a^{\mu\nu}-c(t)a^{00}+\frac{c(t)}{2}g^{00}g_{\mu\nu}a^{\mu\nu}\nonumber \\
	 -&\frac{M_{2}(t)}{4}\left(1+g^{00}\right)^{2}g_{\mu\nu}a^{\mu\nu}+M_{2}(t)\left(1+g^{00}\right)a^{00} \bigg.\nonumber\\
     +& \frac{\Gamma(t)}{3H}\left(1 + g^{00} \right)\left( a^{00} + g_{\mu\nu} a^{\mu\nu}\right)-\xi_{\mu\nu}a^{\mu\nu}\bigg].
\end{align}
The first step is reintroducing $\pir$ by performing a retarded St\"uckelberg trick. Each of the terms in $S_{\rm eff}$ is susceptible to receiving new terms. Bearing in mind that any tensor transforms as (method II)
\begin{align}
	A^{00}&\rightarrow \partial_{\mu}(t+\pir)\partial_{\nu}(t+\pir) A^{\mu\nu}\,,\\
	A^{0i}&\rightarrow \partial_{\mu}(t+\pir) A^{\mu i}\,,\\
	A^{ij}&\rightarrow A^{ij},
\end{align}
while any function of time becomes
\begin{equation}
	f(t)\to f(t+\pir),
\end{equation}
we can now reintroduce the $\pir$ contributions. While the Einstein-Hilbert term
\begin{equation}
	S_{\rm EH} \supset \frac{\Mpl^{2}}{2}\int \dd^4x \sqrt{-g}G_{\mu\nu}a^{\mu\nu},
\end{equation}
remains invariant as there is no explicit time dependence and the integrand is a scalar under retarded time diffeomorphisms, the cosmological constant term transforms as
\begin{equation}
	S_{\Lambda(t)}\supset\int \dd^4x \sqrt{-g}\left[\frac{1}{2}\Lambda(t+\pir)g_{\mu\nu}a^{\mu\nu}\right].
\end{equation}
Similarly, the kinetic term receives
\begin{align}
	S_{c(t)}\supset&\int \dd^4x \sqrt{-g}\bigg[-c(t+\pir)\partial_{\mu}\left(t+\pir\right)\partial_{\nu}\left(t+\pir\right)a^{\mu\nu} \nonumber \\
	+&\frac{1}{2}c(t+\pir)\partial_{\alpha}\left(t+\pir\right)\partial_{\beta}\left(t+\pir\right)g^{\alpha\beta}g_{\mu\nu}a^{\mu\nu}\bigg].
\end{align}
The speed of sound contributions become
\begin{align}
	S_{\cs}\supset&\int \dd^4x \sqrt{-g}\frac{M_{2}(t+ \pir)}{2}\left\{-\frac{1}{2}\left[1+\partial_{\alpha}\left(t+\pir\right)\partial_{\beta}\left(t+\pir\right)g^{\alpha\beta}\right]^{2}g_{\mu\nu}a^{\mu\nu}\right. \nonumber\\
	+&2\left[1+\partial_{\alpha}\left(t+\pir\right)\partial_{\beta}\left(t+\pir\right)g^{\alpha\beta}\right]\partial_{\mu}\left(t+\pir\right)\partial_{\nu}\left(t+\pir\right)a^{\mu\nu}\bigg\}.
\end{align}
At last, the dissipation operator transforms as
\begin{align}
    S_{\Gamma(t)}\supset&\int \dd^4x \sqrt{-g}\frac{\Gamma(t + \pir)}{3H}\bigg\{\left[1+\partial_{\alpha}\left(t+\pir\right)\partial_{\beta}\left(t+\pir\right)g^{\alpha\beta}\right]\left[\partial_{\mu}\left(t+\pir\right)\partial_{\nu}\left(t+\pir\right) + g_{\mu\nu}\right]a^{\mu\nu} \bigg\}.
\end{align}
and the noise operator $\xi_{\mu\nu}a^{\mu\nu}$ does not transform, being fully contracted. 

Thanks to the introduction of the retarded \stuck we can now write the Einstein equations in a covariant form. Introducing the short hand notation
\begin{equation}
	A_{\alpha\beta}=\partial_{\alpha}\left(t+\pir\right)\partial_{\beta}\left(t+\pir\right),
\end{equation}
we obtain the Einstein's equations
\begin{align}
	&\frac{\Mpl^{2}}{2}G_{\mu\nu}+\frac{\Lambda(t+\pir)}{2}g_{\mu\nu}-c(t+\pir)A_{\mu\nu}+\frac{c(t+\pir)}{2}A_{\alpha\beta}g^{\alpha\beta}g_{\mu\nu}\nonumber\\
	&-\frac{M_{2}(t+\pir)}{4}\left(1+A_{\alpha\beta}g^{\alpha\beta}\right)^{2}g_{\mu\nu}+M_{2}(t+\pir)\left(1+A_{\alpha\beta}g^{\alpha\beta}\right)A_{\mu\nu}\nonumber \bigg. \\
    &+ \frac{\Gamma(t + \pir)}{3H}\left(1+A_{\alpha\beta}g^{\alpha\beta}\right)\left(A_{\mu\nu}+ g_{\mu\nu}\right) =\xi_{\mu\nu}.
\end{align}
Each term has to be expanded to linear order in perturbations, leading to 
\begin{align}
	\frac{\Mpl^{2}}{2}G_{\mu\nu}\quad \rightarrow \quad & \frac{\Mpl^{2}}{2}\delta G_{\mu\nu}\,,\\ 
	\frac{\Lambda(t+\pir)}{2}g_{\mu\nu}\quad \rightarrow \quad & \frac{\Lambda}{2}\delta g_{\mu\nu}+\frac{\dot{\Lambda}(t)}{2}\pir\bar{g}_{\mu\nu},
\end{align}
for the Einstein-Hilbert term and the cosmological constant, 
\begin{align}
	-c(t+\pir)A_{\mu\nu}\quad \rightarrow \quad& -\dot{c}(t)\pir\delta_{\mu}^{0}\delta_{\nu}^{0}-c(t)\left(\delta_{\mu}^{0}\partial_{\nu}\pir+\partial_{\mu}\pir\delta_{\nu}^{0}\right)\,, \bigg.\\
	\frac{c(t+\pir)}{2}A_{\alpha\beta}g^{\alpha\beta}g_{\mu\nu}\quad \rightarrow \quad& -\frac{\dot{c}(t)}{2}\pir\bar{g}_{\mu\nu}+\frac{c(t)}{2}\left(\delta g^{00}-2\dpir\right)\bar{g}_{\mu\nu}-\frac{c(t)}{2}\delta g_{\mu\nu},
\end{align}
for the kinetic term,
\begin{align}
	-\frac{M_{2}(t+\pir)}{4}\left(1+A_{\alpha\beta}g^{\alpha\beta}\right)^{2}g_{\mu\nu}\quad \rightarrow \quad&0\,, \bigg.\\
	M_{2}(t+\pir)\left(1+A_{\alpha\beta}g^{\alpha\beta}\right)A_{\mu\nu}\quad \rightarrow \quad&M_{2}(t)\left(\delta g^{00}-2\dpir\right)\delta_{\mu}^{0}\delta_{\nu}^{0},
\end{align}
for the speed of sound contributions and
\begin{align}
    \frac{\Gamma(t + \pir)}{3H}\left(1+A_{\alpha\beta}g^{\alpha\beta}\right)\left(A_{\mu\nu}+ g_{\mu\nu}\right) \quad \rightarrow \quad&  \frac{\Gamma(t)}{3H}\left(\delta g^{00}-2\dpir\right) \left(\delta_{\mu}^{0}\delta_{\nu}^{0} + \bar{g}_{\mu\nu} \right)\,, \bigg.
\end{align}
for the dissipation. 

At last, we use the freedom introduced by the \stuck and the retarded spatial diffs to work in the flat gauge
\begin{equation}
	\delta g_{ij}^{\rm scalar}=0\quad,\quad\delta g_{00}=-2\phi=-\delta g^{00}\quad,\quad \delta g_{0i}=a(t)\partial_{i}F=\delta g^{0i}.
\end{equation}
The perturbed Einstein's equations are given by
\begin{align}
	\frac{\Mpl^{2}}{2}\delta G_{00}+\left[\frac{\Lambda(t)-2M_{2}(t)}{2}\right]\delta g_{00}+3Hc(t)\pir-\left[c(t)+2M_{2}\right]\dot{\pi}_r&=\xi_{00}, \\
	\frac{\Mpl^{2}}{2}\delta G_{0i}+\frac{\Lambda(t)-c(t)}{2}\delta g_{0i}-c(t)\partial_{i}\pir&=\xi_{0i},
\end{align}
as stated in the main text. Note that dissipation does not enter the $00$ and $0i$ Einstein's equations and so does not directly affect the constraints.

\subsection*{Equation of motion for $\pir$}

To derive the equation of motion for $\pir$, we first need to consider all the terms that are linear in $\pia$. To do so, we first perform an advanced \stuck on \Eq{eq:th2app} to reintroduce the $\pia$. It leads to 
\begin{align}
	S_{\mathrm{scalar}}=&\int \dd^4x \sqrt{-g} \bigg[-\dot{\Lambda}(t+\pir)\pia-2c(t+\pir)g^{\mu\nu}\partial_{\mu}\left(t+\pir\right)\partial_{\nu}\pia-\dot{c}(t+\pir)A_{\alpha\beta}g^{\alpha\beta}\pia \nonumber\\
	&+2M_{2}(t+\pir)\left(1+A_{\alpha\beta}g^{\alpha\beta}\right)g^{\mu\nu}\partial_{\mu}\left(t+\pir\right)\partial_{\nu}\pia+\frac{\dot{M}_{2}(t+\pir)}{2}\left(1+A_{\alpha\beta}g^{\alpha\beta}\right)^{2}\pia \nonumber \\
	&+2\Gamma(t+\pir)\left(1+A_{\alpha\beta}g^{\alpha\beta}\right)\pia - \xi_\pi \pia\bigg],
\end{align}
where we defined the scalar noise
\begin{align}
    \xi_\pi &= 2( \dot{\xi}_{00} + 3H \xi_{00} ) - \frac{2}{a^2(t)} \left( \xi_{\parallel}  - 3H \Xi_{\parallel}\right),
\end{align}
by decomposing $\xi_{\mu\nu}$ into 
\begin{align}
    \xi_{00}, \qquad \xi_{0i} =  \partial_i \xi_{\parallel}, \qquad \xi_{ij} = \Xi_{\parallel} \delta_{ij} + \left(\partial_i \partial_i - \frac{1}{3}  \delta_{ij} \partial^2 \right)  \Xi^{\mathrm{TT}}.
\end{align}
This leaves a $\pia$ variation of the form
\begin{align}
	&-\dot{\Lambda}(t+\pir)+2\nabla_{\mu}\left[c(t+\pir)g^{\mu\nu}\partial_{\nu}(t+\pir)\right]-\dot{c}(t+\pir)A_{\alpha\beta}g^{\alpha\beta} \nonumber \\
	&-2\nabla_{\mu}\left[M_{2}(t+\pir)\left(1+A_{\alpha\beta}g^{\alpha\beta}\right)g^{\mu\nu}\partial_{\mu}\left(t+\pir\right)\right]+\frac{\dot{M}_{2}(t+\pir)}{2}\left(1+A_{\alpha\beta}g^{\alpha\beta}\right)^{2} \nonumber \\
	&+2\Gamma(t+\pir)\left(1+A_{\alpha\beta}g^{\alpha\beta}\right)=\xi_{\pi}.
\end{align}
We can then expand all the terms to linear order. The cosmological constant simply becomes
\begin{align}
	-\dot{\Lambda}(t+\pir)\rightarrow&-\ddot{\Lambda}(t)\pir,
\end{align}
while the first part of the kinetic term generates
\begin{align}
	2\nabla_{\mu}\left[c(t+\pir)g^{\mu\nu}\partial_{\nu}(t+\pir)\right]=&2\partial_{\mu}\left[c(t+\pir)\right]g^{\mu\nu}\partial_{\nu}\left(t+\pir\right) \nonumber\\
	&+2c(t+\pir)\partial_{\mu}g^{\mu\nu}\partial_{\nu}\left(t+\pir\right) \nonumber \\
	&+2c(t+\pir)g^{\mu\nu}\partial_{\mu}\partial_{\nu}\pir.
\end{align}
Expanding each term leads to 
\begin{align*}
	2\partial_{\mu}\left[c(t+\pir)\right]g^{\mu\nu}\partial_{\nu}\left(t+\pir\right)\rightarrow&-2\ddot{c}(t)\pir+2\dot{c}(t)\left(\delta g^{00}-2\dot{\pir}\right)\,,\\
	2c(t+\pir)g^{\mu\nu}\partial_{\mu}\left[\text{Log}\left(\sqrt{-g}\right)\right]\partial_{\nu}\left(t+\pir\right)\rightarrow&-6H\dot{c}(t)\pir+6Hc(t)\delta g^{00}-c(t)\partial_{0}\delta g^{00}-6Hc(t)\dot{\pi}_r\,,\\
	2c(t+\pir)\partial_{\mu}g^{\mu\nu}\partial_{\nu}\left(t+\pir\right)\rightarrow& ~2c(t)\partial_{\mu}g^{\mu0}\,,\\
	2c(t+\pir)g^{\mu\nu}\partial_{\mu}\partial_{\nu}\pir\rightarrow&+2c(t)\bar{g}^{\mu\nu}\partial_{\mu}\partial_{\nu}\pir    \,.
\end{align*}
The second part of the kinetic term is much simpler and leads to
\begin{align}
	-\dot{c}(t+\pir)A_{\alpha\beta}g^{\alpha\beta}\rightarrow&~\Ddot{c}(t)\pir-\dot{c}(t)\left(\delta g^{00}-2\dot{\pi}_r\right).
\end{align}
Turning our attention to the speed of sound, the first contribution generates
\begin{align}
	-2\nabla_{\mu}\left[M_{2}(t+\pir)\left(1+A_{\alpha\beta}g^{\alpha\beta}\right)g^{\mu\nu}\partial_{\mu}\left(t+\pir\right)\right]\rightarrow&~\frac{2}{a^{3}(t)}\partial_{0}\left[a^{3}(t)M_{2}(t)\left(\delta g^{00}-2\dot{\pi}_r\right)\right],
\end{align}
which is expanded into
\begin{align}
	\frac{2}{a^{3}(t)}\partial_{0}\left[a^{3}(t)M_{2}(t)\left(\delta g^{00}-2\dot{\pi}_r\right)\right]=\left[6HM_{2}(t)+2\dot{M}_{2}(t)\right]\left(\delta g^{00}-2\dot{\pi}_r\right)+2M_{2}(t)\left(\partial_{0}\delta g^{00}-2\ddot{\pi}_r\right).
\end{align}
The second speed of sound contribution has no linear contribution, such that 
\begin{equation}
	\frac{\dot{M}_{2}(t+\pir)}{2}\left(1+A_{\alpha\beta}g^{\alpha\beta}\right)^{2}\rightarrow 0.
\end{equation}
At last, the dissipation operator generates
\begin{equation}
	2\Gamma(t+\pir)\left(1+A_{\alpha\beta}g^{\alpha\beta}\right)\rightarrow 2\Gamma(t)\left(\delta g^{00}-2\dot{\pi}_r\right).
\end{equation}

Gathering these contributions together, we can now write a first equation as
\begin{equation}\label{eq:pre}
	A\ddot{\pi}_r+B\dot{\pi}_r+C\partial_{i}^{2}\pi_r+D\pir+G_{0}\delta g^{00}+G_{1}\partial_{0}\delta g^{00}+G_{2}\partial_{i}\delta g^{i0}=\xi_{\pi},
\end{equation}
with the self dynamics of the scalar controlled by
\begin{align}
	A=&-2\left[c(t)+2M_{2}(t)\right]\,,\\
	B=&-2\dot{c}(t)-4\Gamma(t)-4\dot{M}_{2}-6Hc(t)-12HM_{2}(t)\,,\\
	C=&~2c(t)\,,\\
	D=&-\ddot{c}(t)-\ddot{\Lambda}(t)-6H\dot{c}(t)\,,
\end{align}
and the mixing with gravity by
\begin{align}
	G_{0}=&~\dot{c}(t)+2\Gamma(t)+2\dot{M}_{2}(t)+6Hc(t)+6HM_{2}(t)\,,\\
	G_{1}=&~c(t)+2M_{2}(t)\,,\\
	G_{2}=&~2c(t)\,.
\end{align}
Dividing \Eq{eq:pre} by $A$, we find the main text result presented in \Eq{eq:pireomexp}.



%
\bibliographystyle{JHEP}
\bibliography{biblio}

\end{document}